\RequirePackage{fix-cm}
\PassOptionsToPackage{dvipsnames}{xcolor}
\documentclass[iicol,sn-basic]{sn-jnl}

\catcode`\|=12\relax 

\usepackage{hyperref}       
\usepackage{url}            
\usepackage{booktabs}       
\usepackage{amsfonts}       
\usepackage{multirow, makecell}

\usepackage{amsmath}
\usepackage{amsthm}
\usepackage{amssymb}

\usepackage{siunitx}
\usepackage{array}
\newcolumntype{P}[1]{>{\centering\arraybackslash}p{#1}}
\newcolumntype{M}[1]{>{\centering\arraybackslash}m{#1}}
\usepackage{tabularx}
\usepackage{diagbox}

\usepackage{mathtools}
\usepackage{graphicx}
\usepackage{color,soul}
\usepackage{subcaption}

\usepackage{algorithm}
\usepackage{listings}
\usepackage{xcolor}

\usepackage{textgreek}

\definecolor{ao(english)}{rgb}{0.0, 0.5, 0.0}
\definecolor{electricpurple}{rgb}{0.75, 0.0, 1.0}

\newcommand{\rone}[1]{#1}
\newcommand{\rtwo}[1]{#1}
\newcommand{\rme}[1]{#1}

\usepackage[textsize=tiny]{todonotes}

\definecolor{codegreen}{rgb}{0,0.6,0}
\definecolor{codegray}{rgb}{0.5,0.5,0.5}
\definecolor{codepurple}{rgb}{0.58,0,0.82}
\definecolor{backcolour}{rgb}{0.95,0.95,0.92}

\lstdefinestyle{mystyle}{
    backgroundcolor=\color{backcolour},   
    commentstyle=\color{codegreen},
    keywordstyle=\color{magenta},
    numberstyle=\tiny\color{codegray},
    stringstyle=\color{codepurple},
    basicstyle=\ttfamily\footnotesize,
    breakatwhitespace=false,         
    breaklines=true,                 
    captionpos=b,                    
    keepspaces=true,                 
    numbers=left,                    
    numbersep=5pt,                  
    showspaces=false,                
    showstringspaces=false,
    showtabs=false,                  
    tabsize=2
}

\usepackage{multicol}
\usepackage{bm}
\usepackage{tikz}
\usetikzlibrary{patterns, patterns.meta}
\usetikzlibrary{fadings}
\usepackage{tikz-3dplot}
\usetikzlibrary{calc}

\makeatletter
\DeclareRobustCommand{\pder}[1]{%
  \@ifnextchar\bgroup{\@pder{#1}}{\@pder{}{#1}}}
\newcommand{\@pder}[2]{\frac{\partial#1}{\partial#2}}
\makeatother

\makeatletter
\DeclareRobustCommand{\der}[1]{%
  \@ifnextchar\bgroup{\@der{#1}}{\@der{}{#1}}}
\newcommand{\@der}[2]{\frac{d#1}{d#2}}
\makeatother

\newcommand{\mb}[1]{\mathbf{#1}}

\newcommand*\diff{\mathop{}\!\mathrm{d}}

\newcommand{\mrm}[1]{\mathrm{#1}}

\begin{document}
\small

\title[optimization]{Topology Optimization for the Full-Cell Design of Porous Electrodes in Electrochemical Energy Storage Devices}
\author[1]{Hanyu Li}
\author[1]{Giovanna Bucci}
\author[1]{Nicholas W. Brady}
\author[1]{Nicholas R. Cross}
\author[1]{Victoria M. Ehlinger}
\author[1]{Tiras Y. Lin}
\author[1]{Miguel Salazar de Troya}
\author[1]{Daniel Tortorelli}
\author[1]{Marcus A. Worsley}
\author*[1]{Thomas Roy}\email{roy27@llnl.gov}

\affil[1]{Lawrence Livermore National Laboratory - 7000 East Ave, Livermore, CA 94550, USA}

\raggedbottom

\abstract{In this paper, we introduce a density-based topology optimization framework to design porous electrodes for maximum energy storage. We simulate the full cell with a model that incorporates electronic potential, ionic potential, and electrolyte concentration. The system consists of three materials, namely pure liquid electrolyte and the porous solids of the anode and cathode, for which we determine the optimal placement.
We use separate electronic potentials to model each electrode, which allows interdigitated designs.
\rtwo{As a result}, a penalization is required to ensure that the anode and cathode do not touch, i.e., causing a short circuit.
We compare multiple 2D designs generated for different fixed conditions, e.g. material properties. A 3D design with complex channel and interlocked structure is also created. All optimized designs are far superior to the traditional monolithic electrode design with respect to energy storage metrics. We observe up to a 750\% increase in energy storage for cases with slow effective ionic diffusion within the porous electrode.}


\maketitle

\section{Introduction}
Electrochemical energy storage devices provide a shift away from fossil fuels by enabling electric vehicles and supporting the adoption of intermittent renewable energy sources \citep{Chu_2012, Chu_2016, Gur_2018}.
Batteries and capacitors are examples of such devices that are ubiquitous in modern technologies and improving their performance is crucial for the green energy transition.

Electrochemical charge storage mechanisms can be broadly grouped into two types: charge separation and Faradaic charge-transfer reactions \citep{simon2014batteries}.
In the former, electrostatic double-layer capacitors (EDLCs) store energy by forming an electric double layer at the interface between the electrodes and electrolyte.
This differs from the latter wherein electrons are transferred via reduction and oxidation (i.e., \emph{redox}) reactions.
Such reactions are the principal energy storage mechanism of batteries, which utilize slow kinetics and therefore lower power density but enable much higher energy density than EDLCs.
As a middle ground, pseudocapacitors use fast redox kinetics, resulting in smaller energy density than batteries but larger than EDLCs, while also demonstrating high power density ~\citep{simon2014batteries}.

\begin{figure}[htb!]
    \centering
    \begin{tikzpicture}[>=latex,scale=0.72]
        \draw[fill=blue!15] (0,0) -- (8,0) -- (8,4) -- (0,4) -- (0,0);
        \draw[preaction={fill, red!15}, pattern=dots, pattern color=red!40] (0,4) -- (0,2.5) -- (8,2.5) -- (8,4) -- (0,4);
        \draw[preaction={fill, red!15}, pattern=dots, pattern color=red!40] (0,0) -- (0,1.5) -- (8,1.5) -- (8,0) -- (0,0);
        \draw[line width=1.2pt] (0,0) -- (8,0);
        \node (a) at (4.,-0.25) {\small Current collector (positive)};
        \draw[line width=1.2pt] (0,4) -- (8,4);
        \node (b) at (4.,4.25) {\small Current collector (negative)};
        \node at (4,3.25) {\small Cathode};
        \node at (4,2) {\small Electrolyte/separator};
        \node at (4,0.75) {\small Anode};
        \draw[dashed] (7,0) -- (7,-0.5) -- (9.5,-0.5) -- (9.5,4.5) -- (7,4.5) -- (7,4);
        \draw[fill=white] (8.5,1) -- (10.5,1) -- (10.5,3) -- (8.5,3) -- (8.5,1);
        \node at (9.5, 2) {\begin{tabular}{c}
            \small Power \\
            \small source / \\ 
            \small load
        \end{tabular}};
        \node at (8.8,4.2) {$e^-$};
        \node at (8.8,-0.2) {$e^-$};
    \end{tikzpicture}
    \caption{Traditional monolithic electrochemical energy storage device.}
    \label{fig:diag}
\end{figure}
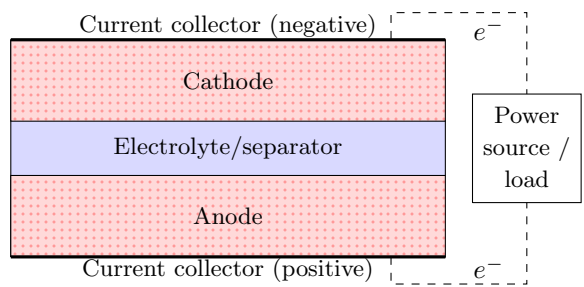

Batteries and capacitors share a common design: two porous electrodes, namely the anode and the cathode, are immersed in an electrolyte \rone{(either liquid or solid)} and separated by a membrane or simply an electrolyte-filled gap.
The traditional monolithic porous electrode cell illustrated in Fig. \ref{fig:diag} consists of two slabs with a uniform porosity and surface area to volume ratio.
The electrodes are typically assemblies of micron-scale conductive particles.
Electrochemical reactions occur over the interface between the electrode-particles and the electrolyte.
As such, cell-design criteria primarily focus on maximizing the surface area to volume ratio while minimizing resistances related to electron or ion transport \citep{wang20083d} to enable higher currents, power outputs, and energy storage.

Simply maximizing electrochemical surface area in the attempt to increase energy density often deteriorates ionic transport within the electrode by replacing the pore network for ion transport with an extremely tortuous path \citep{wang20083d}.
Such paths restrict ionic penetration into the monolithic electrodes and thus reduce their effectiveness. 
Fortunately, promising alternatives to these monolithic designs are emerging; e.g. electrode designs with geometric features, multiple lengthscales and/or materials, and spatially varying porosity \citep{beck_inertially_2021, trembacki2019, zekoll2018, gao2023stable, yang2022gradient}. 
These new designs are made possible thanks to new advanced manufacturing processes \citep{chu2021printing, soares2021AM, chandrasekaran2023aerogels}.
It is this nearly complete control over the electrode design that raises the question of what these designs should be.

One approach to increase the energy density consists in increasing the thickness of conventional slab electrodes and reducing the number of cells needed in a stack. This approach reduces the weight associated with non-active materials such as current collectors, foils and separators \citep{heubner2021tool}.
However, electrode areal capacity is limited by the ion transport within the tortuous electrode structure.
Typical commercial electrodes have a thickness in the range of 100-150 \unit{\micro\meter} \citep{heubner2021tool, xu2019thickness, kuang2019thickreview} \rone{and typical Li-ion battery electrodes have a thickness in the range of 50-100 \unit{\micro\meter} \citep{zhu2022architecture}.}
Further increasing the electrode thickness (and the areal capacity) leads to diminishing returns in volumetric energy density for charge/discharge rates relevant to electric-vehicle applications~\citep{singh2015thick,zheng2012comprehensive,gallagher2015optimizing}.
\rone{As a result, designing thick electrodes with shapes less impacted by ionic transport becomes critical.}

Thicker electrodes with three-dimensional features, in contrast to conventional slabs, have been previously fabricated and tested~\citep{long2004three,long20203d}, demonstrating the benefits of providing structure and low-tortuosity transport pathways that facilitate ion transport in the electrolyte.
A categorization of the various three-dimensional designs has been proposed by \citet{hung2022three}: 1) the electrodes may be individually shaped and separated by a planar separator, which is either solid or porous, or 2) completely intertwined with one another. 
In particular, an interdigitated design, where alternating planar fins or rods protruding from the cathode and anode separated by the electrolyte, has demonstrated increased power and energy performance \citep{hung2021modeling,zadin2011finite}.
Similar three-dimensional interpenetrating structures have been investigated for microbatteries and wearable device applications \citep{lyu2021, zeng2014wear, wang2017fiber}.
\rone{Architected electrodes have also been applied to other electrochemical devices. In recent studies, so-called ``shaped electrodes" exhibited higher performance than standard electrodes, for example in the case of fuel cells with grooved electrodes \citep{lee2023grooved} and water electrolyzers with gradient tapered array electrodes \citep{dong2022overall}.}
The non-planar structures improve performance via increased surface area to volume ratios to promote reaction and increased ionic diffusion through complex pathways to promote mass transport \citep{trembacki2019, chen2020hole}, therefore enabling high power and fast charging lithium-ion batteries \citep{chen2020hole, zekoll2018}.
However, because the current distribution and concentration profiles in these interdigitated and interpenetrating electrodes are highly non-uniform \citep{hart20033, trembacki2019}, their performance becomes strongly dependent on the electrode and electrolyte properties, e.g. specific surface area and conductivity, thus motivating the need for a systematic design optimization paradigm.

Computational techniques such as topology optimization offer a promising opportunity to design optimal porous electrodes.
Density-based topology optimization was initially formulated as a mass distribution problem in which the volume fraction field is optimized to maximize the stiffness of a linear elastic structure subject to a mass constraint \cite{bendsoe}.
Topology optimization has since been adapted to design electrochemical devices.
\cite{yaji2018topology, chen2019computational, lin2022topology} design the channels that transport the electrolyte fluid to the porous electrodes in redox flow batteries.
Similarly, \cite{behrou2019topology, wang2023enhancing, qu2023design} design channels in fuel cells to minimize electrical and pumping power losses while maximizing power density.
Heterogeneous porosity fields have also been optimized for lithium-ion batteries \citep{ramadesigan2010optimal, golmon2012multiscale, xue2016lithium}, and redox flow batteries \citep{Beck_2021}.

Topology optimization has been used in \cite{roy2022tope} to design redox porous electrodes and EDLC electrodes.
This study, which only considers a half cell, i.e., a single electrode, generates designs for a wide range of fixed dimensionless groups encapsulating material parameters, electrode length scale, and operating conditions.
The best performance improvement over monolithic designs was achieved when ionic transport within the porous material is slow. 
\cite{batista2023design} used a similar optimization framework to design a supercapacitor electrode, which was then fabricated using additive manufacturing.
The optimized electrode exhibited nearly twice the volumetric capacitance over a lattice structured electrode.
\rme{Partly inspired by the topology optimization results, a recent study by \cite{lin2024shape} investigates the conditions under which electrode shapes have the largest impact on the performance.}
The models used to predict the response in these previous studies assumed no concentration gradients; the model used in a more recent optimization study by \cite{alizadeh2023numerical} included mass accumulation and entropy effects.
In this work, we optimize both electrodes simultaneously using a full-cell model that includes mass transport. 

A typical feature with density-based topology optimization is the use of the material volume fraction field as the control.
As such, regions in the optimized designs contain mixtures of materials, which is not possible, i.e., we must have so-called $0-1$ designs in which a distinct material resides at each location.
The size of these mixed ``intermediate'' material regions is minimized by using clever penalization schemes in the topology optimization problem formulation.
For structural optimization, the penalization is relatively straightforward because of the explicit trade-off between the structural stiffness and mass cf. e.g. the Simple Isotropic Material with Penalization (SIMP) \citep{bendsoe}.
For electrochemical energy storage systems, the trade-off between ion/electron transport and chemical activity is less obvious, thus special care is required when devising penalization schemes.
Schemes that concurrently increase ohmic loss and reduce energy storage in intermediate material regions have proven effective \citep{roy2022tope,batista2023design}.
Developing effective penalization schemes for topology optimization can be very tedious.
However those that conquer this battle are rewarded with a very effective design tool.

In this manuscript, we use topology optimization to design full-cell electrochemical energy storage devices.
In Sect.~\ref{sec:topopt}, we review topology optimization concepts, and describe the boundary propagation method required for the two-electrode optimization.
Next, in Sect.~\ref{sec:govern}, we introduce the simulation model, specifically the Nernst-Planck equation for a binary electrolyte with redox reactions and electrostatic double layer capacitance.
Implementation and solver details are provided in Sect.~\ref{sec:implementation}.
Following that, in Sect.~\ref{sec:results}, the optimization problem is formulated and solved to generate designs for various key dimensionless groups encapsulating device length scale and material parameters such as conductivity and reaction rate.
We also compare the performances of two optimized designs against traditional monolithic electrodes to demonstrate the effectiveness of the optimization.
Lastly, we showcase a 3D design featuring complex channels and interlocked features.
We conclude in Sect.~\ref{sec:conclusion} with some final discussion.


\section{Topology optimization}
\label{sec:topopt}
\begin{figure}[htb!]
    \centering
    \begin{tikzpicture}[>=latex,scale=0.82]
        \draw[fill=blue!15] (0,0) -- (8,0) -- (8,4) -- (0,4) -- (0,0);
        \draw[preaction={fill, red!15}, pattern=dots, pattern color=red!40] (0,4) -- plot [smooth] coordinates {(0,3) (1.5,3.7) (2,2.5) (3,3.1) (3.4, 3.6) (4,2.4) (4.8,2.8) (5.5,2.3) (6.2,2.9) (8,3)} -- (8,4);
        \draw[preaction={fill, red!15}, pattern=dots, pattern color=red!40] (0,0) -- plot [smooth] coordinates {(0,1.5) (1.1,2.4) (1.9,1) (2.8,1.6) (3.4, 2.1) (4,0.9) (4.8,1.3) (5.5,0.8) (6.2,1.4) (8,1.5)} -- (8,0);
        \draw[line width=1.2pt] (0,0) -- (8,0);
        \node (a) at (4.,-0.25) {\small Current collector $\Gamma_\mathrm{a}$};
        \draw[line width=1.2pt] (0,4) -- (8,4);
        \node (b) at (4.,4.25) {\small Current collector $\Gamma_\mathrm{c}$};
        \node at (5.2,3.5) {\small Cathode ($\chi=1$)};
        \node[align=left] at (5,1.8) {\small Electrolyte/ \\ Separator};
        \node at (6.8,2) {\small ($\chi=0$)};
        \node at (3.4,0.5) {\small Anode ($\chi=1$)};
    \end{tikzpicture}
    \caption{Two-dimensional design space of a full cell with porous electrodes.}
    \label{fig:designspace2D}
\end{figure}
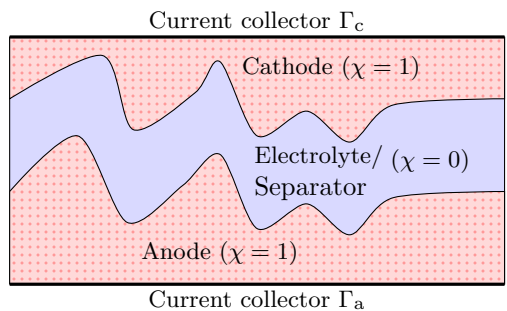
\noindent
Our topology optimization problem concerns the material indicator field $\chi$ in cell domain $\Omega$ defined such that $\chi(\mb{x}) = 1$ indicates that the solid porous electrode material occupies $\mb{x}$, whereas $\chi(\mb{x})=0$ indicates that the separator or pure electrolyte occupies $\mb{x}$, cf. Fig.~\ref{fig:designspace2D}.
Over this domain we solve the model equations $F(\chi,\Phi_1, \Phi_2, c)=0$ to be introduced in Sect.~\ref{sec:govern}, for the electronic potential $\Phi_1$, ionic potential $\Phi_2$, and salt concentration $c$. 
These pieces are used to define the porous electrodes topology optimization problem, i.e.
\begin{equation}
\label{eq:original_opt_problem}
\begin{aligned}
    &\min_{\chi\colon\Omega\to\{0,1\}} \theta_0(\chi) = \int_{\Omega} \pi(\chi,\Phi_1, \Phi_2,c) \diff V, \\
    &\text{s.t. }\theta_i(\chi) = \int_{\Omega} g_i(\chi,\Phi_1, \Phi_2,c) \diff V \leq 0,\; i=1,\dots,m,\\ 
    &\text{and}\ F(\chi,\Phi_1, \Phi_2,c) = 0.
\end{aligned}
\end{equation}
Here, $\theta_0$ is the cost function to be minimized and $\theta_i$ are the $m$ design constraint functions, with local functions $\pi$ and $g_i$ to be defined in Sect.~\ref{sec:optproblem}.

Since the electrode phase is comprised of separate anode and cathode materials, a three-phase (three-material) optimization formulation would be natural \citep{sigmund1997design, watts2016n}.
This three-phase approach requires a second material indicator field that would delineate the anode and cathode phases.
However, since the cathode and anode are in contact with different boundaries $\Gamma_\mathrm{c}$ and $\Gamma_\mathrm{a}$ (see Fig.~\ref{fig:designspace2D}), we instead use a boundary propagation method \citep{sa2022bdprop} for this delineation, cf. Sect.~\ref{sec:bdprop}.
Thus, we optimize with respect to a single field $\chi$.

In the context of density-based topology optimization \citep{bendsoe}, the discrete nature of the design field $\chi$ prevents the application of gradient-based optimization algorithms.
Because of this, and at the risk of not obtaining $0-1$ designs, we replace $\chi \colon \Omega\to \{0,1\}$ with a continuous volume fraction variable $\rho \colon \Omega\to [0,1]$.
Additionally, special care is required to ensure that the optimization problem is well-posed and produces a $0-1$ design.
These concerns are remedied using filtering detailed in Sect.~\ref{sec:filter}, and material penalization described in Sect.~\ref{sec:optproblem}.

\subsection{Filtering}
\label{sec:filter}
Topology optimization problems are often ill-posed.
Their ``solutions'' are characterized by non-converging sequences of structures with highly oscillatory material distributions.
One option to obtain a well-posed problem imposes a feature length scale on the design by filtering $\rho$ to obtain the smoothed field $\tilde\rho$, which is used to define the structures of material distribution \citep{bruns2001, bourdin2001}.
In our problem, we use the diffusion-reaction PDE filter described in \cite{lazarov2011filters} wherein $\tilde\rho$ solves
\begin{equation}
    \label{eq:filter}
    \begin{aligned}
        -r^2\nabla^2 \tilde{\rho}+\tilde{\rho} &= \rho \quad &&\text{in }\Omega, \\
        r^2\nabla \tilde{\rho} \cdot \mb{n} &= 0 \quad &&\text{on }\partial\Omega,
    \end{aligned}
\end{equation}
for the given design $\rho$.
Fine scale oscillations in $\rho$ are mollified in $\tilde\rho$; more mollification occurs as the filter radius $r$ increases.

\begin{figure}[!htbp]
    \centering
    \includegraphics[width=\columnwidth]{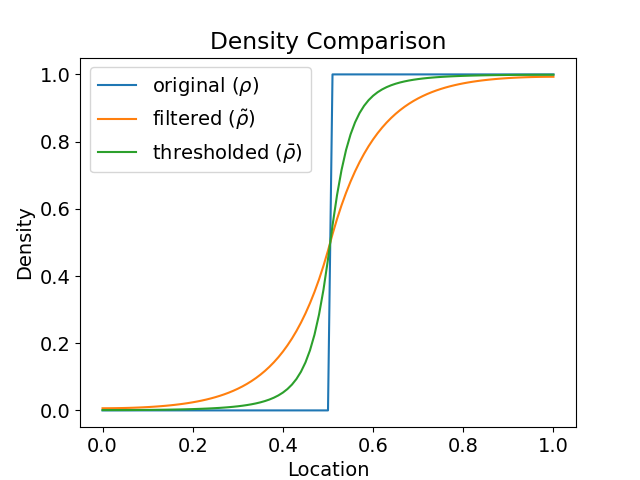}
    \caption{Comparison between original, filtered, and thresholded densities.}
    \label{fig:density_comp}
\end{figure}
Unfortunately, the filtered design $\tilde\rho$ contains large transitional regions where $\tilde\rho\neq \{0,1\}$ and our original design problem only allows for $\chi(\mb{x})\in\{0,1\}$ regions, i.e., $0-1$ designs.
To remedy this, the filtered distribution $\tilde{\rho}$ is thresholded to define the field $\bar\rho$ such that $\bar\rho(\mb{x}) = \{0,1\}$ throughout $\Omega$.
We can do this by equating $\bar\rho = H(\tilde\rho - 0.5)$ where $H$ is the unit Heaviside function.
However, the lack of differentiability of $H$ is not conducive to our gradient-based optimization.
As such, we replace $H$ with a smoothed approximation so that
\begin{equation}
    \bar{\rho} = H_{b,k}(\tilde\rho) = \frac{\tanh(b k) + \tanh\left(b (\tilde{\rho} - k)\right)}{\tanh(bk) + \tanh\left(b (1 - k)\right)}.
\end{equation}
Here, $b$ controls the sharpness of the $0-1$ transition and $k$ controls the cutoff such that
\begin{equation*}
    \lim_{b \rightarrow \infty} H_{b,k}(\tilde\rho) = H(\tilde\rho - k)
\end{equation*}
\citep{guest2004achieving,sigmund2007morphology,sigmund2013topology,xu2010volume}.
In this way, we obtain ``nearly'' $0-1$ designs such that $\bar\rho(\mb{x}) \approx \{0,1\}$ throughout $\Omega$.
\rtwo{The difference between $\rho$, $\tilde{\rho}$, and $\bar{\rho}$ is illustrated in Fig. \ref{fig:density_comp},} wherein we use $b = 4$ and $k=0.5$ for a mild transformation.
Although mild, it combines well with the continuous boundary propagation method to identify the extent of the cathode and anode regions.
Aggressive thresholding, i.e., larger $b$ values, leads to premature convergence to poor local optima.
Because of this, continuation techniques on $b$ are often employed to improve optimization convergence and obtain nearly $0-1$ designs.
In this work, we do not use continuation approaches, since the combination of mild thresholding and the interpolation described in \ref{sec:optproblem} lead to nearly $0-1$ designs.

\subsection{Continuous boundary propagation}
\label{sec:bdprop}
\begin{figure}[!htbp]
    \centering
    \includegraphics[width=\columnwidth]{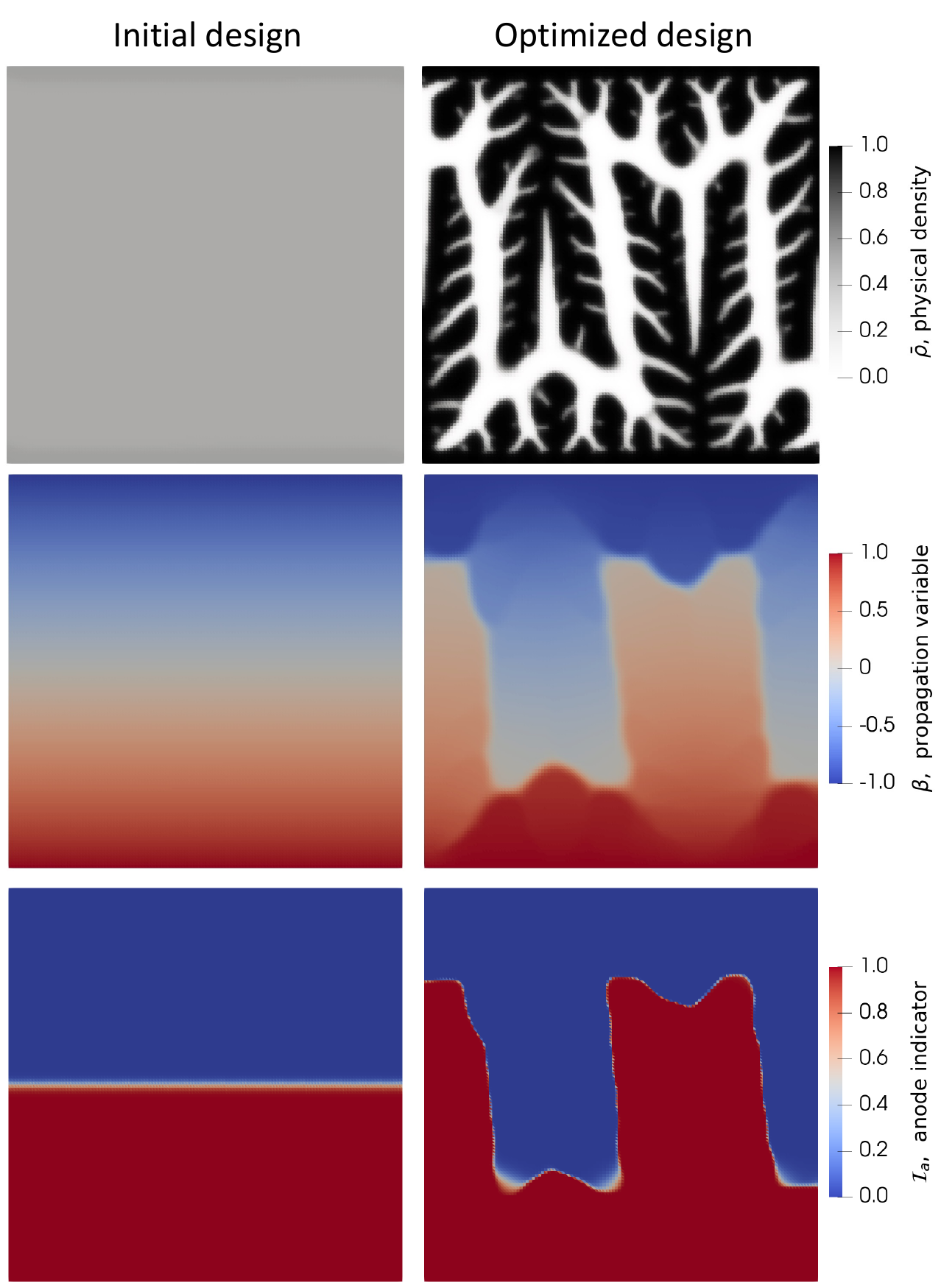}
    \caption{Physical density $\bar{\rho}$ (top) with its corresponding $\beta$ (middle) and anode indicator $\mathcal{I}_a$ (bottom) for the initial design (left) and optimized design (right).}
    \label{fig:electrode_indicator}
\end{figure}
\noindent
We can use $\bar\rho(\mb{x})=\{0,1\}$ to determine if electrolyte or electrode resides at $\mb{x}$, but we do not know if the electrode is the anode or cathode.
This is important because the anode and cathode may be manufactured with different materials and because we need to separate the anode and cathode regions to avoid a short circuit.
Both of these concerns are resolved by using the continuous boundary propagation approach \citep{sa2022bdprop}.

The boundary propagation approach solves the diffusion-reaction problem cf. Fig. \ref{fig:electrode_indicator} for $\beta$:
\begin{equation}
    \label{eq:bd_prop}
    \begin{aligned}
        \bar{\rho} \nabla\cdot\nabla\beta - (1-\bar{\rho})\beta &= 0 \quad &&\text{in } \Omega, \\
        \beta &= 1 \quad &&\text{on }\Gamma_\mathrm{a}, \\
        \beta &= -1 \quad &&\text{on }\Gamma_\mathrm{c}, \\
        \nabla\beta\cdot\mb{n} &= 0 \quad &&\text{on }\partial\Omega/\Gamma_\mathrm{c}\cup\Gamma_\mathrm{a}.
    \end{aligned}
\end{equation}
In electrode regions ($\bar{\rho} = 1$), the equation reduces to $\nabla\cdot\nabla\beta=0$, and in electrolyte regions ($\bar{\rho} = 0$), it reduces to $\beta = 0$.
By prescribing $\beta = 1$ and $\beta = -1$ over the anode and cathode current collector boundaries, namely $\Gamma_a$ and $\Gamma_c$, we are assured that $0\leq \beta \leq 1$ in the anode and $-1 \leq \beta \leq 0$ in the cathode.
The resulting $\beta$ is transformed into differentiable indicator functions for the anode $\mathcal{I}_\mathrm{a}$ and cathode $\mathcal{I}_\mathrm{c}$ regions such that
\begin{equation}
\label{eq:indicators}
\begin{aligned}
    \mathcal{I}_\mathrm{a}(\beta) =& H_{b,k}\left(\frac{\beta + 1}{2}\right), \\
    \mathcal{I}_\mathrm{c}(\beta) =& H_{b,k}\left(\frac{-\beta + 1}{2}\right),
\end{aligned}
\end{equation}
with $b=100$ and $k=0.5$.

Examples of $\bar{\rho}$ distributions, their corresponding $\beta$ propagation, and anode indicator are illustrated in Fig. \ref{fig:electrode_indicator}. 
To summarize, $\bar{\rho}$ distinguishes between electrode and electrolyte regions, and
$\beta$ produces the indicator fields $\mathcal{I}_{\mrm{a}}$ and $\mathcal{I}_{\mrm{c}}$ to distinguish between anode and cathode regions.
The indicator fields are also used to ensure model consistency as discussed in Sect.~\ref{sec:separate_phi1}.

\section{Governing equations}
\label{sec:govern}
In this section, we introduce the governing equations to model both redox and capacitive charging effects in electrochemical energy storage devices with \rone{a liquid electrolyte} \citep{johnson1971desalt,doyle1993galvano,newman1962analysis,newman1975theory}.
Extending the two-potential system described in our previous study \citep{roy2022tope}, we solve the Nernst--Planck equation for the ion concentration fields, which affect the charge-transfer and adsorption kinetics, and the electrolyte conductivity.
Therefore, accounting for their mass transport improves the predictability of our designs.
Battery models may also include ionic diffusion in the solid particles \citep{doyle1993galvano, brosa2022continuum}, however, for simplicity we ignore solid diffusion, which is a reasonable assumption when the electrode particles are of the order of 1 $\mu$m in size. 
\rtwo{Many battery models also have separate void fractions for active and binder materials, however we assume a single solid material.}
Additionally, we assume that the anode and cathode have identical \rtwo{porosity and electronic conductivity.}

\subsection{Physical model}
We consider a binary electrolyte in which the cation participates in the redox reaction that produces current density $i_n$.
In the battery, this reaction with cation $A^{z_+}$ and electron $e^-$ has the form 
\begin{equation}
    s_+ A^{z_+}+ne^- \rightleftharpoons s_+ A,
\end{equation}
where $n$ is the number of electrons transferred, $z_+$ is the cation charge number, and $s_+=n/z_+$ is the cation stoichiometric number.
\rtwo{For the sake of simplicity, we neglect any side reactions in our model.}
Both cations and anions are adsorbed by double-layer charging, which produces current density $i_c$.
The Nernst--Planck equation for the cation ($c_+$) and anion ($c_-$) species concentrations are as follows:
\begin{equation}
  \label{eq:species_constitutive}
  \begin{aligned}
      \frac{\partial \left(\epsilon c_+\right)}{\partial t} =& - \nabla\cdot \mb{N}_+ + \frac{dq_+}{dq}\frac{1}{z_+ F}a i_c + \frac{s_+}{n F} a i_n &&\text{in }\Omega, \\
      \frac{\partial \left(\epsilon c_-\right)}{\partial t} =& - \nabla\cdot \mb{N}_- + \frac{dq_-}{dq}\frac{1}{z_- F}a i_c &&\text{in }\Omega,
  \end{aligned}
\end{equation}
where $\mathbf{N}_i$ is the ionic flux for species $\textit{i}=\pm$, $\epsilon$ is porosity, $F$ is the Faraday constant (\qty{96485}{\coulomb\per\mole}), and $a$ is the specific surface area, which is zero in the liquid electrolyte region. 
The variation in the amount of adsorbed cations/anions is captured by $\dfrac{dq_i}{dq}$, where $q_i$ is the adsorbed cation/anion surface charge density in the double layer, and $q$ is the total surface charge density in the double layer.
Assuming the absence of cosorption of anions and cations, $\frac{dq_+}{dq}$ = 1 and $\frac{dq_-}{dq}$ = 0 for a cation responsive electrode, and $\frac{dq_+}{dq}$ = 0 and $\frac{dq_-}{dq}$= 1 for an anion responsive electrode \citep{johnson1971desalt}.

We now define constitutive equations for $i_n$, $i_c$, and $\mb{N}_i$.
The redox and capacitive current densities are defined as
\begin{equation}
\begin{aligned}
    i_n &= i_0 \left[\exp\left(\frac{\alpha_a F}{RT} \eta \right) - \exp\left(\frac{-\alpha_c F}{RT} \eta \right)\right],\\
    i_c &= C_d \frac{\partial \left(\Phi_1 - \Phi_2\right)}{\partial t},
\end{aligned}
\end{equation}
where $R$ (\qty{8.3145}{J.K^{-1}.mol^{-1}}) is the ideal gas constant, $T$ is the temperature, $C_d$ is the specific capacitance, and $\alpha_a$ and $\alpha_c$ are the anodic and cathodic charge transfer coefficients, respectively.
For simplicity, we assume $\alpha=\alpha_a = \alpha_c$.
\rone{We model the exchange current density to be a function of concentration as} $i_0 = F k_{\mrm{rxn}} c^{\alpha}$, where $k_{\mrm{rxn}}$ is the reaction rate and $c$ is the salt concentration defined below.
The overpotential $\eta = \Phi_1 - \Phi_2 - U_0$ is  obtained from the electronic potential $\Phi_1$, the ionic potential $\Phi_2$, and the equilibrium potential $U_0$,
\rtwo{which is defined with respect to a reference material, e.g. the anode material, making $U_0$ the difference between the cathode and the anode equilibrium potentials.}
Note that battery models typically treat $U_0$ as a function of solid surface concentration \citep{doyle1993galvano}; for simplicity, we assume $U_0$ is constant.
Therefore, the Nernstian loss, which is defined by the change in the equilibrium potential as a function of the concentration, is neglected.
To account for the effect of concentration on ion adsorption, we define $C_d = C_{d,0} \left(\frac{c}{c_0}\right)^{\alpha}$, where $C_{d,0}$ is the reference specific capacitance and $c_0$ is the initial concentration.
This nonstandard addition to the model prevents nonphysical extreme capacitive current density in the designs where high reaction rates could lead to ion depletion.

The ionic flux for an individual species $c_i$ is calculated by
\begin{equation}
  \label{eq:ionic_flux}
  \mb{N}_i = -z_i u_i F c_i \nabla \Phi_2 - D_i \nabla c_i.
\end{equation}
In the above, $u_i$ is the effective ion mobility and $D_i$ is the effective molecular diffusion coefficient of the species, which satisfy the Nernst-Einstein relationship
\begin{equation}
  \label{eq:nernst_einstein}
  D_i = RT u_i.
\end{equation}
The stoichiometric coefficient and ion charge number are related by
\begin{equation}
    \label{eq:stoichiometry}
    z_i = \frac{n}{s_i}.
\end{equation}

Both relations \eqref{eq:nernst_einstein} and \eqref{eq:stoichiometry} are useful for nondimensionalization of the governing system.
Assuming electroneutrality, we have the constraint
\begin{equation}
  \label{eq:electroneutrality}
  \sum_i z_i c_i = 0 ,
\end{equation}
which is also equivalent to
\begin{equation}
  \label{eq:neutrality_equ}
  \sum_i z_i \nu_i = 0,
\end{equation}
where $\nu_i$ is the number of moles of cations or anions per mole of electrolyte. For a binary electrolyte, we can then define the salt concentration as
\begin{equation}
  \label{eq:concentration}
  c=\frac{c_+}{\nu_+}=\frac{c_-}{\nu_-} .
\end{equation}

Considering electroneutrality, equations \eqref{eq:species_constitutive} are combined into one conservation equation for the salt concentration (see \cite{newman2012electrochemical} for details).
Combining that with the charge conservation in the electronic and ionic potentials, we obtain the governing field equations that solve for $\Phi_1$, $\Phi_2$, and $c$ \citep{newman1975theory}:
\begin{subequations}
  \label{eq:govern_system}
  \begin{align}
    -\nabla\cdot (\sigma \nabla \Phi_1 ) &= -a\left(i_n + i_c\right) &&\text{in }\Omega, \label{eq:electric}\\
    -\nabla\cdot (\kappa \nabla \Phi_2) &- z_+ \nu_+ F\nabla\cdot \left((D_{+} - D_{-}) \nabla c\right) \nonumber &&\\& = a\left(i_n + i_c\right) &&\text{in }\Omega, \label{eq:ionic}\\
    \frac{\partial (\epsilon c)}{\partial t} - \nabla\cdot (&D \nabla c) = \frac{s_+}{F n \nu_+} t_{-} a i_n \nonumber &&\\+ \frac{1}{F z_+ \nu_+} &\left(t_{-} \frac{dq_+}{dq} - t_{+} \frac{dq_-}{dq}\right)a i_c &&\text{in }\Omega, \label{eq:species}
  \end{align}
  subject to the boundary and initial conditions:
  \begin{align}
    \Phi_1 &= 0 \quad &&\text{on }\Gamma_\mathrm{a}, \label{eq:bd1}\\
    \Phi_1 &= \zeta t  \quad &&\text{on }\Gamma_\mathrm{c}, \label{eq:bd2}\\
    \nabla \Phi_1 \cdot \mb{n} &=0\quad &&\text{on }\partial \Omega \setminus (\Gamma_\mathrm{a}\cup\Gamma_\mathrm{c}) \label{eq:bd3}, \\
    \nabla \Phi_2 \cdot \mb{n} &=0\quad &&\text{on }\partial \Omega, \label{eq:bd4} \\
    \nabla c \cdot \mb{n} &= 0 \quad  &&\text{on }\partial\Omega, \label{eq:bd5} \\
    \Phi_1(\mb{x},t=0) &= 0 \quad &&\text{in } \Omega, \label{eq:ic1} \\
    \Phi_2(\mb{x},t=0) &= 0 \quad &&\text{in } \Omega, \label{eq:ic2} \\
    c(\mb{x},t=0) &= c_0 \quad &&\text{in } \Omega, \label{eq:ic3}
  \end{align}
\end{subequations}
for $t\in (0,T_{\mrm{f}}]$, where $\sigma$ is the effective electronic conductivity, $\kappa$ is the effective ionic conductivity, and $D$ is the effective salt diffusion coefficient.
The dimensionless quantity $t_+$ denotes the transference number of the cation; the transference number of the anion is related by $t_- = 1 - t_+$.
The boundary conditions describe a linearly increasing potential in time on the cathode current collector, with scan rate $\zeta$.

\rone{These boundary conditions correspond to a linear sweep voltammetry experiment. While this is not a typical operating condition for a battery, the choice was motivated by the fact that a voltage-controlled system is more conducive to topology optimization. Firstly, the terminal analysis time $T_{\mrm{f}}$ of the simulation is fixed for each design, whereas in a galvanostatic simulation, $T_{\mrm{f}}$ is the time required to reach a voltage limit, and it depends on the physical response of the system, which varies for each design. This adds an additional complexity to the sensitivity analysis, which we presently wish to avoid. Furthermore, fixing the voltage range between $0$ and $\zeta T_{\mrm{f}}$ results in more predictable responses, which is convenient when solving these coupled nonlinear equations over a wide range of design and physical parameters. We will incorporate fixed-current (galvanostatic) scenarios in our future studies, which will consider specific material choices.}

In the electrode region, the effective transport properties, namely effective electronic conductivity $\sigma$, effective ionic conductivity $\kappa$, and the effective salt diffusion coefficient $D$ are influenced by the tortuosity of the pores.
The Bruggeman correlation \citep{bruggeman1935berechnung} quantifies this tortuosity as a power law of porosity with exponent $\frac{3}{2}$ for packed spheres, and thus the effective transport properties are calculated as
\begin{equation}
\label{eq:bruggeman}
    \sigma = (1-\epsilon)^{\frac{3}{2}}\sigma_0, \quad \kappa = \epsilon^{\frac{3}{2}}\kappa_0, \quad D = \epsilon^{\frac{3}{2}} D_0,
\end{equation}
where the $0$ subscripts represents the bulk property values.
Specifically, $\kappa_0$ and $D_0$ refer to the ionic conductivity and diffusion coefficient in pure electrolyte ($\epsilon = 1$), and $\sigma_0$ refers to the electronic conductivity in non-porous solid material ($\epsilon = 0$).
Note that $D_0$ and $\kappa_0$ are computed from the individual ion species as
\begin{equation}
    D_0 = D_{0_+} D_{0_-} \frac{z_+ - z_-}{z_+ D_{0_+} - z_- D_{0_-}},
\end{equation}
\begin{equation}
  \label{eq:kappa}
  \kappa_0 = F^2 c (z^2_+ \nu_+ u_{0_+} + z^2_- \nu_- u_{0_-}).
\end{equation}
The values for bulk conductivities, i.e., $\sigma_0$ and $\kappa_0$, and porosity $\epsilon$ are readily available from experimental data.
The bulk diffusion coefficient $D_0$, transference numbers $t_i$, and specific surface area $a$ are more difficult to measure.

Experimental data suggests that the Bruggeman correlation overestimates ionic conductivity in porous electrodes with certain particle arrangements \citep{TJADEN201644, tjaden2018tortuosity}.
Therefore, we also consider a modified correlation that reduces the ionic conductivity and diffusion coefficient inside the porous electrode by equating $\kappa = f_m \epsilon^{\frac{3}{2}}\kappa_0$ and $D_i = f_m \epsilon^{\frac{3}{2}} D_{0,i}$, where $f_m$ is a correction coefficient that takes values in the range $[0.02, 0.79]$, depending on the electrode microstructure \citep{Madabattula_2020}.

\subsection{Nondimensionalization}
To facilitate the analysis, we nondimensionalize the system described by \eqref{eq:govern_system} by defining the dimensionless quantities:
\begin{equation}
  \label{eq:dimless_variable}
  \begin{aligned}
    &\hat{\Phi}_1 = \frac{F}{RT}\Phi_1, &\hat{c} &= \frac{c}{c_0}, &\hat{t} &= \frac{D_0}{L^2}t,\\
    &\hat{\Phi}_2 = \frac{F}{RT}(\Phi_2 + U_0), &\hat{\mb x} &= \frac{\mb x}{L}. &&
  \end{aligned}
\end{equation}
The dimensionless location $\hat{\mb x}$ leads to the nondimensional domain $\hat \Omega$ and boundaries $\hat \Gamma_\mathrm{a}$ and $\hat\Gamma_\mathrm{c}$, as well as gradient $\hat \nabla$.
We also introduce the dimensionless specific surface area $\hat a$, effective diffusion coefficient $\hat D$, and effective electronic conductivity $\hat \sigma$, which are design-dependent quantities we will describe in Sect.~\ref{sec:interp}.
Using \eqref{eq:nernst_einstein} and \eqref{eq:stoichiometry} the dimensionless field equations read
\begin{subequations}
  \label{eq:govern_system_dimless}
  \begin{align}
    -\hat{\nabla}\cdot \left(\hat \sigma \hat{\nabla}\hat{\Phi}_1\right) &= -\lambda \hat{a} \left(\delta_r \hat{i}_n + \delta_c \hat{i}_c\right) &&\text{in }\hat{\Omega}, \label{eq:electric_dimless}\\
    -\hat{\nabla}\cdot \left(\hat D \hat{c}\hat{\nabla} \hat{\Phi}_2\right) -& \left(\frac{t_{+}}{z_+} + \frac{t_{-}}{z_-}\right) \hat{\nabla}\cdot\left(\hat D \hat{\nabla}\hat{c}\right) && \nonumber \\
     &= (1-\lambda) \hat{a}\left(\delta_r \hat{i}_n +\delta_c \hat{i}_c\right) &&\text{in }\hat{\Omega}, \label{eq:ionic_dimless}\\
    \frac{\partial (\epsilon \hat{c})}{\partial \hat{t}} -\hat{\nabla}\cdot(\hat D \hat{\nabla}\hat{c}) &= \frac{z_+ z_-}{t_{+} t_{-} (z_- - z_+)} (1-\lambda) &&\nonumber \\
    \hat{a} t_{-}\bigg( \delta_r \hat{i}_n + &\left(\frac{dq_+}{dq} - \frac{t_{+}}{t_{-}} \frac{dq_-}{dq}\right) \delta_c \hat{i}_c\bigg)  &&\text{in }\hat{\Omega},\label{eq:species_dimless}
  \end{align}
  subject to the boundary and initial conditions
  \begin{align}
    \hat{\Phi}_1 &= 0 \quad &&\text{on }\hat{\Gamma}_\mathrm{a}, \label{eq:bc1_dimless} \\
    \hat{\Phi}_1 &= \xi \hat{t} \quad &&\text{on }\hat{\Gamma}_\mathrm{c}, \label{eq:bc2_dimless} \\
    \hat\nabla \hat\Phi_1 \cdot \mb{n} &=0\quad &&\text{on }\partial \hat\Omega \setminus (\hat\Gamma_\mathrm{a}\cup\hat\Gamma_\mathrm{c}) \label{eq:bc3_dimless}, \\
    \hat{\nabla} \hat{\Phi}_2 \cdot \mb{n} &=0\quad &&\text{on }\partial \hat{\Omega}, \label{eq:bd4_dimless} \\
    \hat{\nabla} \hat{c} \cdot \mb{n} &= 0 \quad  &&\text{on }\partial\hat{\Omega}, \label{eq:bd5_dimless} \\
    \hat{\Phi}_1(\hat{\mb{x}},\hat{t}=0) &= 0 \quad &&\text{in } \hat{\Omega}, \label{eq:ic1_dimless} \\
    \hat{\Phi}_2(\hat{\mb{x}},\hat{t}=0) &= 0 \quad &&\text{in } \hat{\Omega}, \label{eq:ic2_dimless} \\
    \hat{c}(\hat{\mb{x}},\hat{t}=0) &= \hat{c}_0 \quad &&\text{in } \hat{\Omega}, \label{eq:ic3_dimless}
  \end{align}
\end{subequations}
for $\hat t \in (0, \hat{T}_{\mrm{f}}]$ with $\hat{T}_{\mrm{f}}=D_0 T_{\mrm{f}} / L^2$.
In the above we introduce the dimensionless current densities
\begin{equation}
\begin{aligned}
    \hat{i}_n =& \hat{c}^{\alpha} \left[e^{\alpha (\hat{\Phi}_1 - \hat{\Phi}_2)} - e^{-\alpha (\hat{\Phi}_1 - \hat{\Phi}_2)}\right],\\
    \hat{i}_c =& \hat{c}^{\alpha} \frac{\partial \left(\hat{\Phi}_1 - \hat{\Phi}_2\right)}{\partial \hat{t}}.
\end{aligned}
\end{equation}
and dimensionless ionic conductivity $\hat{\kappa}=\hat{c}\hat{D}$.

For future reference we also introduce the independent dimensionless groups as
\begin{equation}
  \label{eq:dimless_groups}
  \begin{aligned}
    &\delta_{r} = \frac{a_0 i_0 L^2 F}{RT} \left(\frac{1}{\sigma_0} + \frac{1}{\kappa_0}\right), &  t_{+} &=\frac{z_+ u_{+}}{z_+ u_{+} - z_- u_{-}}, \\
    &\delta_c = a_0 C_{d,0} D_0 \left(\frac{1}{\sigma_0} + \frac{1}{\kappa_0}\right), & \lambda &= \frac{\kappa_0}{\sigma_0 + \kappa_0},\\
    &\xi = \frac{\zeta L^2 F}{D_0 RT}. &
  \end{aligned}
\end{equation}
with the following meanings
\begin{itemize}
    {\small \item $\delta_r$ and $\delta_c$ represent the ratios between the kinetic resistance and ohmic resistance, for redox reaction and capacitive charging, respectively;}
    {\small \item $\lambda$ is the ratio between the ionic conductivity and the total conductivity;}
    {\small \item $t_+$ denotes the transference number of cation; the anion transference number is related by $t_- = 1 - t_+$;}
    {\small \item $\xi$ is the scan rate to diffusion timescale ratio.}
\end{itemize}
In \eqref{eq:dimless_groups}, the ionic conductivity $\kappa_0$, the exchange current density $i_0$, and specific capacitance $C_{d,0}$ are computed with the initial concentration $c_0$.

We also define $\delta = \delta_r + \delta_c$ as the ratio between total kinetic resistance and ohmic resistance, and $\gamma = \delta_r/\delta$ as the fractional contribution to $\delta$ from the redox reaction, so that
\begin{equation}
  \label{eq:tot_frac}
  \begin{aligned}
      \delta &= \frac{a_0 D_0 F}{RT} \left(\frac{1}{\sigma_0} + \frac{1}{\kappa_0}\right) \left(\frac{i_0 L^2}{D_0} + \frac{C_{d,0} RT}{F}\right), \\
      \gamma &= \frac{i_0 L^2/D_0}{i_0 L^2/D_0 + C_{d,0} RT/F}.
  \end{aligned}
\end{equation}
Hence, $\gamma$ can be varied while keeping $\delta$ constant to have different reaction compositions for the same total kinetic resistance.
In the literature, the Wagner number, $\mathrm{Wa} = 1/\delta$, is often used to analyze the limiting kinetics and current distribution.

\subsection{Electrode-specific electronic potential}
\label{sec:separate_phi1}
In a full-cell model, the electronic potential only exists in the electrode regions (where $\bar\rho\to 1$).
System \eqref{eq:govern_system_dimless} models $\hat{\Phi}_1$ on the entire domain $\Omega$, by assigning the dimensionless electronic conductivity to a very small positive value in the separator region (where $\bar\rho\to 0$) so that $\hat{\sigma}\hat{\nabla}\hat{\Phi}_1 \approx 0$ in the electrolyte.
However, because of filtering, $\bar\rho$ is not exactly zero the electrolyte region and hence the interpolated conductivity may not be negligible, which allows some electronic current to bypass the ionic phase, creating a short circuit between the electrodes.
To eliminate the short circuit effect, the optimized designs will position the electrodes far apart, so that $\bar\rho$ approaches zero in the center of the electrolyte region.
Consequently, the optimization produces suboptimal designs, with no possibility for creating interdigitated structures that are known to be effective ~\citep{hung2022three, long2004three}.

To prevent a short circuit and allow for interdigitation, we represent $\hat{\Phi}_1$ via two fields: $\hat{\Phi}_{1,\mrm{a}}$ in the anode and $\hat{\Phi}_{1,\mrm{c}}$ in the cathode and use the indicator functions from \eqref{eq:indicators} to define their effective subdomains.
The electronic potential equation \eqref{eq:electric_dimless} is thus replaced by the two equations
\begin{subequations}
  \label{eq:electric2}
  \begin{align}
    -\hat{\nabla}\cdot \Big(\mathcal{I}_\mathrm{a} &\hat\sigma \hat{\nabla}\hat{\Phi}_{1,\mathrm{a}}\Big)&&\nonumber \\
    &= -\lambda \hat{a} \left(\delta_r \hat{i}_{n,\mathrm{a}} + \delta_c \hat{i}_{c,\mathrm{a}}\right) \quad &&\text{in }\hat{\Omega}, \label{eq:anode_electric_dimless}\\
    -\hat{\nabla}\cdot \Big(\mathcal{I}_\mathrm{c} &\hat\sigma \hat{\nabla}\hat{\Phi}_{1,\mathrm{c}}\Big)&&\nonumber \\
    &= -\lambda \hat{a} \left(\delta_r \hat{i}_{n,\mathrm{c}} + \delta_c \hat{i}_{c,\mathrm{c}}\right) \quad &&\text{in }\hat{\Omega}, \label{eq:cathode_electric_dimless}
  \end{align}
  with associated boundary conditions
  \begin{align}
    \hat{\Phi}_{1,\mathrm{a}} &= 0 \quad &&\text{on }\hat{\Gamma}_\mathrm{a}, \label{eq:bc2_separate_dimless} \\
    \hat\nabla \hat\Phi_{1,\mathrm{a}} \cdot \mb{n} &=0\quad &&\text{on }\partial \hat\Omega \setminus \hat\Gamma_\mathrm{a} \label{eq:bc3_separate_dimless}, \\
    \hat{\Phi}_{1,\mathrm{c}} &= \xi \hat{t} \quad &&\text{on }\hat{\Gamma}_\mathrm{c}, \label{eq:bc1_separate_dimless} \\
    \hat\nabla \hat\Phi_{1,\mathrm{c}} \cdot \mb{n} &=0\quad &&\text{on }\partial \hat\Omega \setminus \hat\Gamma_\mathrm{c} \label{eq:bc4_separate_dimless}, 
  \end{align}
\end{subequations}
where the anodic currents are computed as
\begin{equation}
    \begin{aligned}
        \hat{i}_{n,\mathrm{a}} =& \mathcal{I}_\mathrm{a} \hat{c}^{\alpha} \left[e^{\alpha (\hat{\Phi}_{1,\mathrm{a}} - \hat{\Phi}_2)} - e^{-\alpha (\hat{\Phi}_{1,\mathrm{a}} - \hat{\Phi}_2)}\right],\\
        \hat{i}_{c,\mathrm{a}} =& \mathcal{I}_\mathrm{a} \hat{c}^{\alpha} \frac{\partial \left(\hat{\Phi}_{1,\mathrm{a}} - \hat{\Phi}_2\right)}{\partial \hat{t}}.
    \end{aligned}
\end{equation}
The cathodic currents $\hat{i}_{n,\mathrm{c}}$ and $\hat{i}_{c,\mathrm{c}}$ are computed similarly by replacing $\mathcal{I}_\mathrm{a}$ and $\hat{\Phi}_{1,\mathrm{a}}$ with $\mathcal{I}_\mathrm{c}$ and $\hat{\Phi}_{1,\mathrm{c}}$.
For all other terms defined above, the total current densities are evaluated as $\hat{i}_n = \hat{i}_{n,\mathrm{a}} + \hat{i}_{n,\mathrm{c}}$, and $\hat{i}_c = \hat{i}_{c,\mathrm{a}} + \hat{i}_{c,\mathrm{c}}$, and the electronic potential as $\hat{\Phi}_1 = \mathcal{I}_\mathrm{a} \hat{\Phi}_{1,\mathrm{a}} + \mathcal{I}_\mathrm{c} \hat{\Phi}_{1,\mathrm{c}}$.

\subsection{Design-dependent material parameters}
\label{sec:interp}
In our topology optimization, the model parameters such as the porosity, the dimensionless surface area, electronic conductivity, and diffusivity, vary spatially as a function of the material indicator field $\bar \rho$. 
The porosity and dimensionless surface area are given by interpolations as
\begin{equation}
\label{eq:ahat}
    \begin{aligned}
        \epsilon =& \epsilon_M + \bar{\rho}(\epsilon_N - \epsilon_M),\\
        \hat{a} =& \bar{\rho}^q,
    \end{aligned}
\end{equation}
where $\epsilon_M$ is the porosity in the electrolyte/separator, $\epsilon_N$ is the porosity in the electrode, and $q \geq 1$ is a penalization parameter.
The porosity $\epsilon$ is volume averaged while the specific area $\hat{a}$ is penalized.
As such, intermediate regions where $\bar\rho(\hat{\mb{x}}) \neq \{0,1\}$ have artificially smaller $\hat{a}$, thus reducing current densities.
Because of this, the optimized designs predominantly have $\bar\rho (\hat{\mb{x}})= \{0,1\}$, which approximates our original problem in \eqref{eq:original_opt_problem}, i.e., $\chi\approx \bar\rho$.
Meanwhile, the dimensionless electronic conductivity $\hat{\sigma}$ and diffusion coefficients $\hat{D}$ are computed from the Bruggeman correlation \eqref{eq:bruggeman} and interpolations
\begin{equation}
\label{eq:sdhat}
    \begin{aligned}
        \hat{\sigma} =& \bar{\rho}^p (1-\epsilon_N)^{\frac{3}{2}}, \\
        \hat{D} =& \epsilon_M^{\frac{3}{2}} + \bar{\rho}^p \left(\epsilon_N^{\frac{3}{2}}-\epsilon_M^{\frac{3}{2}}\right),
    \end{aligned}
\end{equation}
where the parameter $p \geq 1$ is similarly used to penalize ionic and electronic transport in intermediate material regions.
Note that $\epsilon_M$ is not involved in the interpolation for $\hat{\sigma}$ since the electrolyte has negligible electronic conductivity.
For the modified Bruggeman correlation, the $\epsilon_N^\frac{3}{2}$ is replaced by $f_m\epsilon_N^\frac{3}{2}$ in the formula for $\hat{D}$.
In this paper, we choose $\epsilon_M=1$, $\epsilon_N=0.5$, and $f_m =0.02$, \rtwo{which is the most extreme value in \cite{Madabattula_2020}.}
The choice of the penalization parameters $p$ and $q$ is discussed in Sect.~\ref{sec:optproblem}.
 
\subsection{Energy metrics}
We list quantities used to construct the optimization cost and constraint functions, namely energy input $E_{\mrm{in}}$, kinetic energy $E_{\mrm{kin}}$, and ohmic loss $E_{\mrm{ohm}}$,
\begin{subequations}
  \label{eq:energy_quantities}
  \begin{align}
    E_{\mrm{in}} =& \frac{1}{\lambda} \int_0^{\hat{T}_{\mrm{f}}}\int_{\partial\hat{\Omega}} \hat\sigma  \hat{\nabla}\hat{\Phi}_1 \cdot \mb{n}\hat{\Phi}_1 \diff\hat{s} \diff\hat{t}, \label{eq:energy_input}\\
    E_{\mrm{kin}} =& \delta_r \int_0^{\hat{T}_{\mrm{f}}}\int_{\hat{\Omega}} \hat{a}\hat{i}_n(\hat{\Phi}_1-\hat{\Phi}_2)\diff\hat{x} \diff\hat{t}\nonumber \\
    + &\delta_c \int_0^{\hat{T}_{\mrm{f}}}\int_{\hat{\Omega}}\hat{a}\hat{i}_c(\hat{\Phi}_1 - \hat{\Phi}_2)\diff\hat{x} \diff\hat{t},\label{eq:kinetic_energy}\\
    E_{\mrm{ohm}}=&\frac{1}{\lambda} \int_0^{\hat{T}_{\mrm{f}}}\int_{\hat{\Omega}} \hat\sigma  \hat{\nabla}\hat{\Phi}_1 \cdot \hat{\nabla}\hat{\Phi}_1 \diff\hat{x} \diff\hat{t}\nonumber \\
    + &\frac{1}{1-\lambda} \int_0^{\hat{T}_{\mrm{f}}}\int_{\hat{\Omega}}\hat D \hat{c} \hat{\nabla} \hat{\Phi}_2 \cdot \hat{\nabla} \hat{\Phi}_2 \diff\hat{x} \diff\hat{t}\nonumber \\
    + &\frac{1}{1-\lambda}\left(\frac{t_{+}}{z_+} + \frac{t_{-}}{z_-}\right)\int_0^{\hat{T}_{\mrm{f}}}\int_{\hat{\Omega}} \hat D \hat{\nabla}\hat{c} \cdot \hat{\nabla} \hat{\Phi}_2 \diff\hat{x} \diff\hat{t}\label{eq:ohmic_loss},
    \end{align}
\end{subequations}
which all have positive values when charging.
The detailed derivation of these expressions is presented in Appendix ~\ref{sec:energy_balance}.
The kinetic energy, henceforth referred to as the stored energy, includes both the energy stored in the chemical reaction and the dissipation due to finite charge-transfer kinetics.
The energy storage depends on the thermodynamic potential, which is characteristic of a specific material system.
To keep the analysis general, we do not specify the thermodynamic potential and instead infer its value via the dimensionless ionic potential, which naturally groups kinetic loss with kinetic storage.

The optimization problem focuses on reducing the ohmic losses, because this can be lowered by shaping the electrodes to facilitate ion and electron transport across the cell.
By reducing the ohmic losses, a larger current output can be extracted from the cell, for a given applied voltage.
While a larger current also causes larger kinetic losses (due to the activation barrier for redox reaction) we expect the cell efficiency to be dominated by transport losses.

\subsection{Optimization problem setup}
\label{sec:optproblem}

In this section, we introduce the general optimization strategy and several techniques devised to improve the convergence and effectiveness of the optimization algorithm.
Given the complexity of the physical problem, we need to carefully define the cost and constraint functions and the initial design to effectively traverse the design space and find optimized designs.

\subsubsection{Parameter penalizations}
\label{sec:penalizations}
To promote $0-1$, i.e., $\bar\rho(\hat{\mb{x}}) \approx \{0,1\}$ designs, we use different interpolations for the design-dependent quantities $\hat a$, $\hat \sigma$, and $\hat D$ in the cost and constraint functions cf. \eqref{eq:energy_quantities}, as compared to the primal analysis cf. \eqref{eq:govern_system_dimless} and \eqref{eq:electric2}.
\begin{figure}[!htbp]
    \centering
    \includegraphics[width=\columnwidth]{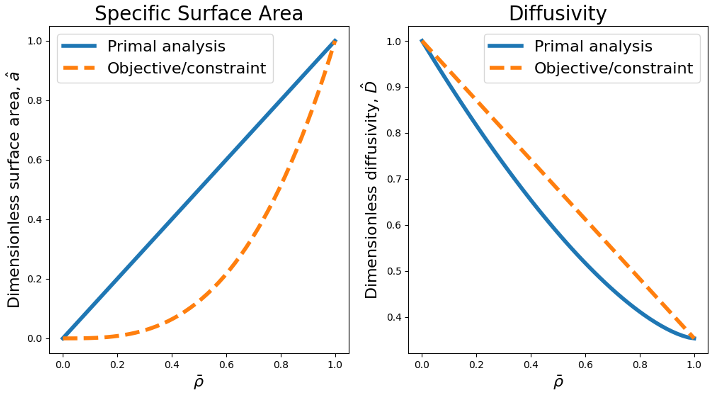}
    \caption{Different penalization for specific surface area and ionic conductivity for the primal analysis and cost/constraint computation to produce $0-1$ designs.}
    \label{fig:penalization}
\end{figure}
In the cost and constraint computations, $\hat a$ is interpolated with $q=3$, whereas in the primal analysis $q=1$, cf. \eqref{eq:ahat}.
Reversely, $\hat{\sigma}$ and $\hat{D}$ use $p=1$ in the cost and constraint computations and $p=1.5$ in the primal analysis.
The different interpolations adopted for the primal analysis and the cost/constraint computations are visualized in Fig.~\ref{fig:penalization}.
The $q=3$ penalization for the specific surface area and $p=1.5$ for the conductivity disproportionately reduces the energy storage and disproportionately increases the ohmic loss in the intermediate material regions.
These penalizations lead to inefficient use of intermediate material regions and hence the size of such regions in the optimized designs is small.
\rone{This is desirable because intermediate material regions cannot be fabricated and the energy balance presented in Appendix \ref{sec:energy_balance} is only valid for $0-1$ designs and hence our model predications would be inaccurate.}

\subsubsection{Cost and Constraint}
\label{sec:obj_constraint}
Maximizing the kinetic energy $E_{\mrm{kin}}$ defined in \eqref{eq:energy_quantities} is an intuitive choice for the (negative) cost function, as a proxy for maximizing energy stored.
Unfortunately, this choice does not consider the ohmic loss and thus does not benefit from the \eqref{eq:sdhat} diffusivity/conductivity interpolation scheme.
In response, we formulate the optimization problem as follows
\begin{multline}
    \max_{\rho\colon\Omega \to [0,1]} \theta_0 = \frac{1}{2}\left(E_{\mrm{kin}} + E_{\mrm{in}} - E_{\mrm{ohm}}\right) \\ \text{subject to } \theta_1 = \frac{E_{\mrm{ohm}}}{E_{\mrm{in}}} \leq \eta_{\max},
    \label{eq:original_opt}
\end{multline}
i.e., we maximize energy storage and ensure a minimum energy efficiency of $(1-\eta_{\max})$.

According to \eqref{eq:original_opt}, the design can be optimized by either increasing $E_{\mrm{kin}}$ or decreasing $E_{\mrm{ohm}}$.
And hence the interpolations of $\hat a$, $\hat{\sigma}$, and $\hat{D}$ contribute disproportionately lower $E_{\mrm{kin}}$ and disproportionately higher $E_{\mrm{ohm}}$ in the intermediate material regions of the cost and constraint function integrands, relative to the primal analysis integrands.
As such the optimizer generates predominantly $0-1$ designs.

\rone{Another intuitive choice would be to maximize power, which has been done for electrochemical power source devices such as fuel cells and flow batteries \citep{alizadeh2023mixed}.
Although we focus on energy, our future studies will also consider power and the trade-off between them.}

\rtwo{Note that, unlike classical structural topology optimization problems, there is no constraint on the volume of the electrode material since both electrode and electrolyte-only phases can be beneficial to energy storage.
Indeed, in the absence of the efficiency constraint, the volumes of the electrode and electrolyte-only phases are determined implicitly by the governing physics.
This is clearly observed in \cite{roy2022tope, lin2024shape}, where more electrode phase leads to more active material (increasing energy stored), while more electrolyte-only phase leads to faster ion transport (reducing ohmic loss).
The inclusion of concentration variations in this current study introduces additional complexity into the design process due to ion depletion.
}

\begin{figure}[!htbp]
    \centering
    \includegraphics[width=\columnwidth]{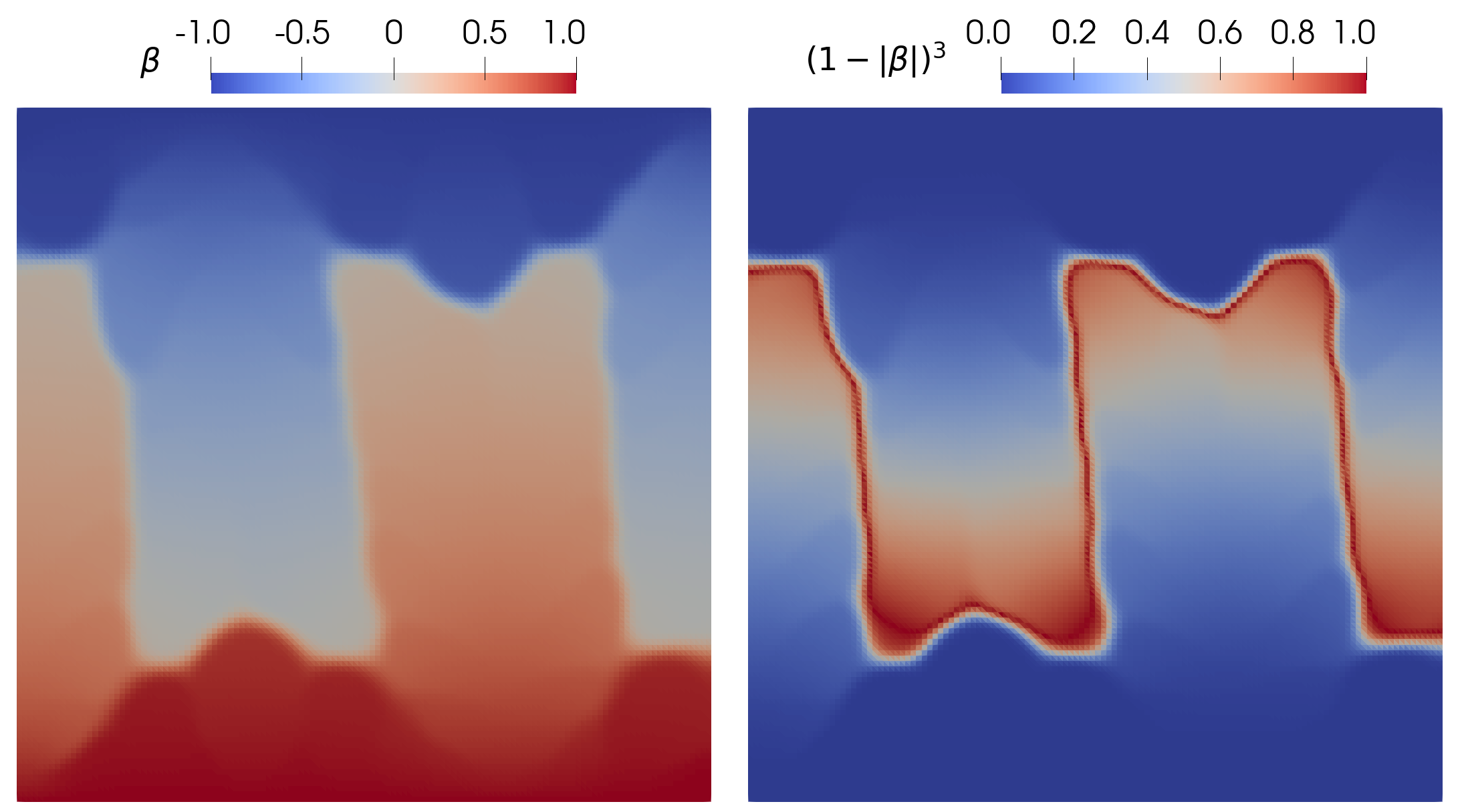}
    \caption{Interdigitated electrodes with near uniform $\beta\approx \pm 1$ distribution bounded by a thin separator layer with large $(1-|\beta|)^3$ values.}
    \label{fig:short_circuit_indicator}
\end{figure}

According to \eqref{eq:electric2}, the only relation between the two electronic potentials is through the reaction terms.
Consequently, a physical connection between the anode and cathode is not simulated as a short circuit (electronic current bypassing the reaction).
To prevent such connections, we define the short circuit intensity
\begin{equation}
    \label{eq:short_circuit_penalty}
    \mathcal{I}_{\mrm{SC}} = \int_{\Omega} \left(1-|\beta|\right)^3\bar{\rho}\diff\hat{x}
\end{equation}
The term $(1-|\beta|)^3$ takes its maximum value when $\beta = 0$.
In the absence of a short circuit, in the electrodes where $\bar\rho \approx 1$, we have a near uniform $\beta = \pm 1$ distribution.
The exception occurs in the separator layer where $\beta$ experiences a sharp transition as illustrated in Fig.~\ref{fig:short_circuit_indicator}.
A ``small'' short circuit intensity requires $\bar\rho \approx 0$ when $\beta\neq \pm 1$, thus ensuring the electrolyte fills the gap between the electrodes.
We weakly enforce this short circuit intensity constraint via the penalty method, i.e., we now solve the optimization problem
\begin{multline}
    \min_{\rho\colon\Omega \to [0,1]} \tilde{\theta}_0 = \frac{1}{\theta_0} + w_{\mrm{SC}} \mathcal{I}_{\mrm{SC}} \\ \text{subject to } \theta_1 = \frac{E_{\mrm{ohm}}}{E_{\mrm{in}}} \leq \eta_{\max},
    \label{eq:min_problem}
\end{multline}
where $w_{\mrm{SC}}=1$ is the penalty weight used in our study.
Note that the minimization of $-\theta_0$, is less robust than $\frac{1}{\theta_0}$ since it requires case-specific scaling values for $\theta_0$.
However, for \eqref{eq:min_problem} to work properly, $E_{\mrm{ohm}} < E_{\mrm{in}} + E_{\mrm{kin}}$ must be satisfied for the initial design, so that the cost $\tilde{\theta}_0$ is always positive, and hence minimized by maximizing $\theta_0$.
If we initially have $\theta_0 < 0$, then $\tilde{\theta}_0$ is minimized as $\theta_0 \rightarrow 0^-$, which completely changes the optimization trajectory and hence designs.

It is well known that the optimized designs may depend on the initial design.
To this end, we solve the optimization problem using three uniform initial designs $\rho(\mb{x})\in \{0.45, 0.5, 0.55\}$ and choose the best overall design.
\rone{In a well-posed optimization problem, a minor change in the initial design only slightly alters the optimized design. Thus, this is an ad-hoc means of ensuring that we have a well-posed optimization problem. This is also a cheap way to prevent numerical artifacts such as ``islands" of electrode material. These islands appeared in two cases for $\rho=0.5$, in which case we picked the optimized designs from the best of $\rho = \{0.45, 0.55\}$.}

From the initial uniform design, the constraint on $\theta_1$ increases the efficiency of the system by placing more electrode material closer to the current collectors to reduce ohmic losses.
This configuration creates a gap between the two electrodes, thus avoiding the appearance of a short circuit in the design.
We set the constraint bound $\eta_{\max}=\theta_1^{(0)}\,\Sigma$, where $\Sigma\in (0.0,1.0]$ and $\theta_1^{(0)}$ denote the ohmic loss ratio of the initial design.
We generate designs for different combinations of dimensionless parameters, with both original and modified Bruggeman correlation.
$\Sigma = 0.4$ is used for all $\delta = 0.5$, $\lambda = 0.1$, and three-dimensional designs; $\Sigma = 0.5$ is used for all other designs.

\begin{figure*}
    \centering
    \includegraphics[width=\linewidth]{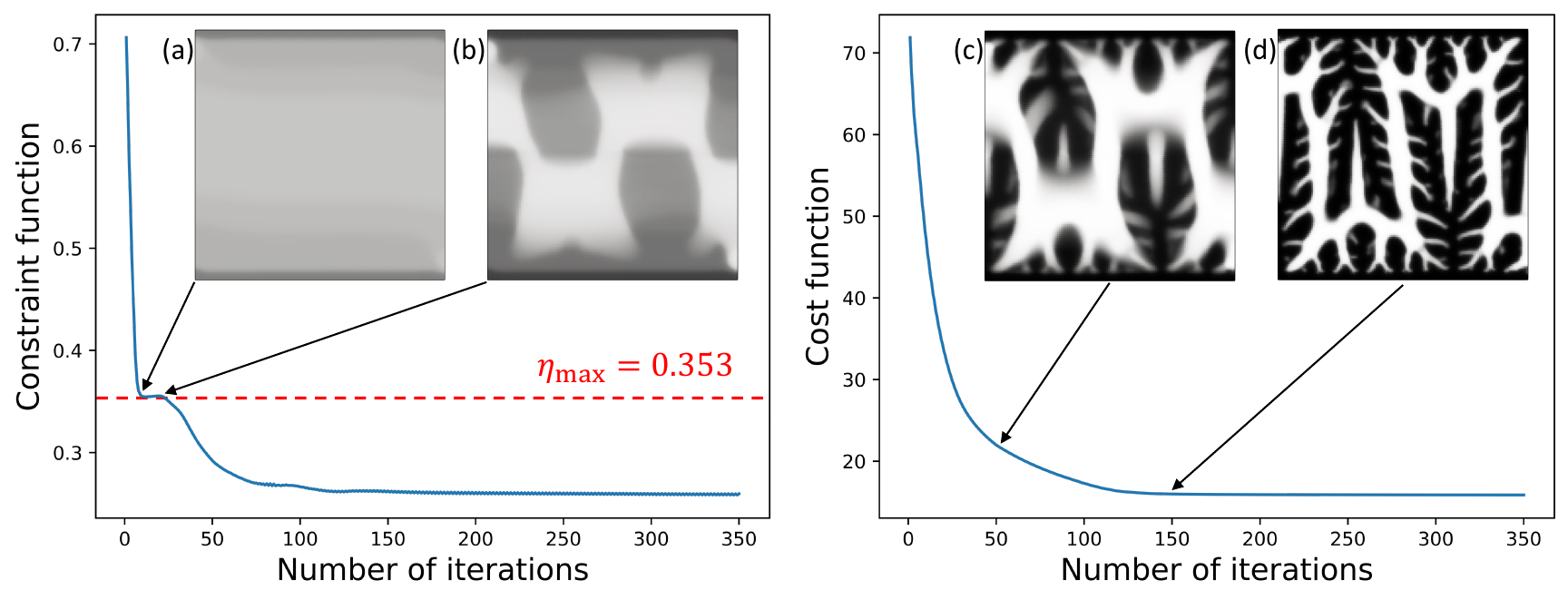}
    \caption{Cost and constraint function history for $\delta=2, \gamma=0.5$, $\lambda=0.01$ with modified Bruggeman correlation. The inserts are the designs at iteration (a) 8, (b) 20, (c) 50, and (d) 150. Only one symmetric half of the design is illustrated to save space.}
    \label{fig:convergence_history}
\end{figure*}
A typical optimization convergence history is illustrated in Fig. \ref{fig:convergence_history}.
During early design iterations, the constraint drives the optimization by separating the electrodes and reducing the amount of electrode material, which otherwise leads to increased reaction current and ohmic loss.
While eliminating short circuits is necessary, a tight constraint, i.e., small $\Sigma$, hinders the development of the cell morphology.
Therefore, we deactivate this constraint after the bulk design features have formed, which for our study is after iteration 150.
The ensuing iterations refine the shape rather than the overall layout.

\subsubsection{Case specific treatments}
\label{sec:special_treatment}
Certain cases within our parameter sweep require extra care.
1) For cases with the original Bruggeman correlation, i.e., $f_m = 1$, large effective ionic conductivity/diffusivity within the porous electrode allows ions to travel deeper into the pore network, resulting in better material utilization as discussed in Sect.~\ref{sec:performance}.
Accordingly, during early design iterations, the optimizer positions a considerable amount of electrode material in the middle region to benefit from the largest potential drop.
Unfortunately this causes a strong connection between the electrodes that cannot be removed in the later design iterations.
To remedy this, the optimization adopts a continuation strategy wherein the initial design uses the modified Bruggemann correlation with $f_m = 0.02$ which increases to the target $f_m = 1$ after iteration 50.
2) For cases with large $\delta$, $\theta_0 < 0$ for the initial design, which is undesirable as stated above.
We remedy this by invoking yet another continuation strategy where the initial design uses a smaller $\delta$ such that $E_{\mrm{ohm}} < E_{\mrm{in}} + E_{\mrm{kin}}$ is satisfied.
In our study, the $\delta = 5$ cases start with $\delta = 2$ and increase to its target value after iteration 50 (see Fig. \ref{fig:convergence_history}).

In some extreme situations, namely the $\delta = 5$ with modified Bruggeman correlation cases, large regions of intermediate material appear in the optimized designs.
The small contributions to $E_{\mrm{kin}}$ and $E_{\mrm{ohm}}$ from these regions are overshadowed by the other regions where the current density is substantial.
As such, the penalization scheme is not effective.
To counter this effect, we scale $E_{\mrm{kin}}$ and $E_{\mrm{ohm}}$ after the bulk structures are established (i.e., after iteration 150 when the efficiency constraint is deactivated).
To motivate our scaling scheme, refer to the typical $\beta$ distribution shown in Fig.~\ref{fig:electrode_indicator} (right-hand side) where $|\beta| \approx 0$ around the electrode tips and $|\beta| \approx 1$ near the current collectors.
Accordingly, we consider three different types of scaling, namely $|\beta|$, $1-|\beta|$, and $|\beta|(1-|\beta|)$, to scale down the energy integrands around the electrode peaks, the current collector, and both, respectively.
The scaling is readily incorporated into our strategy with a continuation scheme.
With $|\beta|$ scaling as an example, the energy storage is computed by
\begin{multline}
    \tilde{E}_{\mrm{kin}} = \delta_r \int_0^{\hat{T}_{\mrm{f}}}\int_{\hat{\Omega}} |\beta|^s \hat{a}\hat{i}_n(\hat{\Phi}_1-\hat{\Phi}_2)\diff\hat{x} \diff\hat{t}\\ + \frac{\delta_c}{2} \int_{\hat{\Omega}} |\beta|^s \hat{a}(\hat{\Phi}_1 - \hat{\Phi}_2)^2\diff\hat{x},
\end{multline}
where $s = 0$ for the first 150 iterations whereafter it is increased to $s = 1$.
We use the $|\beta|$ scaling for $E_{\mrm{kin}}$ and $E_{\mrm{ohm}}$ in the $\gamma = \{0.5, 1\}$ cases.
For the $\gamma = 0$ case, we use $|\beta|$ scaling for $E_{\mrm{kin}}$ and $|\beta|(1-|\beta|)$ scaling for $E_{\mrm{ohm}}$.

The short circuit penalty $w_{SC}\mathcal{I}_{SC}$ in \eqref{eq:min_problem} is effective in preventing short circuits in our 2D designs.
However, for the 3D optimization case, the electrodes may contact each other over a small region, thus forming a short circuit.
Some of the strategies described above could be used to prevent such short circuit designs: different initial designs, penalizations, etc.
Instead, motivated by \cite{lazarov2016}, we implement a simple dilation/erosion strategy, which allows us to remove the short circuit.
This strategy applies a one-time adjustment to the design field $\rho$ to eliminate the small contact region.
Namely, we use shifted versions of the electrode indicator functions \eqref{eq:indicators} such that
\begin{equation}
\label{eq:indicators_shift}
    \begin{aligned}
        \tilde{\mathcal{I}}_\mathrm{a} &= H_{100,0.52}\left(\frac{\beta+1}{2}\right), \\
        \tilde{\mathcal{I}}_\mathrm{c} &= H_{100,0.52}\left(\frac{-\beta+1}{2}\right). 
    \end{aligned}
\end{equation}
In this way, regions where $\tilde{\mathcal{I}}_\mathrm{a}= 1$ and $\tilde{\mathcal{I}}_\mathrm{c}= 1$ are reduced in size as compared to those using $\mathcal{I}_{\mrm{a}}=1$ and $\mathcal{I}_{\mrm{c}}=1$ computed via $H_{100,0.5}$.
Consequently, there exists a subdomain $\Omega_\mrm{s}$ such that $\tilde{\mathcal{I}}_\mathrm{a}(\mb{x}) + \tilde{\mathcal{I}}_\mathrm{c}(\mb{x}) = 0$ for all $\mb{x} \in \Omega_{\mrm{s}}$.
Subsequently equating $\rho \coloneqq \rho \, (\tilde{\mathcal{I}}_\mathrm{a} + \tilde{\mathcal{I}}_\mathrm{c})$ ensures $\rho=0$ in $\Omega_{\mrm{s}}$.
Once the short circuit region is removed, it is unlikely to reappear due to the short circuit intensity penalization.
We perform the shifted indicator adjustment after iteration 150.
The subsequent iterations use the original indicator functions $\mathcal{I}_{\mrm{a}}$ and $\mathcal{I}_{\mrm{c}}$.
This strategy is appropriate when the electrode contact region is a small ``grazing" type, while the ``head-on collision" connections are avoided by reducing $\Sigma$.

\section{Implementation}
\label{sec:implementation}
We resolve the governing equations \eqref{eq:govern_system_dimless}, \eqref{eq:electric2}, filtering equation \eqref{eq:filter}, and boundary propagation equation \eqref{eq:bd_prop} with the finite element library \texttt{Firedrake} \citep{FiredrakeUserManual}, which uses \texttt{PETSc} \citep{petsc-user-ref, petsc-web-page} as the backend for the iterative solvers.
The sensitivities of the optimization cost and constraint functions are computed by \texttt{pyadjoint} \citep{mitusch2019dolfin}.
A \texttt{Python} implementation of the MMA algorithm \citep{svanberg1987method}, \texttt{pyMMAopt} \citep{miguel_salazar_de_troya_2021_4456055} is used to solve the optimization problem.
The designs are considered converged after 350 iterations, where the change in the normalized cost function is typically in the magnitude of $10^{-3}$. 

Piecewise linear finite elements are used for spatial discretization of the governing physics equations \eqref{eq:govern_system_dimless} and \eqref{eq:electric2}.
We use a mixed finite element method to solve the filtering equation \eqref{eq:filter} and the continuous boundary propagation problem \eqref{eq:bd_prop}, as detailed in Appendix \ref{sec:mixed}.
The domain is either partitioned into triangular (2D) or tetrahedral (3D) elements.
The two-dimensional domain is a rectangle with distance between the current collectors of nondimensional length 1 and the width of 2.
The additional dimension of the three-dimensional domain is a depth of 2.
We simulate half of the total domain in 2D, and a quarter in 3D, due to the domain symmetry.
The total number of degrees of freedom (DoF) for the 2D and 3D models are $\sim$60k and $\sim$4M, respectively.
Backward Euler is used for temporal discretization with 20 total timesteps for a dimensionless simulation time of $\hat{T}_{\mrm{f}}=1$, which is the smallest number of timesteps that is stable for all studies.
We parallelize the computation via an approximately 20k DoFs/CPU load.

With a transient problem, computing sensitivities with the adjoint method requires storing the primal response for all timesteps.
In this work, we have not encountered any memory bottleneck, but it is an inevitability for larger simulations.
In response, checkpointing schemes such as \cite{revolve} can be utilized.

The finite element system corresponding to \eqref{eq:govern_system_dimless} with the two-electronic potential formulation \eqref{eq:electric2} is solved using Newton's method with linesearch. 
The 2D linearized systems for the Newton updates are solved using a direct method.
For the 3D cases, however, an iterative method with a scalable preconditioner is essential \citep{wathen2015preconditioning}.
The linearized systems have the following block form:
\begin{equation}\label{eq:block}
    \begin{bmatrix}
        A_{11} & A_{12} & A_{13} & A_{14}\\
        A_{21} & A_{22} & A_{23} & A_{24}\\
        A_{31} & A_{32} & A_{33} & A_{34}\\
        A_{41} & A_{42} & A_{43} & A_{44}\\
    \end{bmatrix}
    \begin{bmatrix}
        \Delta \mathbf{\Phi}_{1, \mathrm{c}} \\
        \Delta \mathbf{\Phi}_{1, \mathrm{a}} \\
        \Delta \mathbf{\Phi}_2 \\
        \Delta \mathbf{c}
    \end{bmatrix}
    =
    \begin{bmatrix}
        \mathbf{r}_1 \\
        \mathbf{r}_2 \\
        \mathbf{r}_3 \\
        \mathbf{r}_4 \\
    \end{bmatrix},
\end{equation}
where $\mathbf{r}_i,\ i\in \{1,2,3,4\}$, are the residual vectors corresponding to Eqn. \eqref{eq:cathode_electric_dimless}, \eqref{eq:anode_electric_dimless}, \eqref{eq:ionic_dimless}, and \eqref{eq:species_dimless} that are expressed in terms of the $j\in \{1,2,3,4\}$ response vectors $\mathbf{\Phi}_{1,c}$, $\mathbf{\Phi}_{1,a}$, $\mathbf{\Phi}_2$, and $\mathbf{c}$.
Each block $A_{ij}$ represents the linearization of the finite element residual $\mb r_i$ with respect to vector $j$.

The diagonal blocks $A_{ii}$ result from the discretization
of elliptic or parabolic operators, for which multigrid methods \citep{brandt1977multi} are often ideal preconditioners. We
thus use the preconditioned GMRES \citep{saad1986gmres}, with a block triangular preconditioner of the form
\begin{equation}\label{eq:precon}
P = 
    \begin{bmatrix}
        A_{11} & 0 & 0 &0\\
        A_{21} & A_{22} & 0 & 0\\
        A_{31} & A_{32} & A_{33} & 0\\
        A_{41} & A_{42} & A_{43} & A_{44}\\
    \end{bmatrix}.
\end{equation}
Evaluating the inverse of this matrix requires the inverses of the diagonal blocks $A_{ii}$, which is done approximately using a
single AMG V-cycle \citep{ruge1987algebraic} performed by \texttt{BoomerAMG}
\citep{henson2002boomeramg} from the \texttt{hypre} library \citep{falgout2002hypre}.
The \texttt{PETSc} options for solving the field, the filter, and boundary propagation equations can be found in Appendix \ref{sec:petsc}.
\begin{figure}
    \centering
    \includegraphics[width=0.8\linewidth]{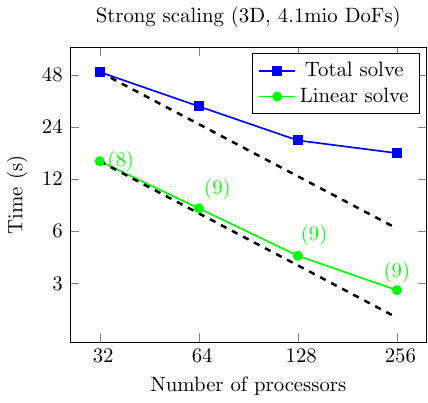}
    \caption{Strong scaling for the 3D solver. The timings show the total solution time and the total linear solver time where the number of GMRES iterations appears in the parentheses.}
    \label{fig:my_label}
\end{figure}

In Fig. \ref{fig:my_label}, we demonstrate the scalability of this preconditioning approach by simulating the first timestep for the initial uniform $\rho=0.5$ design.
We observe that the linear solver of \eqref{eq:block} exhibits almost linear scalability, which indicates that our preconditioning is effective.

\section{Results and discussion}
\label{sec:results}
\par Herein, we introduce the optimize designs for various choices of dimensionless parameters.
The fixed dimensionless parameters among all designs are listed in Table \ref{tab:forward_param}.
\begin{table}[!htbp]
    \centering
    \caption{Dimensionless parameters that are \textbf{fixed} amongst all designs.}
    \label{tab:forward_param}
    \begin{tabular}{lll}
        \hline\noalign{\smallskip}
        Parameter & Quantity & Value  \\
        \noalign{\smallskip}\hline\noalign{\smallskip}
        $\alpha$ & charge transfer coefficient & 0.5 \\
        $t_+$ & cation transference number & 0.5 \\
        $\xi$ & potential scan rate & 1.0 \\
        $\hat{T}_{\mrm{f}}$ & final time & 1.0 \\
        $\epsilon_M$ & electrolyte porosity & 1.0\\
        $\epsilon_N$ & electrode porosity & 0.5\\
        \noalign{\smallskip}\hline\noalign{\smallskip}
    \end{tabular}
\end{table}
Table~\ref{tab:forward_param_variable} lists the three dimensionless parameters that are varied amongst the designs.
The results presented in Sect.~\ref{sec:discussion} show how different combinations of $(\lambda, \delta, \gamma)$ lead to different optimal electrode designs.
\begin{table}[!htbp]
    \centering
    \caption{Dimensionless parameters that are \textbf{varied} amongst the designs.}
    \label{tab:forward_param_variable}
    \begin{tabular}{ll}
        \hline\noalign{\smallskip}
        Parameter & Quantity  \\
        \noalign{\smallskip}\hline\noalign{\smallskip}
        $\lambda$ & ratio between ionic and total conductivity \\
        $\delta$ & ratio between ohmic and kinetic resistance \\
        $\gamma$ & redox contribution to total reaction\\
        \noalign{\smallskip}\hline\noalign{\smallskip}
    \end{tabular}
\end{table}

\newcommand{\firstElement}[1]{%
    \expandafter\firstElementFromTuple#1\relax
}
\def\firstElementFromTuple#1,#2\relax{#1}

\newcommand{\secondElement}[1]{%
    \expandafter\secondElementFromTuple#1\relax
}
\def\secondElementFromTuple#1,#2\relax{#2}

\newcommand{\pecost}[1]{
    \def\mytuple{#1}
    $E_{\mrm{kin}}=\firstElement{\mytuple},\ E_{\mrm{ohm}}=\secondElement{\mytuple}$
}

\newcommand{\figurestabledelgam}[6]{
\providecommand{\plottable}[6]{
    \renewcommand\theadset{\def\arraystretch{.85}}%
    \setlength{\extrarowheight}{2pt}
    \centering
    \noindent\makebox[\textwidth]{
    \begin{tabular}{llll}
        \Xhline{2\arrayrulewidth}
                                        & \multicolumn{3}{c}{$\delta$}                                                               \\
        \cline{3-4}
                                        &                                 & \multicolumn{1}{c}{0.5}           & \multicolumn{1}{c}{5} \\
        \Xhline{2\arrayrulewidth}
        \multirowcell{6}[-12.0em]{$\gamma$}   &
                                        \multirow{2}{*}[-4em]{0.0}      &
                                        \multicolumn{1}{c}{\pecost{##1}} & \multicolumn{1}{c}{\pecost{##2}}                            \\[-1em]
                                        &                                 &  \begin{minipage}[t]{0.03\linewidth} \vspace{0pt}  (a)  \end{minipage}  \begin{minipage}[t]{0.4\linewidth}\vspace{0pt} #1 \end{minipage}                         & \begin{minipage}[t]{0.03\linewidth} \vspace{0pt}  (b)  \end{minipage}  \begin{minipage}[t]{0.4\linewidth}\vspace{0pt} #2 \end{minipage}                      \\[-1em]
                                        & \multirow{2}{*}[-4em]{0.5}
                                        & \multicolumn{1}{c}{\pecost{##3}} & \multicolumn{1}{c}{\pecost{##4}}                          \\[-1em]
                                        &                                 & \begin{minipage}[t]{0.03\linewidth} \vspace{0pt}   (c)  \end{minipage}  \begin{minipage}[t]{0.4\linewidth}\vspace{0pt} #3 \end{minipage}                        & \begin{minipage}[t]{0.03\linewidth} \vspace{0pt}  (d)  \end{minipage}  \begin{minipage}[t]{0.4\linewidth}\vspace{0pt} #4 \end{minipage}                     \\[-1em]
                                        & \multirow{2}{*}[-4em]{1.0}
                                        & \multicolumn{1}{c}{\pecost{##5}} & \multicolumn{1}{c}{\pecost{##6}}                          \\[-1em]
                                        &                                 & \begin{minipage}[t]{0.03\linewidth} \vspace{0pt}   (e)  \end{minipage}  \begin{minipage}[t]{0.4\linewidth}\vspace{0pt} #5 \end{minipage}                        & \begin{minipage}[t]{0.03\linewidth} \vspace{0pt}  (f)  \end{minipage}  \begin{minipage}[t]{0.4\linewidth}\vspace{0pt} #6 \end{minipage}                     \\[-1em]
        \Xhline{2\arrayrulewidth}
    \end{tabular}
    }
}
\plottable
}

\newcommand{\figurestablelamgam}[6]{
\providecommand{\plottable}[6]{
    \renewcommand\theadset{\def\arraystretch{.85}}%
    \setlength{\extrarowheight}{2pt}
    \centering
    \noindent\makebox[\textwidth]{
    \begin{tabular}{llll}
        \Xhline{2\arrayrulewidth}
                                        & \multicolumn{3}{c}{$\lambda$}                                                               \\
        \cline{3-4}
                                        &                                 & \multicolumn{1}{c}{0.01}           & \multicolumn{1}{c}{0.1} \\
        \Xhline{2\arrayrulewidth}
        \multirowcell{6}[-12.0em]{$\gamma$}   &
                                        \multirow{2}{*}[-4em]{0.0}      &
                                        \multicolumn{1}{c}{\pecost{##1}} & \multicolumn{1}{c}{\pecost{##2}}                            \\[-1em]
                                        &                                 &  \begin{minipage}[t]{0.03\linewidth} \vspace{0pt}  (a)  \end{minipage}  \begin{minipage}[t]{0.4\linewidth}\vspace{0pt} #1 \end{minipage}                         & \begin{minipage}[t]{0.03\linewidth} \vspace{0pt}  (b)  \end{minipage}  \begin{minipage}[t]{0.4\linewidth}\vspace{0pt} #2 \end{minipage}                      \\[-1em]
                                        & \multirow{2}{*}[-4em]{0.5}
                                        & \multicolumn{1}{c}{\pecost{##3}} & \multicolumn{1}{c}{\pecost{##4}}                          \\[-1em]
                                        &                                 & \begin{minipage}[t]{0.03\linewidth} \vspace{0pt}   (c)  \end{minipage}  \begin{minipage}[t]{0.4\linewidth}\vspace{0pt} #3 \end{minipage}                        & \begin{minipage}[t]{0.03\linewidth} \vspace{0pt}  (d)  \end{minipage}  \begin{minipage}[t]{0.4\linewidth}\vspace{0pt} #4 \end{minipage}                     \\[-1em]
                                        & \multirow{2}{*}[-4em]{1.0}
                                        & \multicolumn{1}{c}{\pecost{##5}} & \multicolumn{1}{c}{\pecost{##6}}                          \\[-1em]
                                        &                                 & \begin{minipage}[t]{0.03\linewidth} \vspace{0pt}   (e)  \end{minipage}  \begin{minipage}[t]{0.4\linewidth}\vspace{0pt} #5 \end{minipage}                        & \begin{minipage}[t]{0.03\linewidth} \vspace{0pt}  (f)  \end{minipage}  \begin{minipage}[t]{0.4\linewidth}\vspace{0pt} #6 \end{minipage}                     \\[-1em]
        \Xhline{2\arrayrulewidth}
    \end{tabular}
    }
}
\plottable
}

\subsection{Discussion of generated designs}
\label{sec:discussion}
\begin{figure*}[!htbp]
    \figurestabledelgam{
        \includegraphics[width=\linewidth]{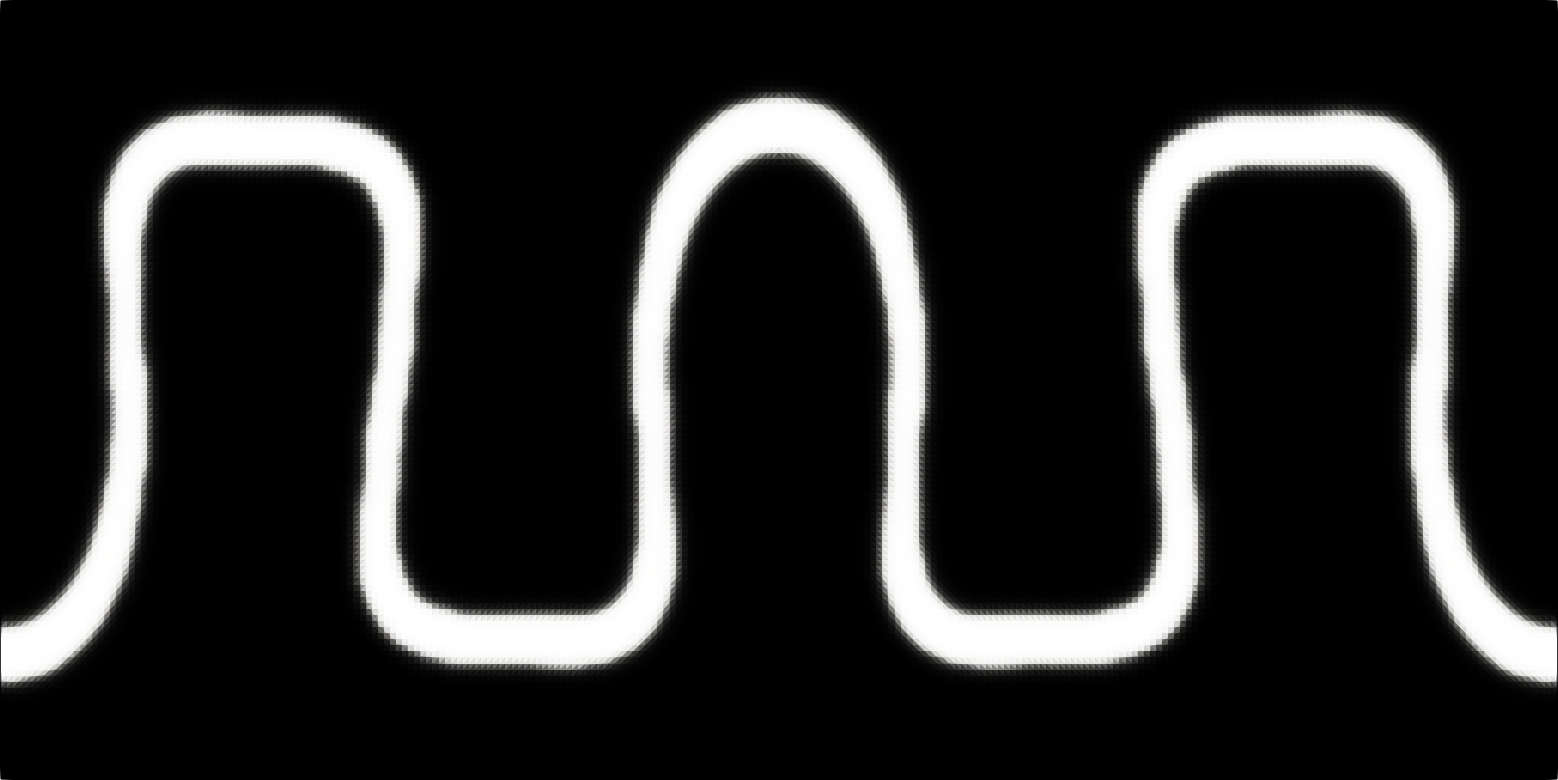}
    }{
        \includegraphics[width=\linewidth]{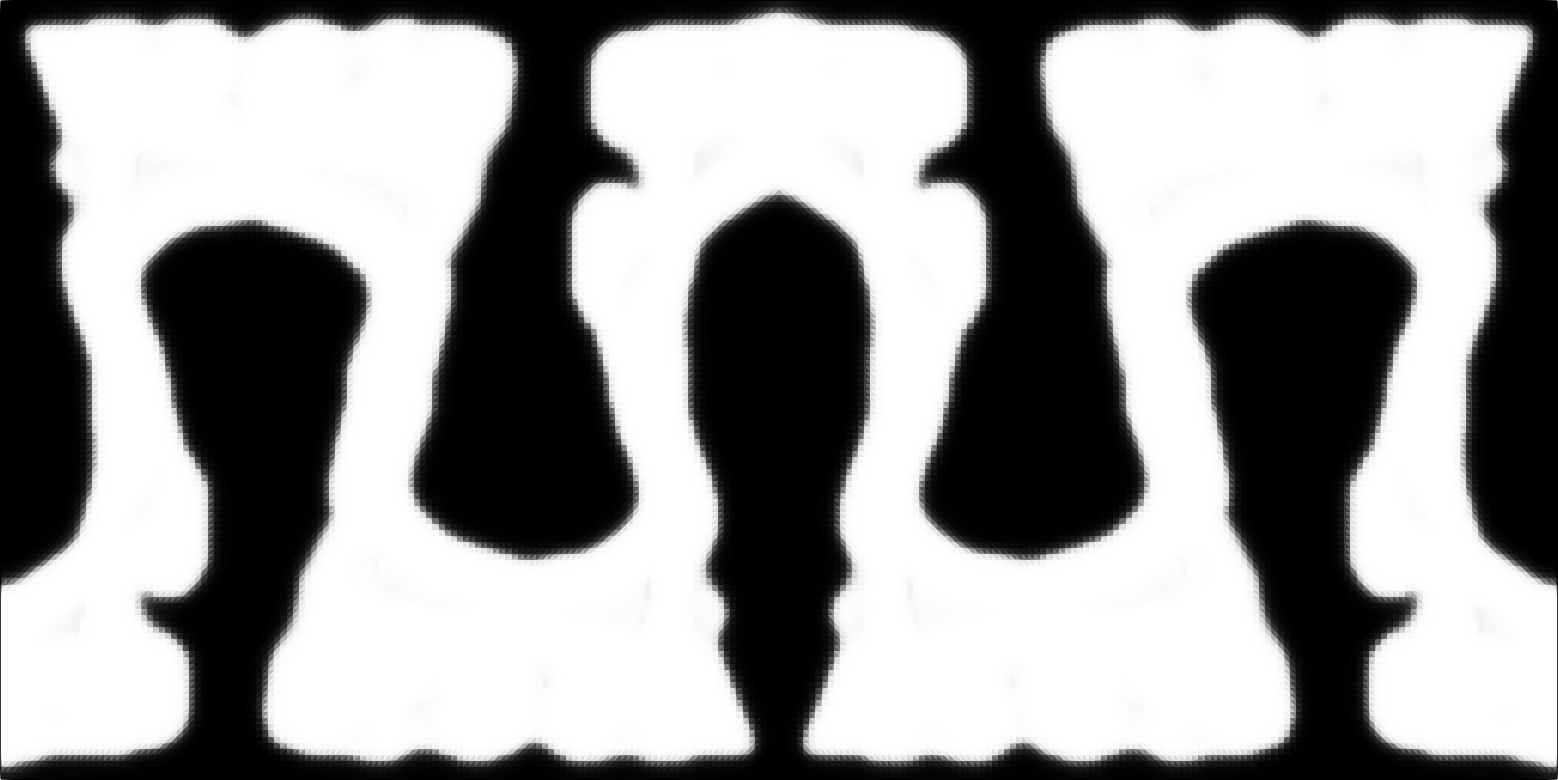}
    }{
        \includegraphics[width=\linewidth]{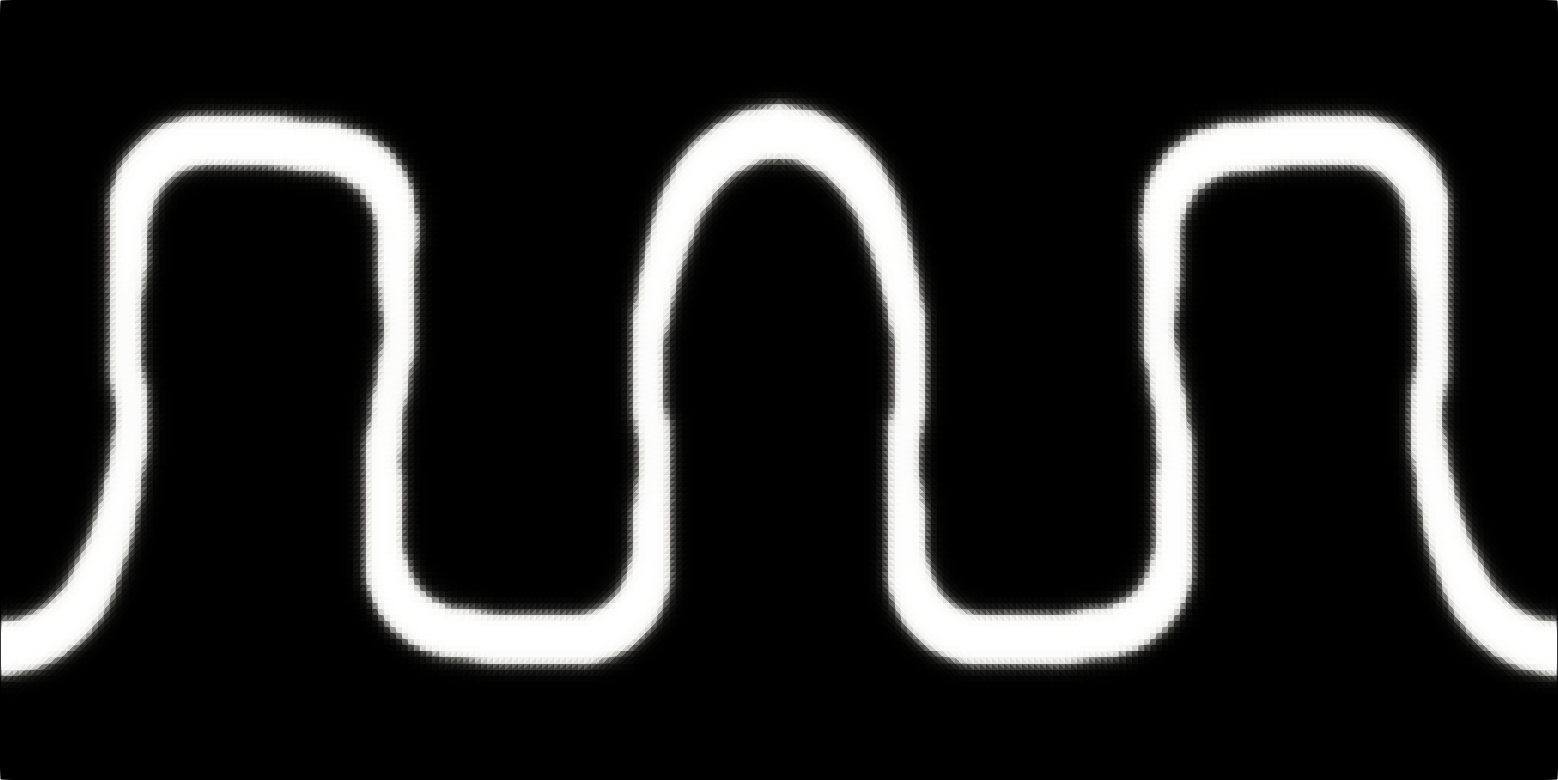}
    }{
        \includegraphics[width=\linewidth]{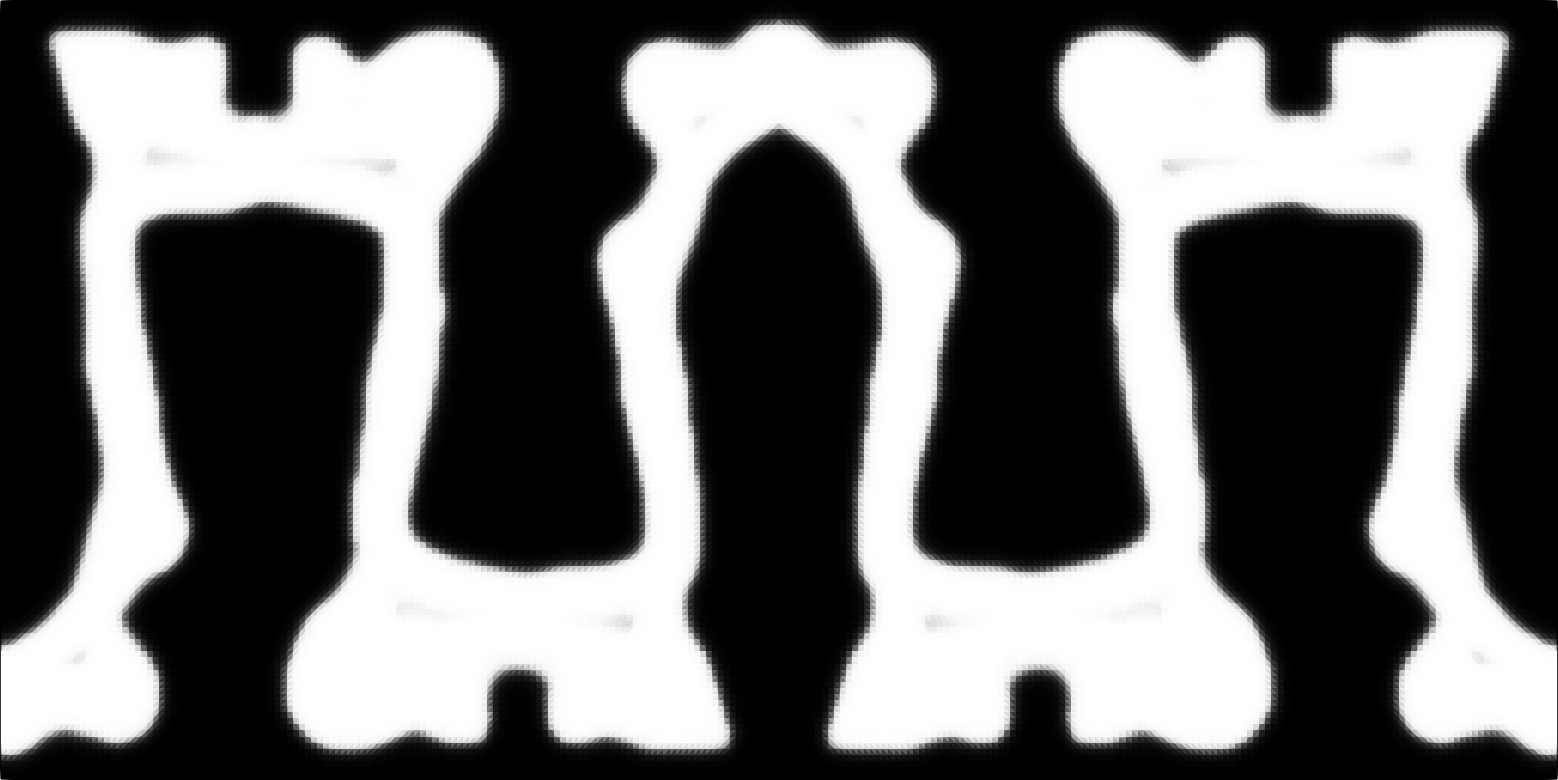}
    }{
        \includegraphics[width=\linewidth]{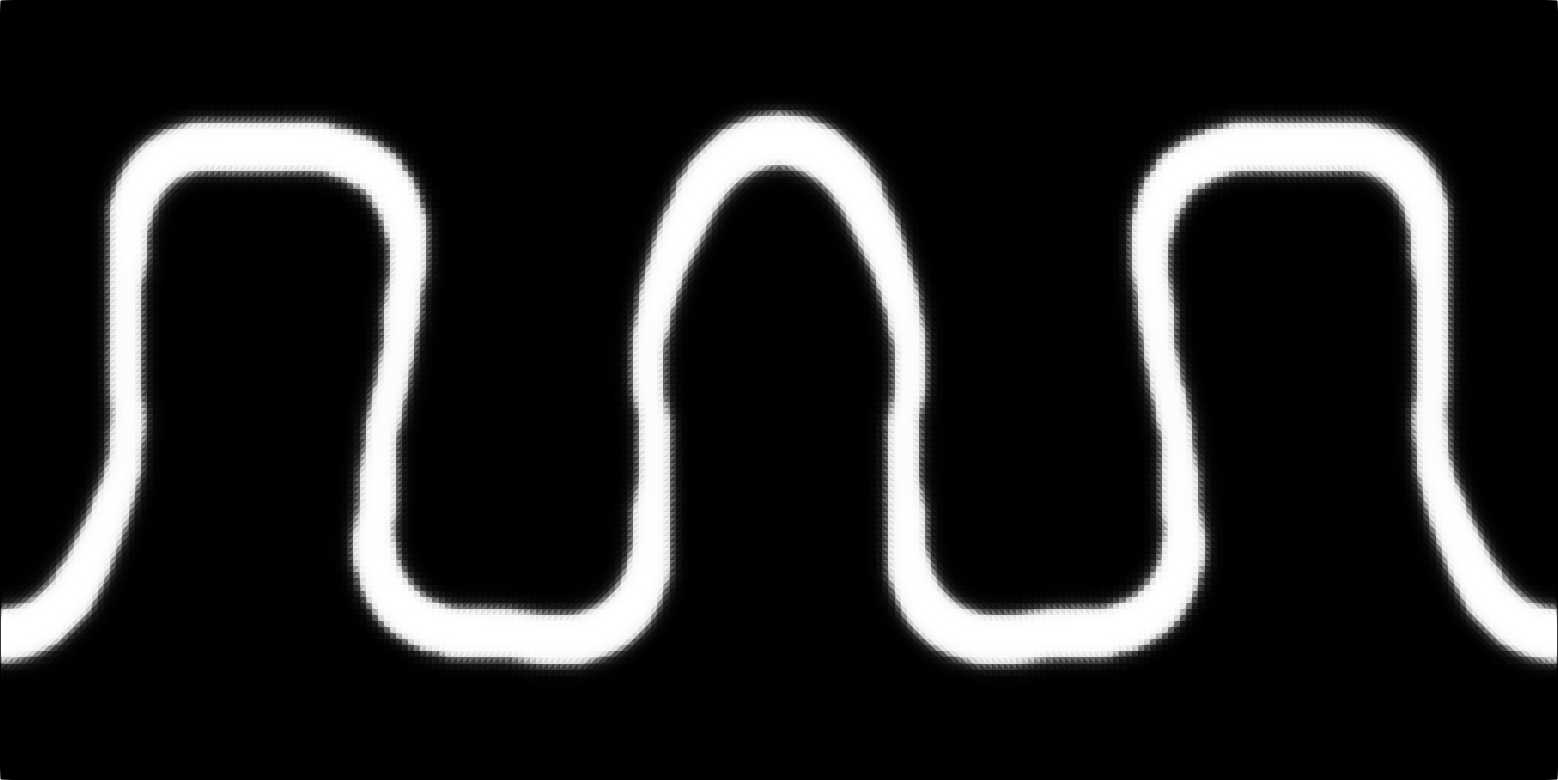}
    }{
        \includegraphics[width=\linewidth]{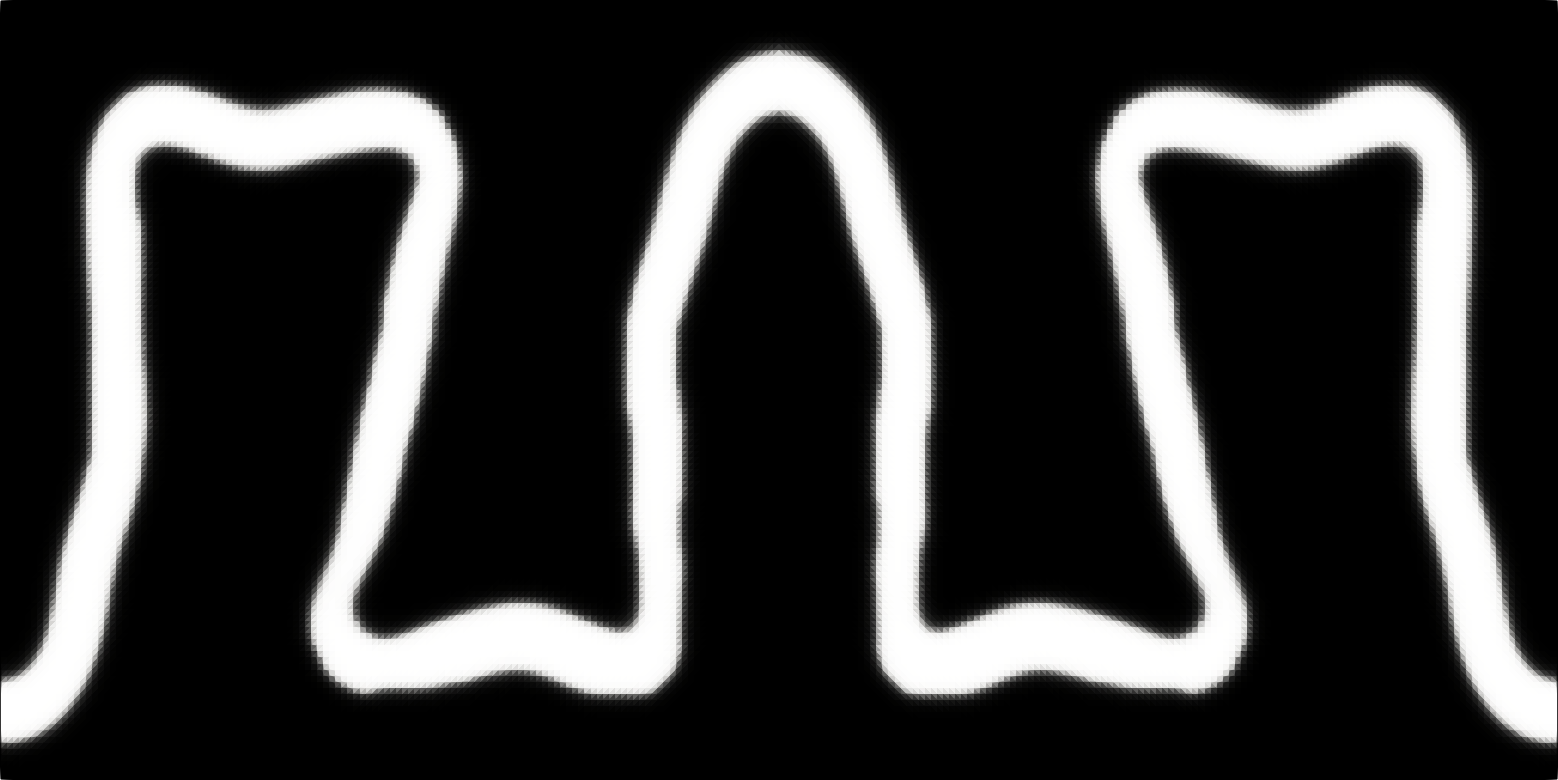}
    }{.0427\,(110\%), .0038\,(39\%)}{.1098\,(231\%), .0403\,(65\%)}{.0383\,(107\%), .0024\,(41\%)}{.1395\,(207\%), .0404\, (56\%)}{.0334\, (103\%), .0015\, (42\%)}{.2254\,(192\%), .0497\,(59\%)}
    {\phantomsubcaption \label{tab:pe_oribrug1a}}
    {\phantomsubcaption \label{tab:pe_oribrug1b}}
    {\phantomsubcaption \label{tab:pe_oribrug1c}}
    {\phantomsubcaption \label{tab:pe_oribrug1d}}
    {\phantomsubcaption \label{tab:pe_oribrug1e}}
    {\phantomsubcaption \label{tab:pe_oribrug1f}}
    \caption{Optimized porous electrode designs modeled with the original Bruggeman correlation, $\lambda=0.01$, and varying $\delta$ and $\gamma$. Black is $\bar{\rho} = 1$; white is $\bar{\rho} =0$.}
    \label{tab:pe_oribrug1}
\end{figure*}

\begin{figure*}[!htbp]
    \centering
    \begin{multicols}{2}
        \centering
        \begin{subfigure}[b]{\linewidth}
            \centering
            Monolithic electrodes \\
            \includegraphics[width=\linewidth]{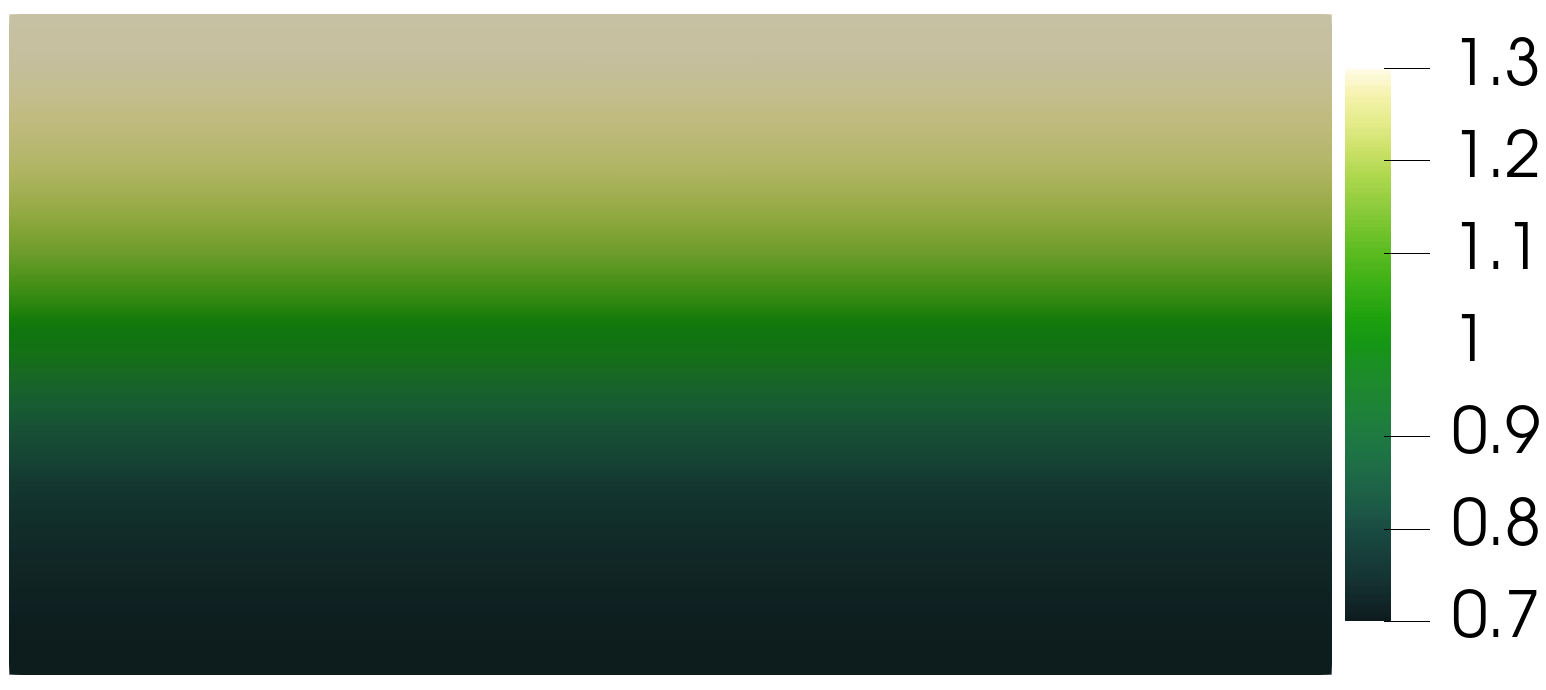}
            \caption{$\hat{c} \in [0.7, 1.3],\ \hat{c}_{\mrm{var}}=0.2161$}
            \label{fig:oribrug_mono_conc}
        \end{subfigure}
        \begin{subfigure}[b]{\linewidth}
            \includegraphics[width=\linewidth]{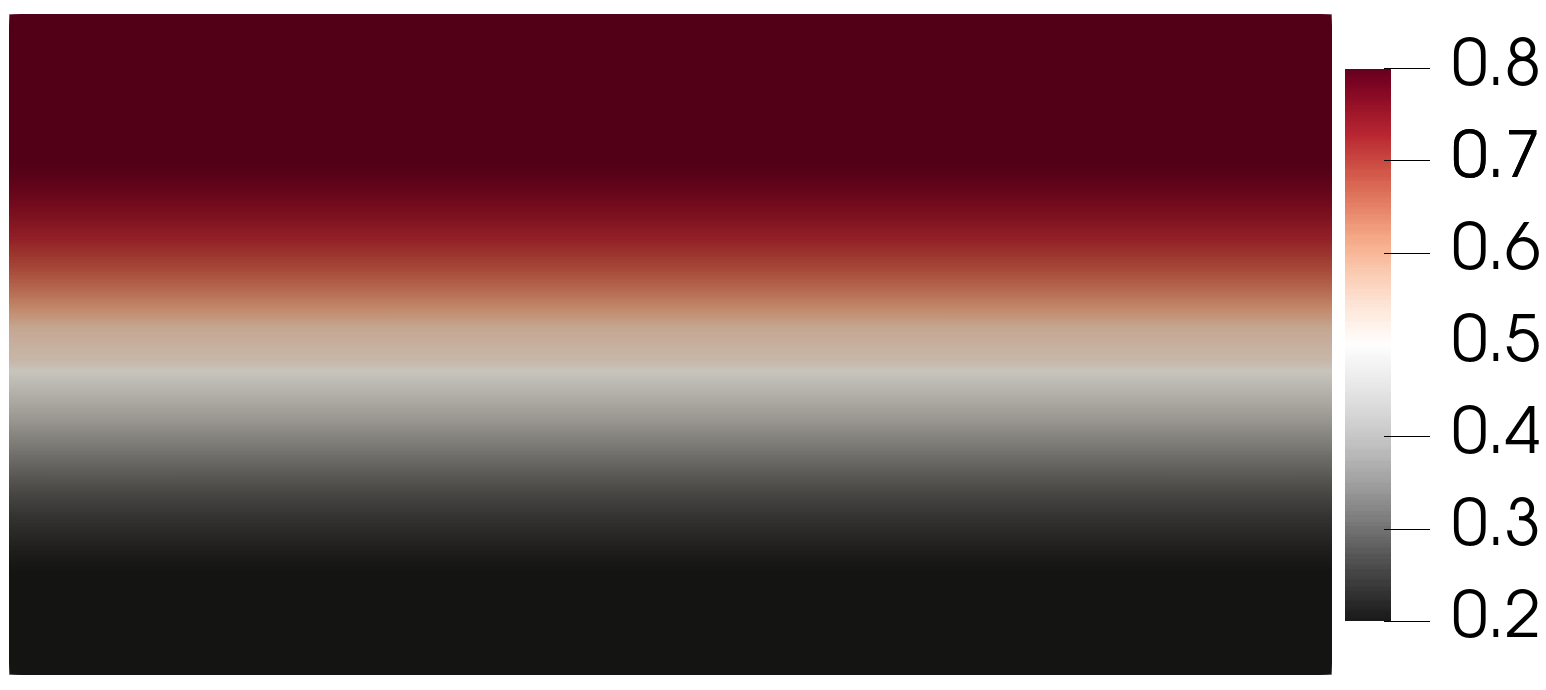}
            \caption{$\hat{\Phi}_2 \in [0.15, 0.85],\ |\hat{\nabla} \hat{\Phi}_2|_{\mrm{avg}} = 0.8280$}
            \label{fig:oribrug_mono_phi2}
        \end{subfigure}
        \begin{subfigure}[b]{\linewidth}
            \centering
            Optimized electrodes \\
            \includegraphics[width=\linewidth]{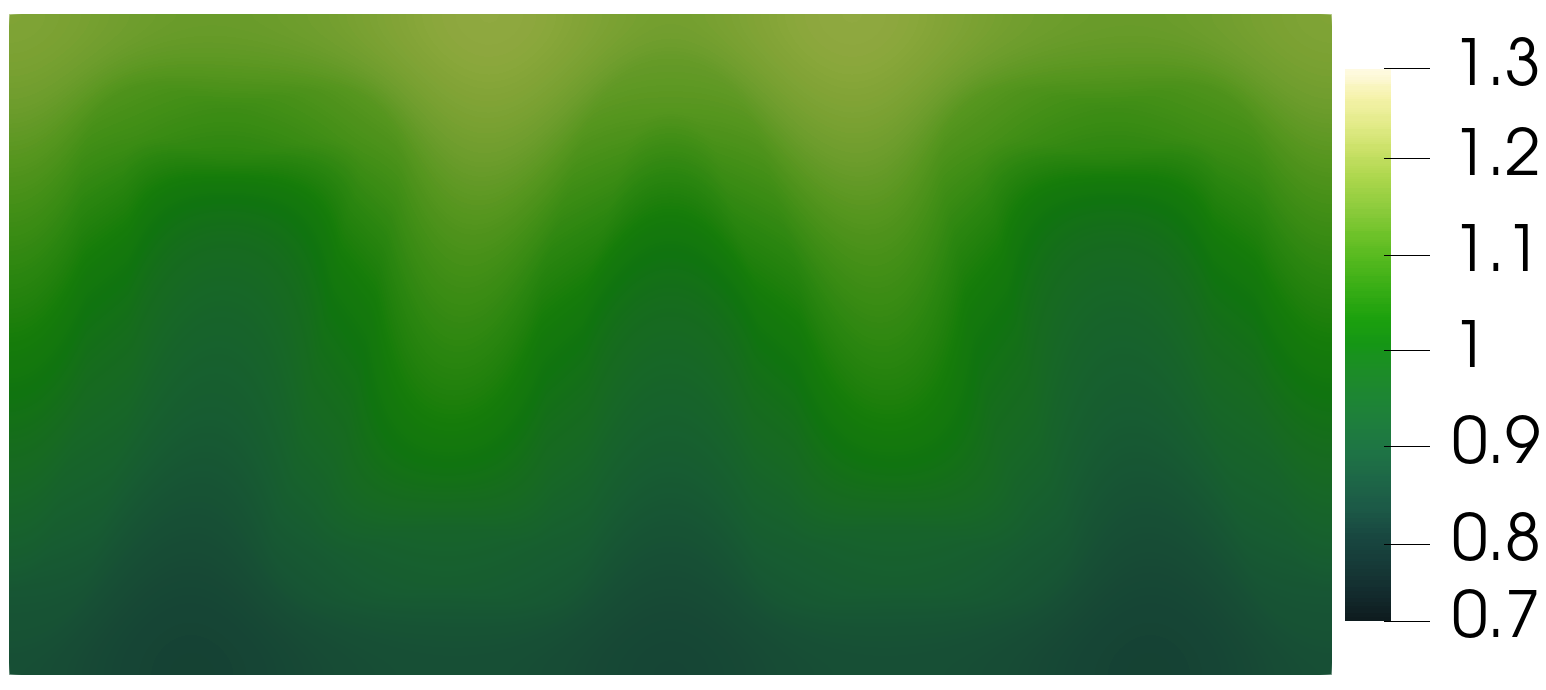}
            \caption{$\hat{c} \in [0.8, 1.2],\ \hat{c}_{\mrm{var}}=0.1015$}
            \label{fig:oribrug_opt_conc}
        \end{subfigure}
        \begin{subfigure}[b]{\linewidth}
            \includegraphics[width=\linewidth]{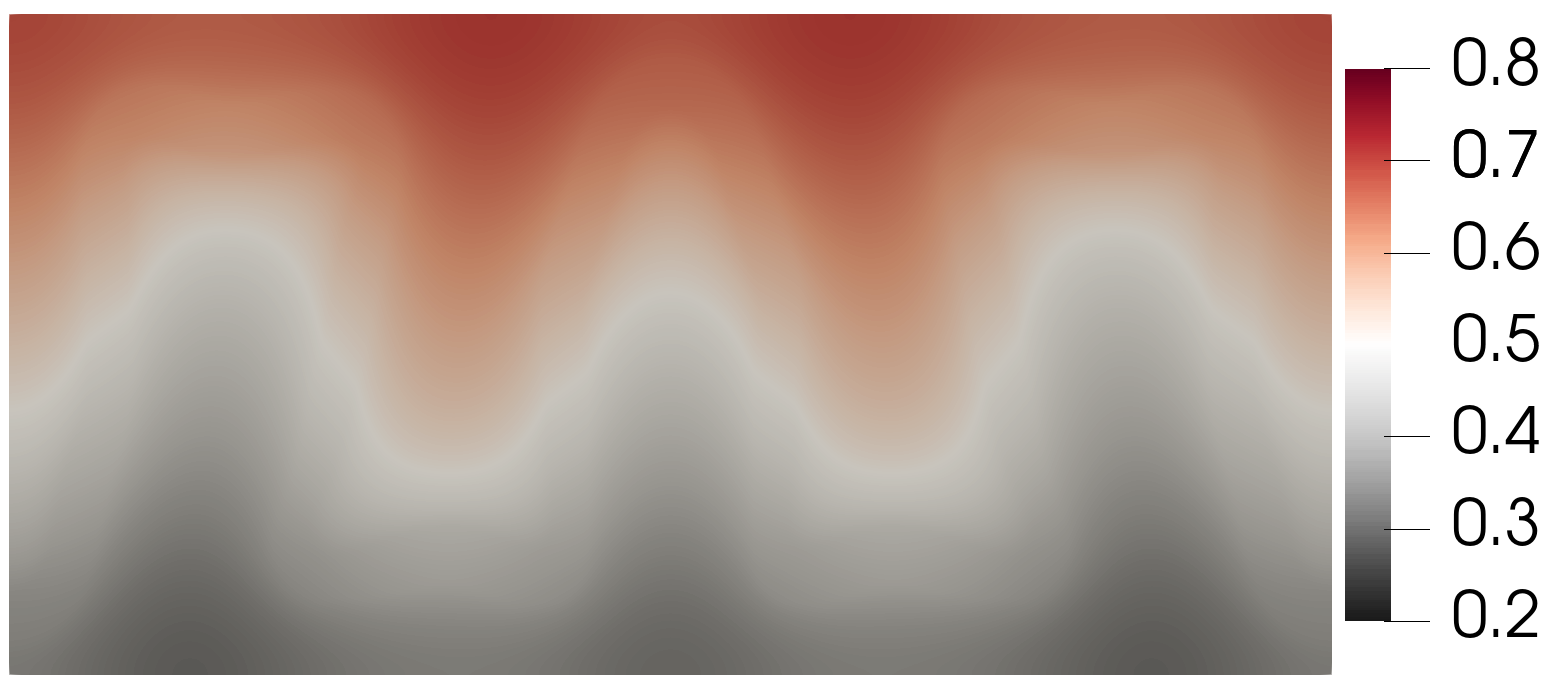}
            \caption{$\hat{\Phi}_2 \in [0.3, 0.7],\ |\hat{\nabla} \hat{\Phi}_2|_{\mrm{avg}} = 0.4932$}
            \label{fig:oribrug_opt_phi2}
        \end{subfigure}
    \end{multicols}
    \setlength{\abovecaptionskip}{-10pt}
    \caption{Concentration (top) and ionic potential (bottom) distributions of monolithic (left) and optimized (right) electrodes with $\delta=5$, $\gamma=1$, $\lambda=0.01$, and the original Bruggeman correlation.}
    \label{fig:oribrug_analysis_comp}
\end{figure*}

\begin{figure*}[!htbp]
    \figurestabledelgam{
        \includegraphics[width=\linewidth]{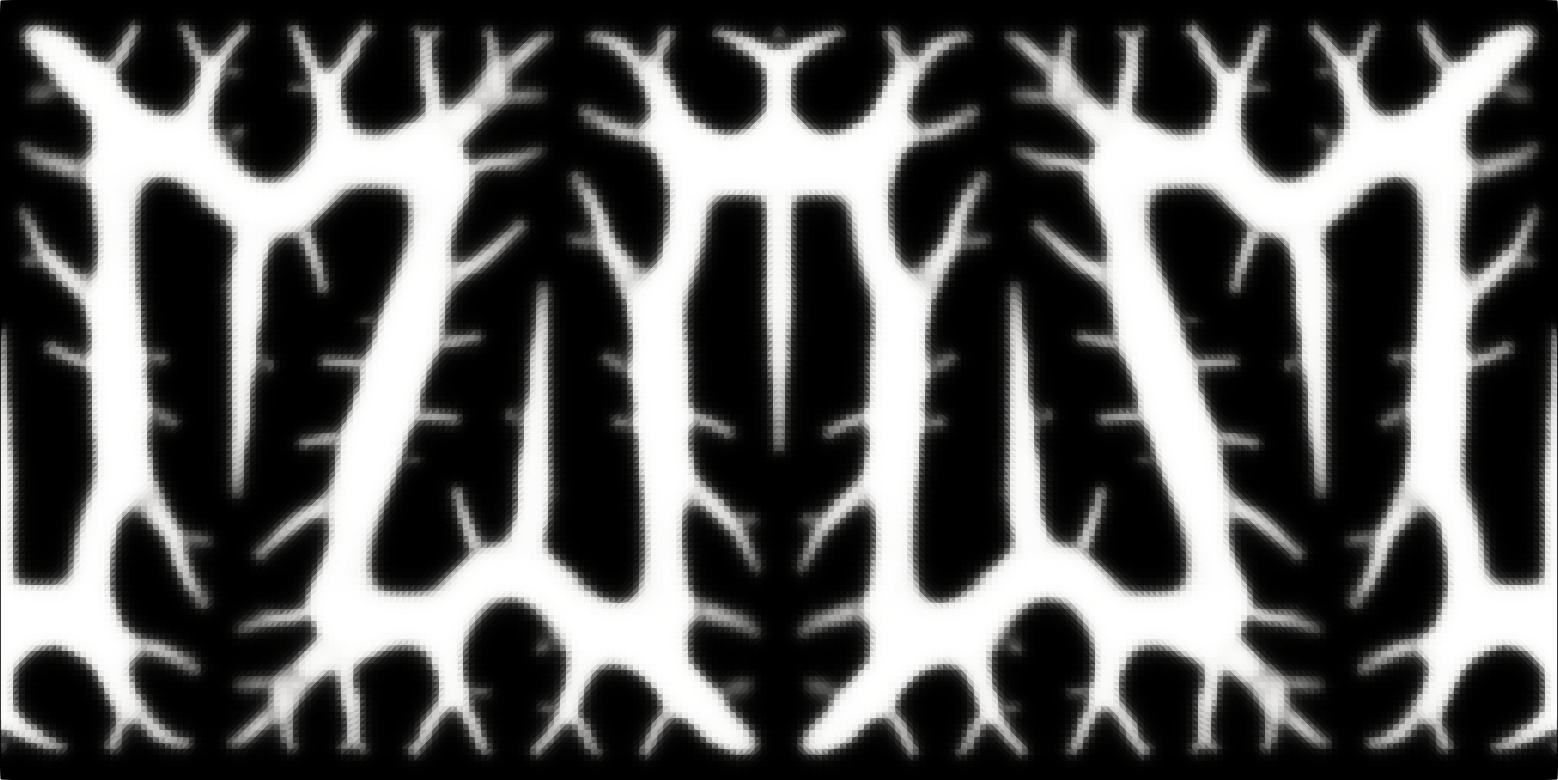}
    }{
        \includegraphics[width=\linewidth]{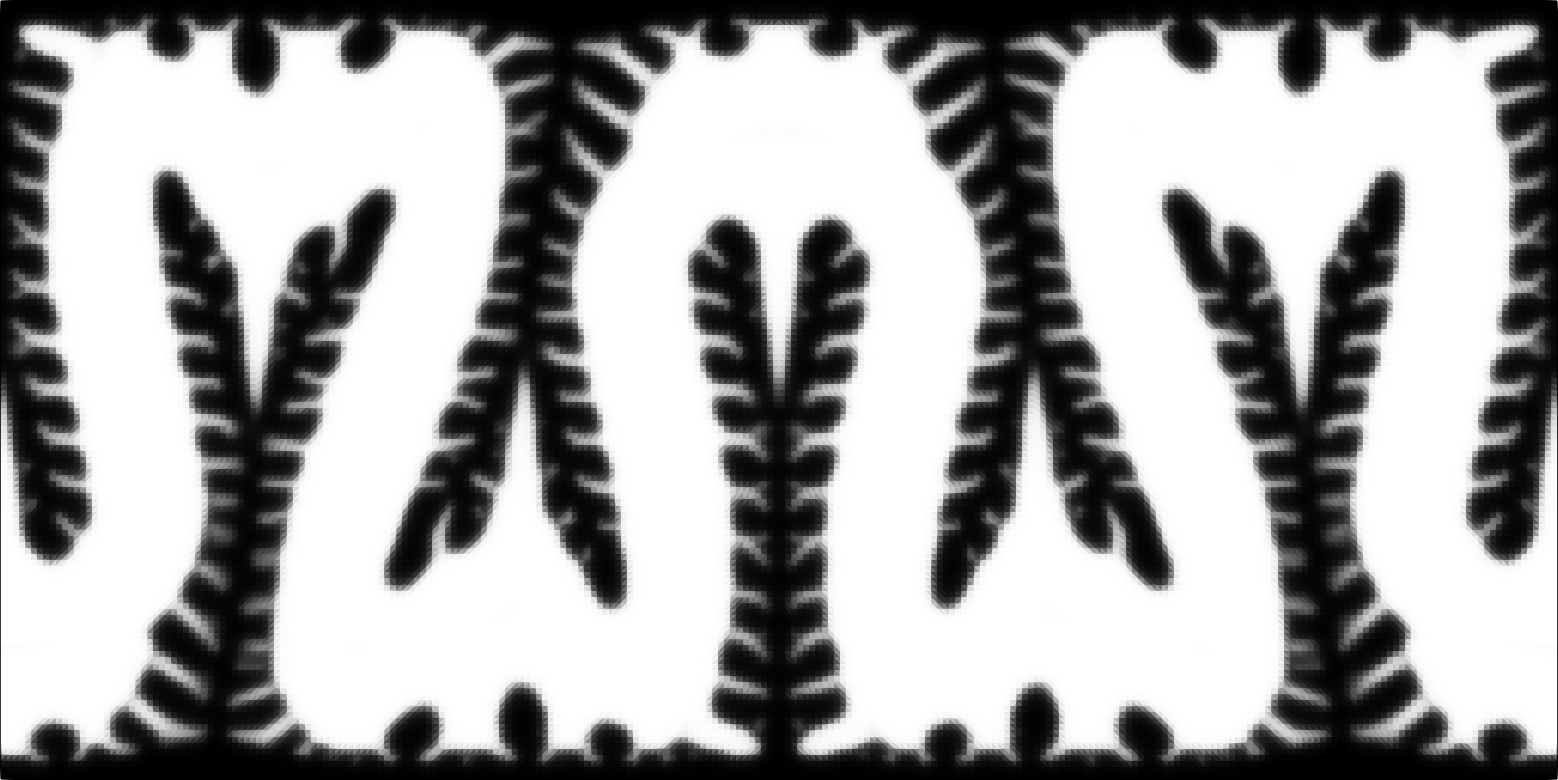}
    }{
        \includegraphics[width=\linewidth]{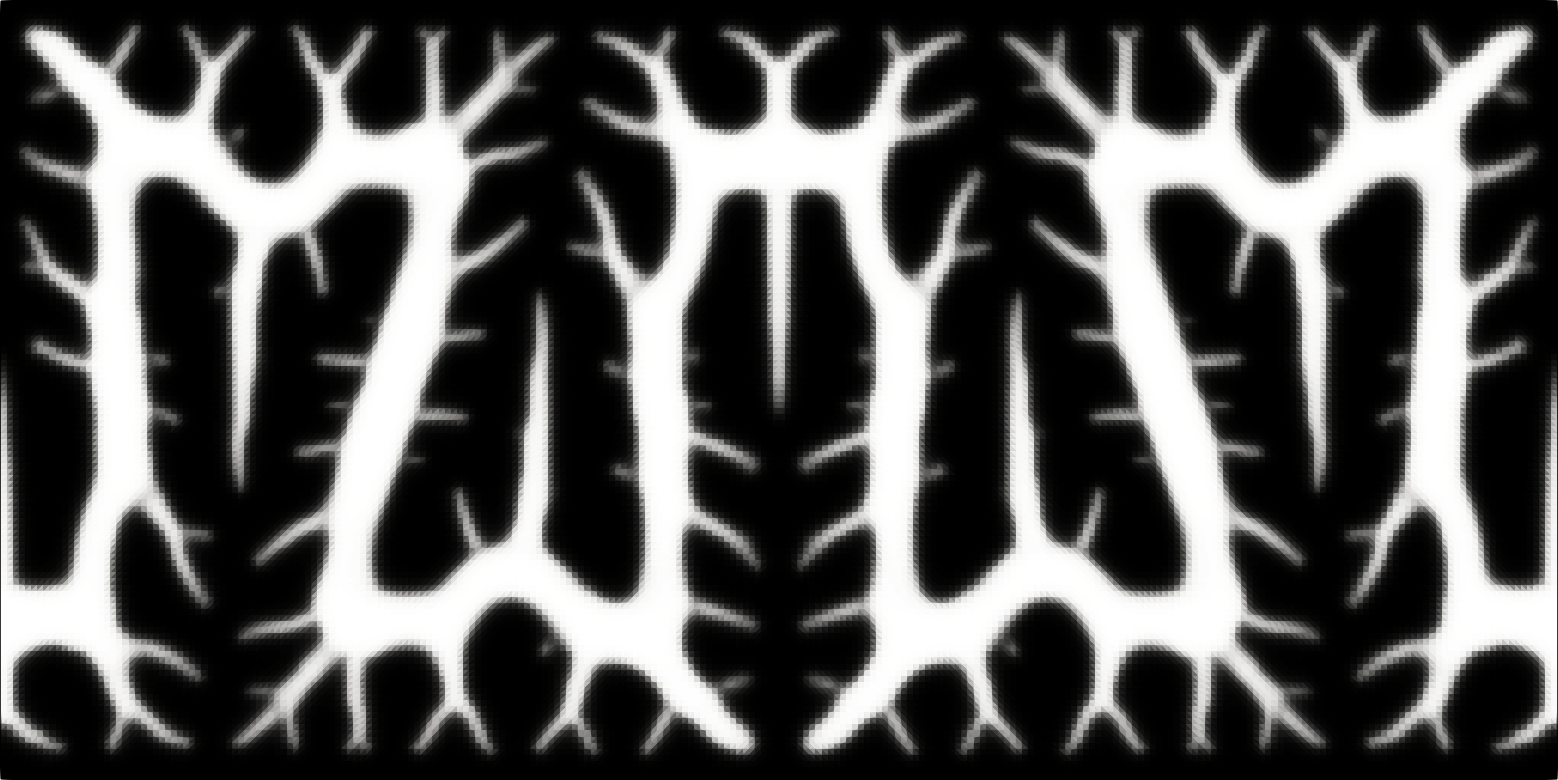}
    }{
        \includegraphics[width=\linewidth]{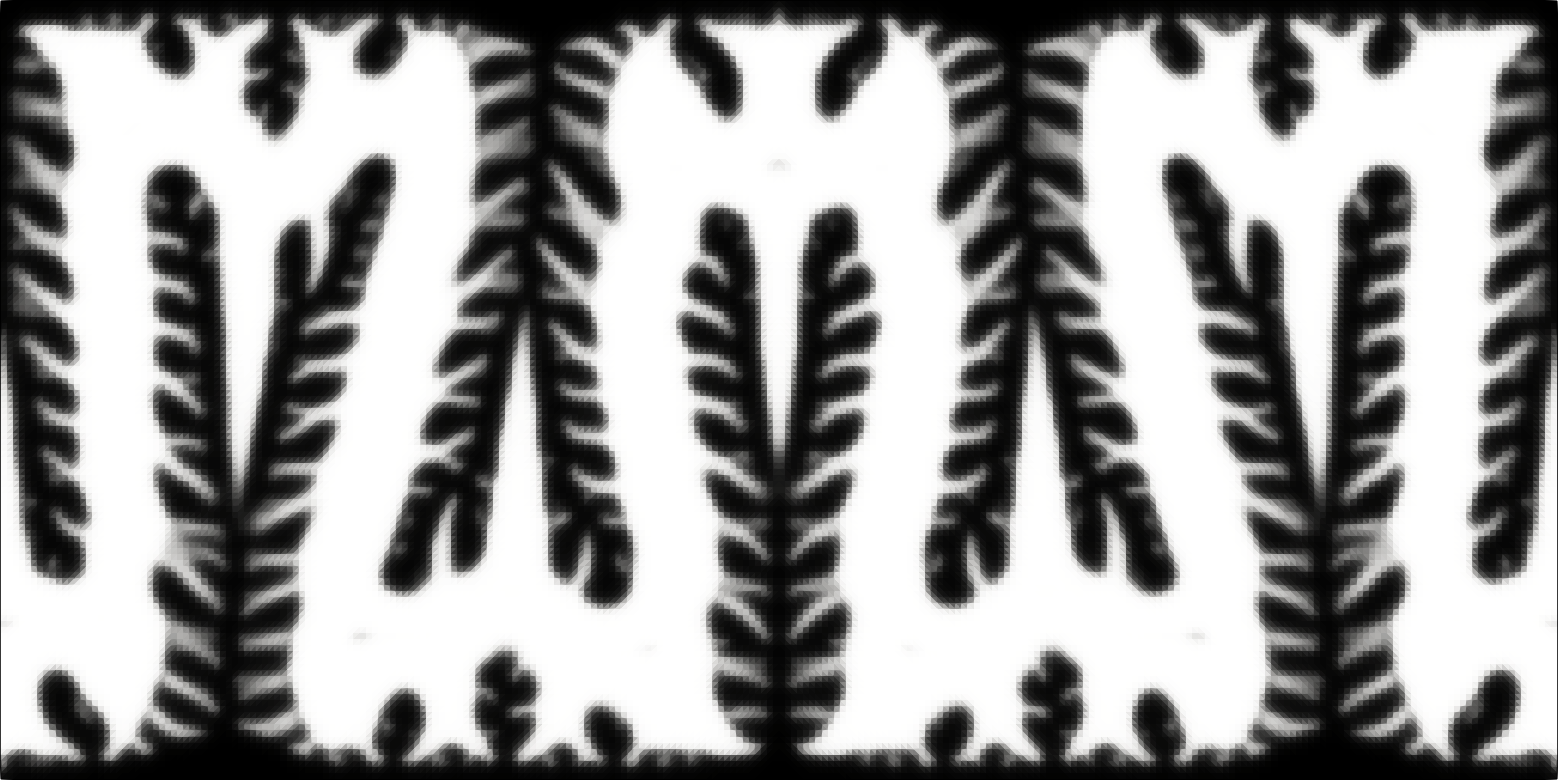}
    }{
        \includegraphics[width=\linewidth]{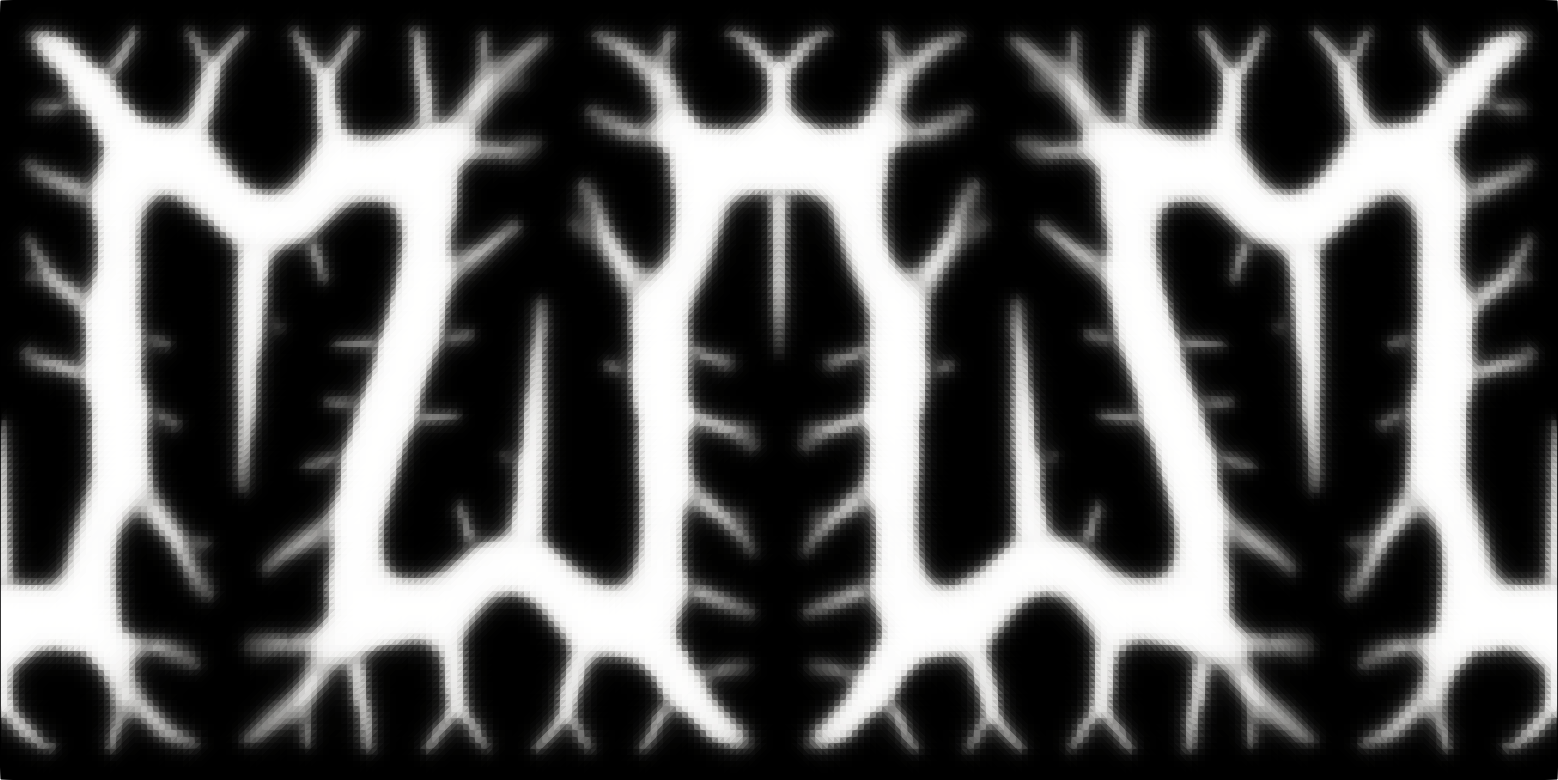}
    }{
        \includegraphics[width=\linewidth]{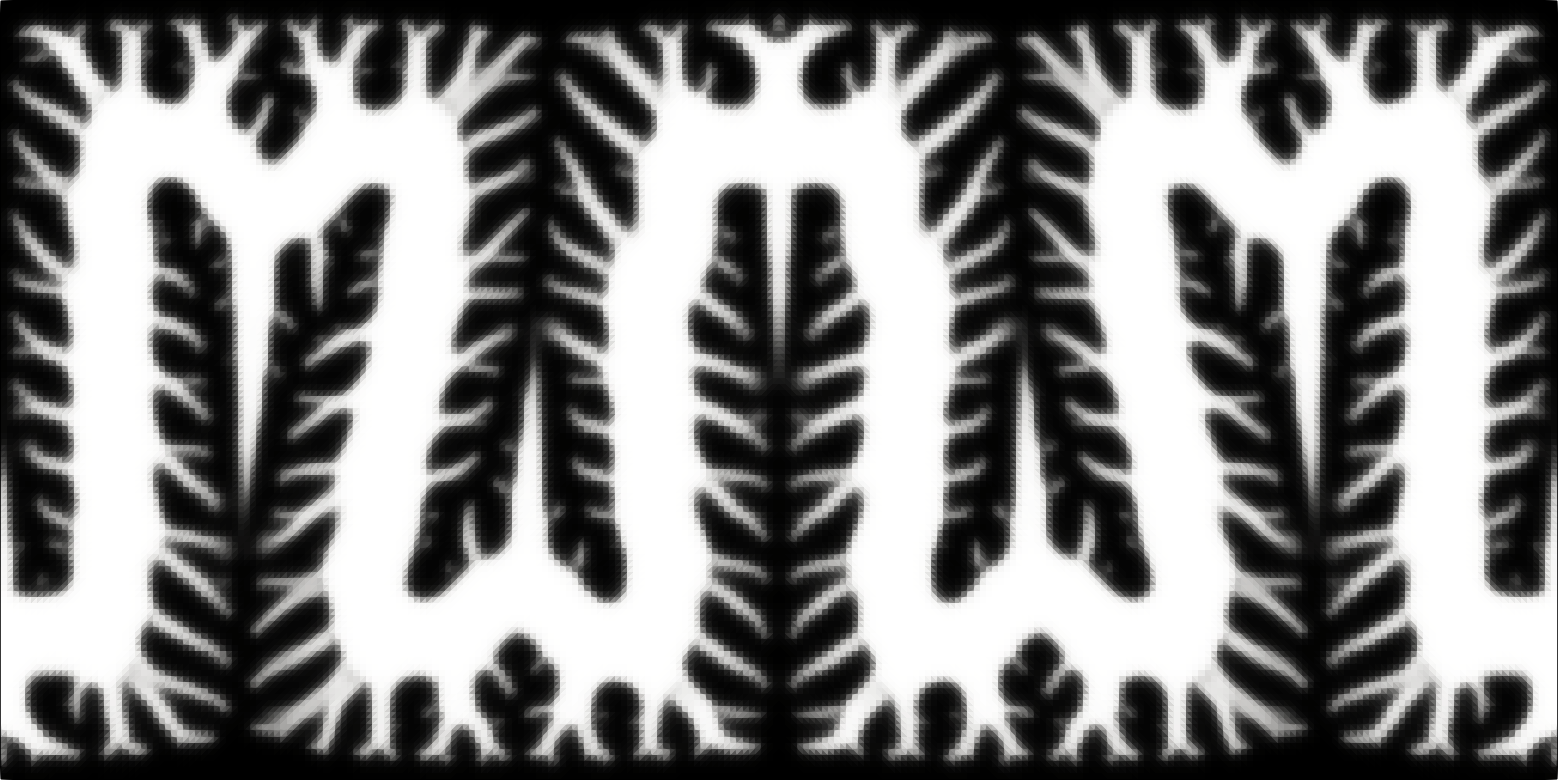}
    }{.0328\,(656\%), .0101\,(204\%)}{.0962\,(786\%), .0465\,(312\%)}{.0291\,(557\%), .0069\,(120\%)}{.1152\,(855\%), .0400\,(267\%)}{.0248\,(452\%), .0048\,(72\%)}{.1330\,(851\%), .0414\,(265\%)} %
    {\phantomsubcaption \label{tab:pe_modbrug1a}}
    {\phantomsubcaption \label{tab:pe_modbrug1b}}
    {\phantomsubcaption \label{tab:pe_modbrug1c}}
    {\phantomsubcaption \label{tab:pe_modbrug1d}}
    {\phantomsubcaption \label{tab:pe_modbrug1e}}
    {\phantomsubcaption \label{tab:pe_modbrug1f}}
    \caption{Optimized porous electrode designs modeled with the modified Bruggeman correlation, $\lambda=0.01$, and varying $\delta$ and $\gamma$. Black is $\bar{\rho} = 1$; white is $\bar{\rho} =0$.}
    \label{tab:pe_modbrug1}
\end{figure*}

\begin{figure*}[!htbp]
    \figurestablelamgam{
        \includegraphics[width=\linewidth]{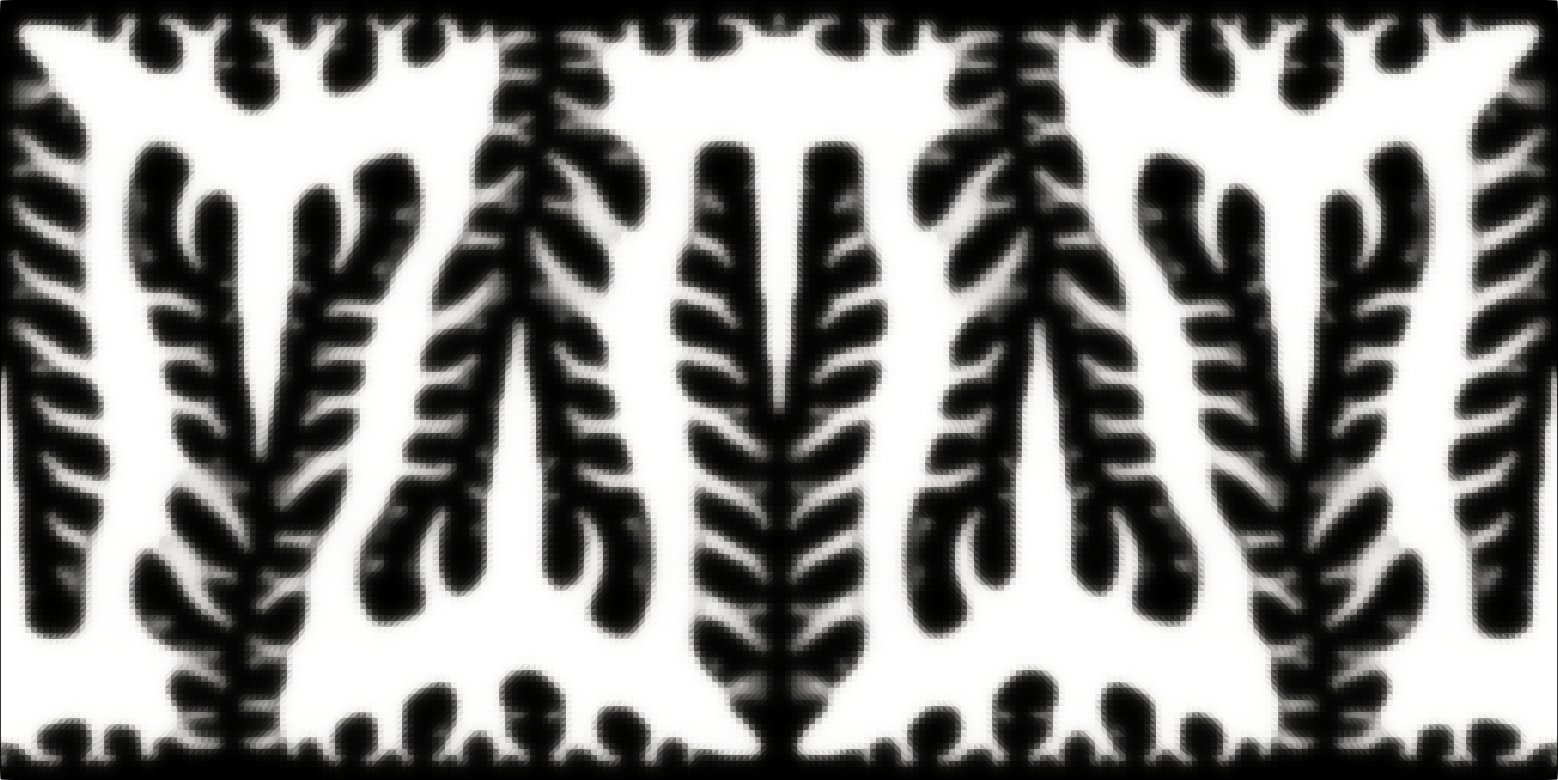}
    }{
        \includegraphics[width=\linewidth]{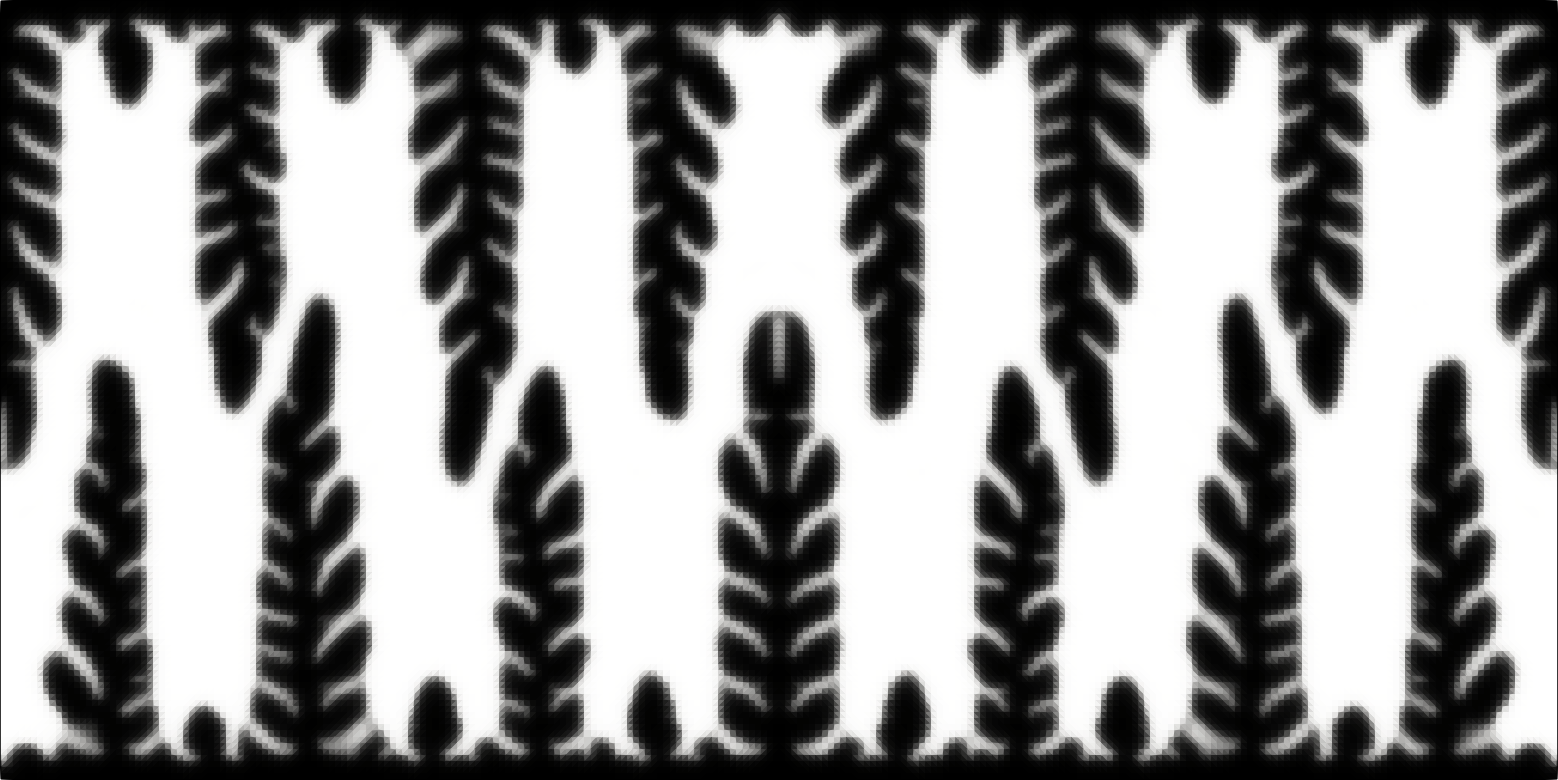}
    }{
        \includegraphics[width=\linewidth]{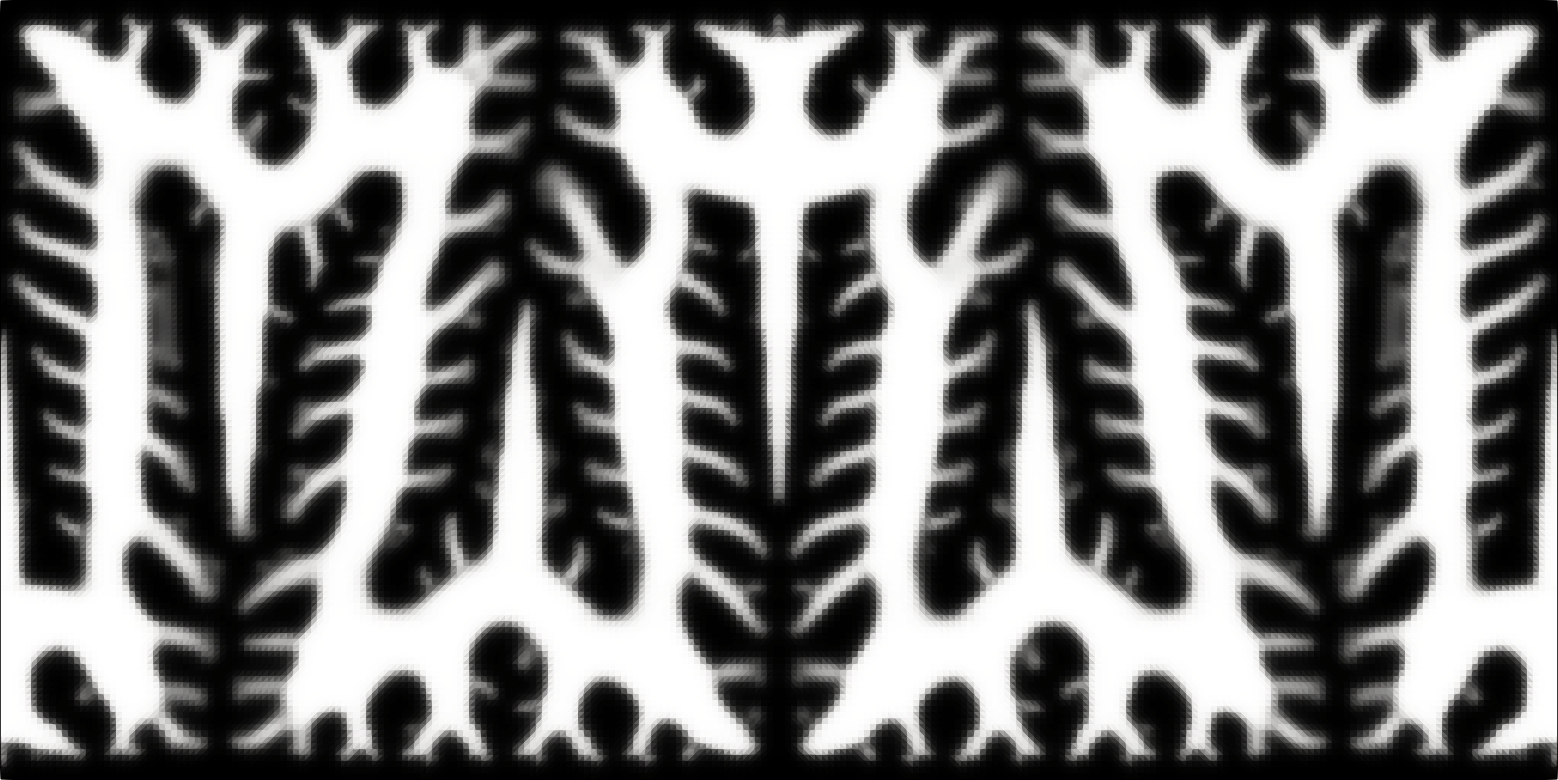}
    }{
        \includegraphics[width=\linewidth]{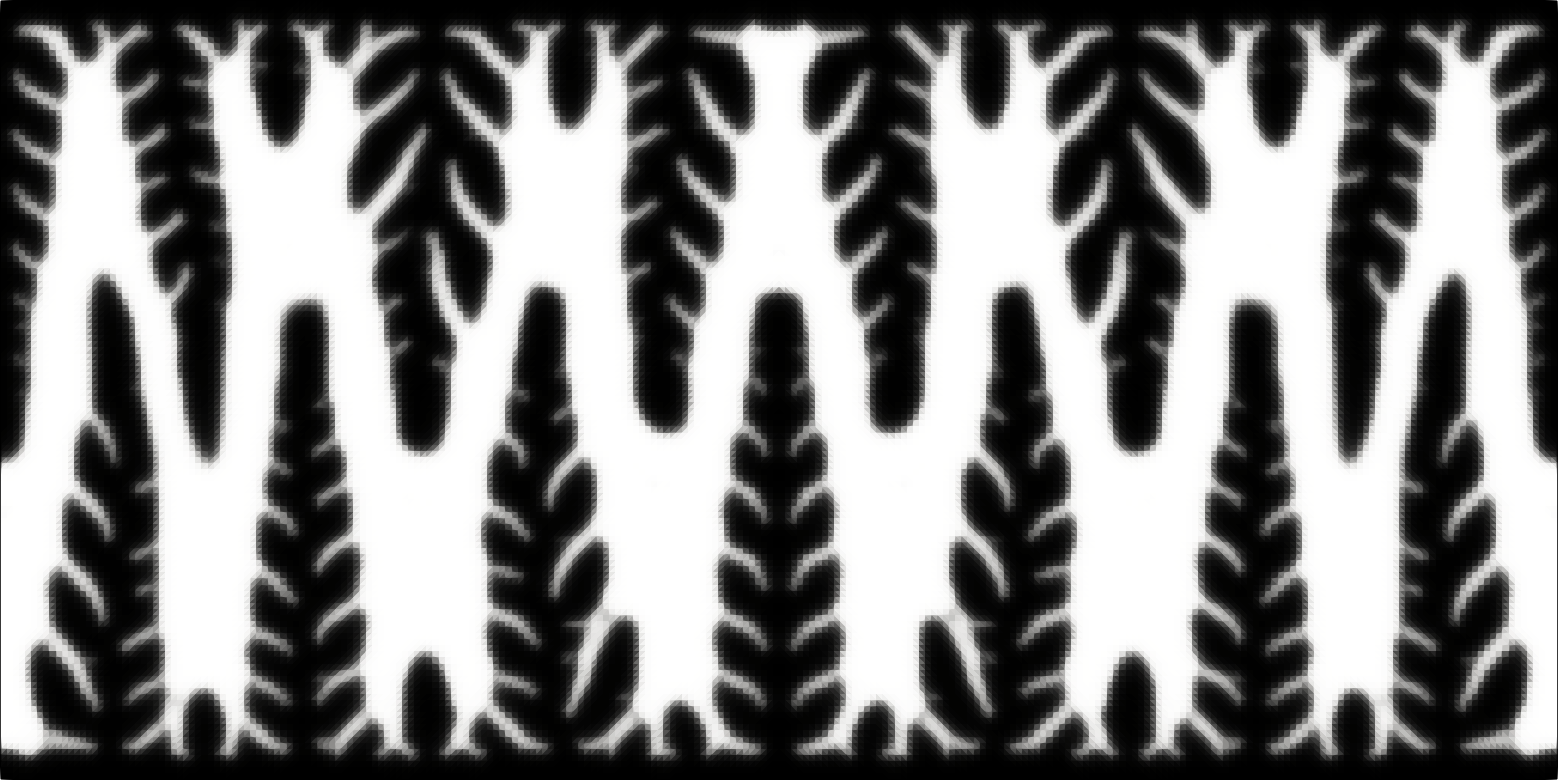}
    }{
        \includegraphics[width=\linewidth]{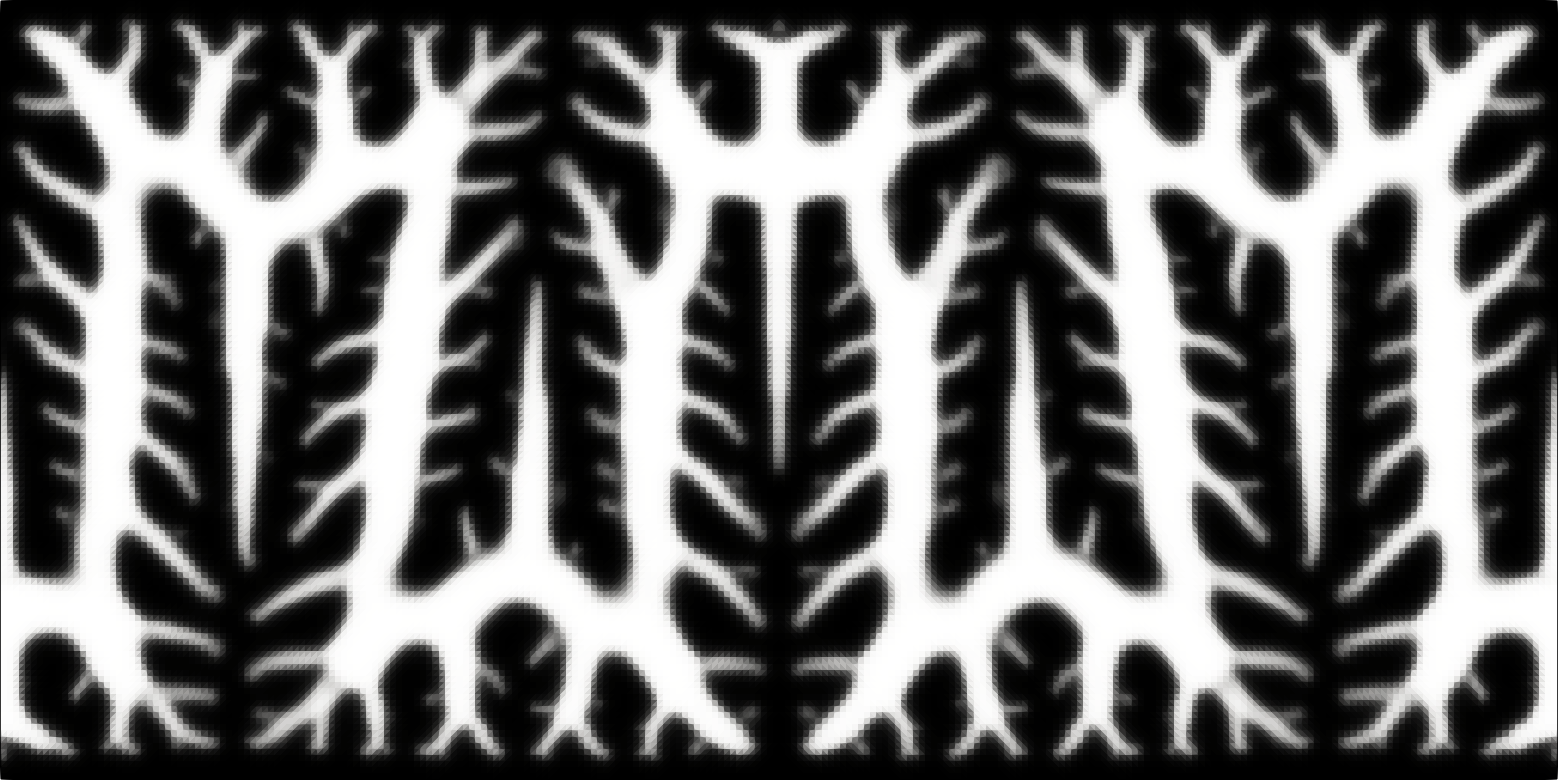}
    }{
        \includegraphics[width=\linewidth]{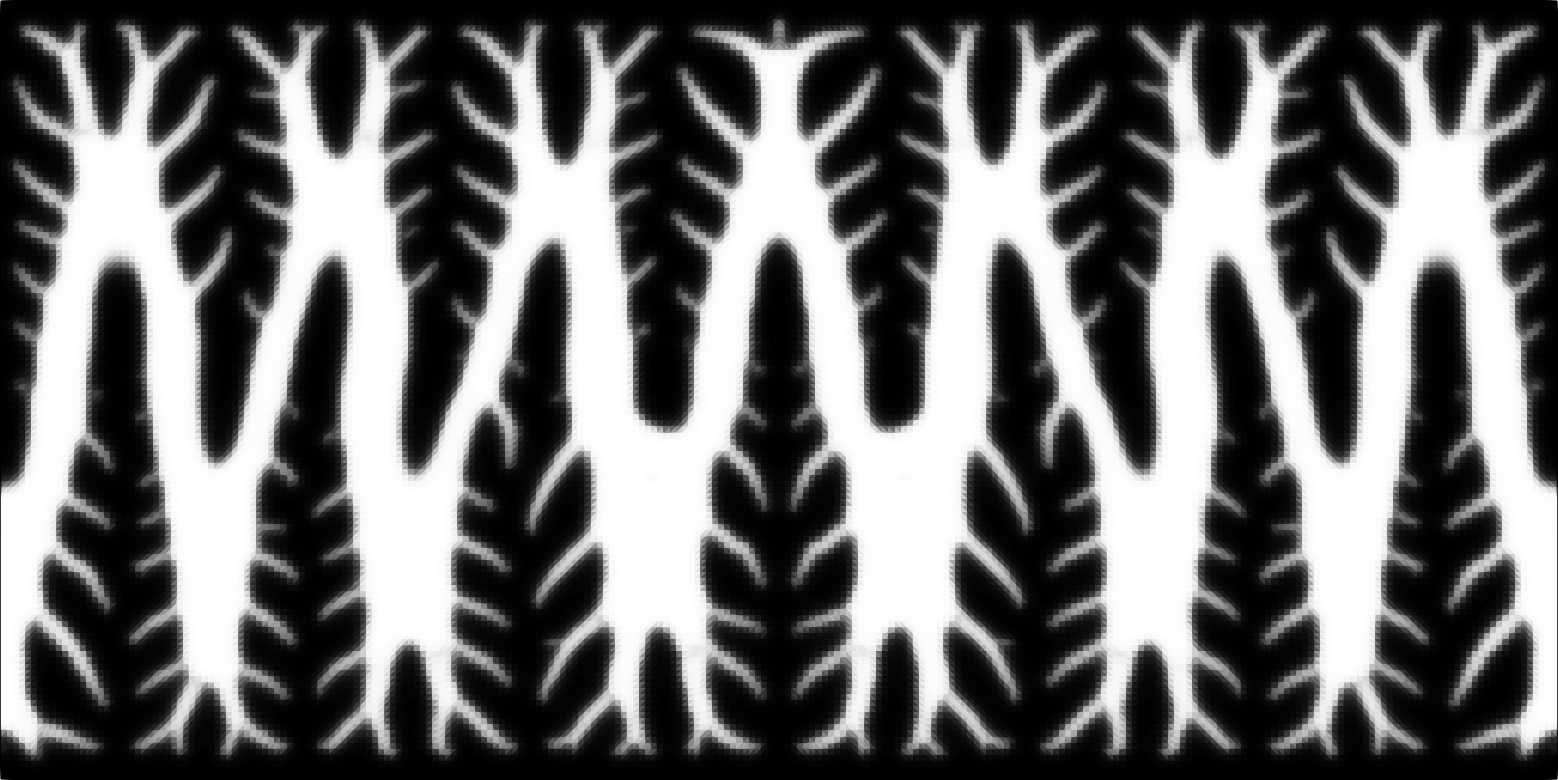}
    }{.0803\,(858\%), .0189\,(165\%)}{.0601\,(617\%), .0167\,(136\%)}{.0761\,(798\%), .0169\,(159\%)}{.0592\,(602\%), .0195\,(173\%)}{.0734\,(741\%), .0153\,(155\%)}{.0572\,(594\%), .0131\,(160\%)} %
    {\phantomsubcaption \label{tab:pe_modbrug2a}}
    {\phantomsubcaption \label{tab:pe_modbrug2b}}
    {\phantomsubcaption \label{tab:pe_modbrug2c}}
    {\phantomsubcaption \label{tab:pe_modbrug2d}}
    {\phantomsubcaption \label{tab:pe_modbrug2e}}
    {\phantomsubcaption \label{tab:pe_modbrug2f}}
    \caption{Optimized porous electrode designs modeled with the modified Bruggeman correlation, $\delta=2$, and varying $\lambda$ and $\gamma$. Black is $\bar{\rho} = 1$; white is $\bar{\rho} =0$.}
    \label{tab:pe_modbrug2}
\end{figure*}

We choose energy storage and ohmic loss as performance metrics and compare the optimized designs to a monolithic design with electrode separation distance of 0.05, as this is the average gap among all optimized designs.
The optimized designs are shown in Fig.~\ref{tab:pe_oribrug1} - ~\ref{tab:pe_modbrug2} with the percentage improvement over the monolithic design.
In the following subsections, we discuss the designs corresponding to various combinations of the three nondimensional parameters, namely $\lambda$, $\delta$, and $\gamma$, described in Table~\ref{tab:forward_param_variable}, and Bruggeman correlation functions (which approximates the tortuosity of the ion-diffusion pathway within the porous electrodes.)

Here, $\lambda$, i.e., the ratio between ionic and total conductivity (defined in ~\eqref{eq:dimless_groups}), affects the macroscale localization of the reaction front.
In cases where ionic conductivity is limiting ($\lambda < 0.5$), the reaction front localizes at the electrode/separator interface.
Conversely, when the electronic conductivity is the limiting factor, ($\lambda > 0.5$), the reaction front shifts towards the current collector.

The parameter $\delta$, i.e., the ratio between ohmic and kinetic resistance (defined in ~\eqref{eq:dimless_groups}), controls the electrode penetration depth.
In the absence of concentration gradients, the relative reaction rate between the back and the front of electrodes can be quantified by \citep{fuller2018electrochemical}
\begin{equation}
\label{eq:relative_rate}
    \text{Relative rate} \approx \frac{1}{\cosh\left(\sqrt{\delta}\right)}.
\end{equation}
For small $\delta$ values, the system is dominated by the kinetic resistance, and results in a uniform or near-uniform distribution of current density, which leads to a uniform reaction throughout the electrode.
In this case, the macroscale geometry of the electrode has less effect on the electrode material utilization, i.e., monolithic designs perform sufficiently well.
On the other hand, for large $\delta$ values, the system is dominated by ohmic resistance and results in a non-uniform (localized) distribution of current density.
As mentioned in the introduction, thick monolothic electrodes perform poorly due to small penetration depth.
Therefore, this case favors designs with complex electrode geometries.

The parameter $\gamma$ quantifies the fractional contribution to the total reaction current from the redox reaction.
When the reaction is capacitive-dominated ($\gamma < 0.5$), the system is more likely to experience ion depletion, since both the cation and the anion are adsorbed by the porous electrodes.
Alternatively, when the reaction is redox-dominated ($\gamma>0.5$), the average concentration is stabilized from ions created by one of the two electrodes.

\subsubsection{Original Bruggeman correlation}
\begin{figure*}[!htbp]
    \figurestablelamgam{
        \includegraphics[width=\linewidth]{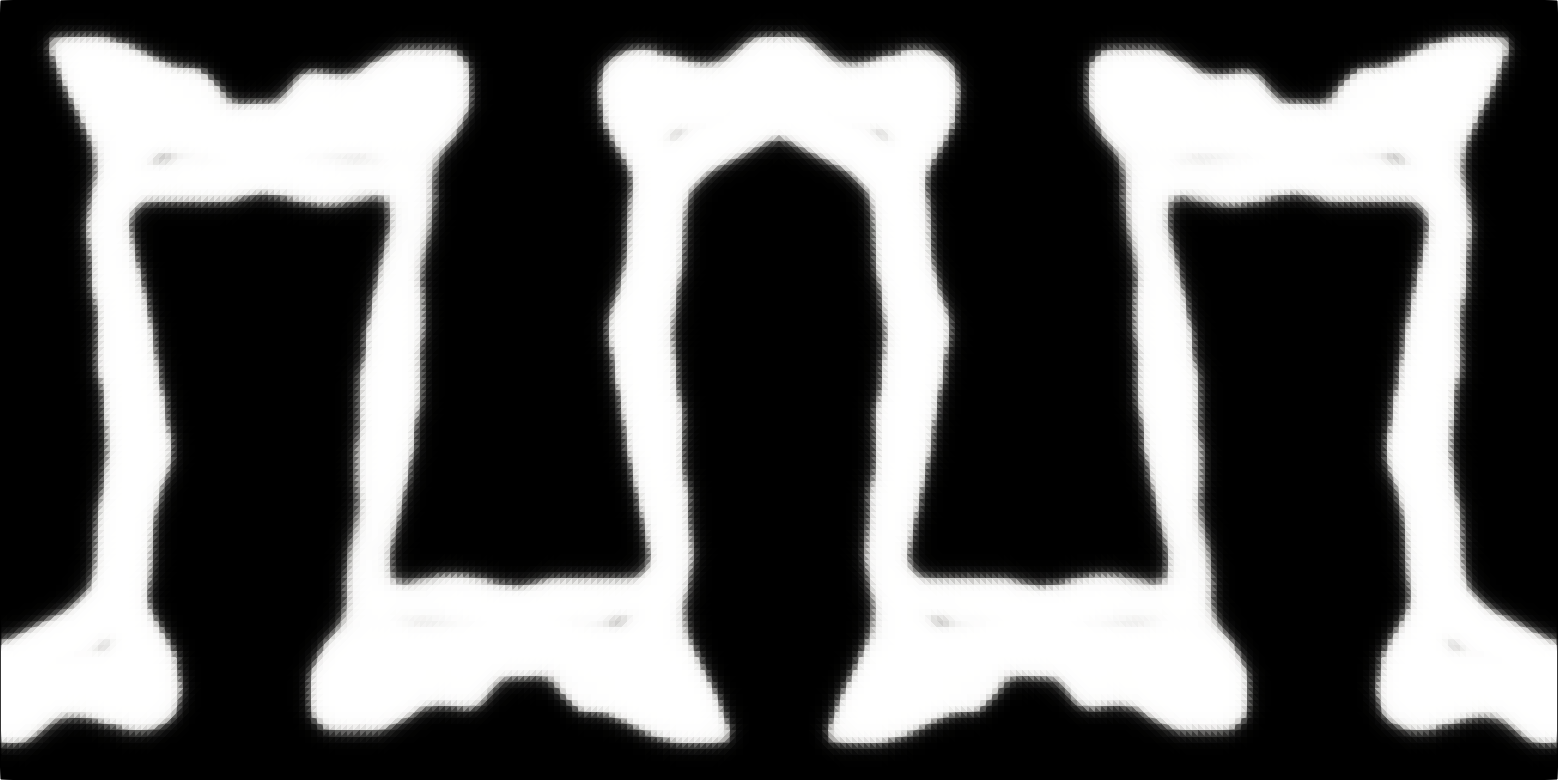}
    }{
        \includegraphics[width=\linewidth]{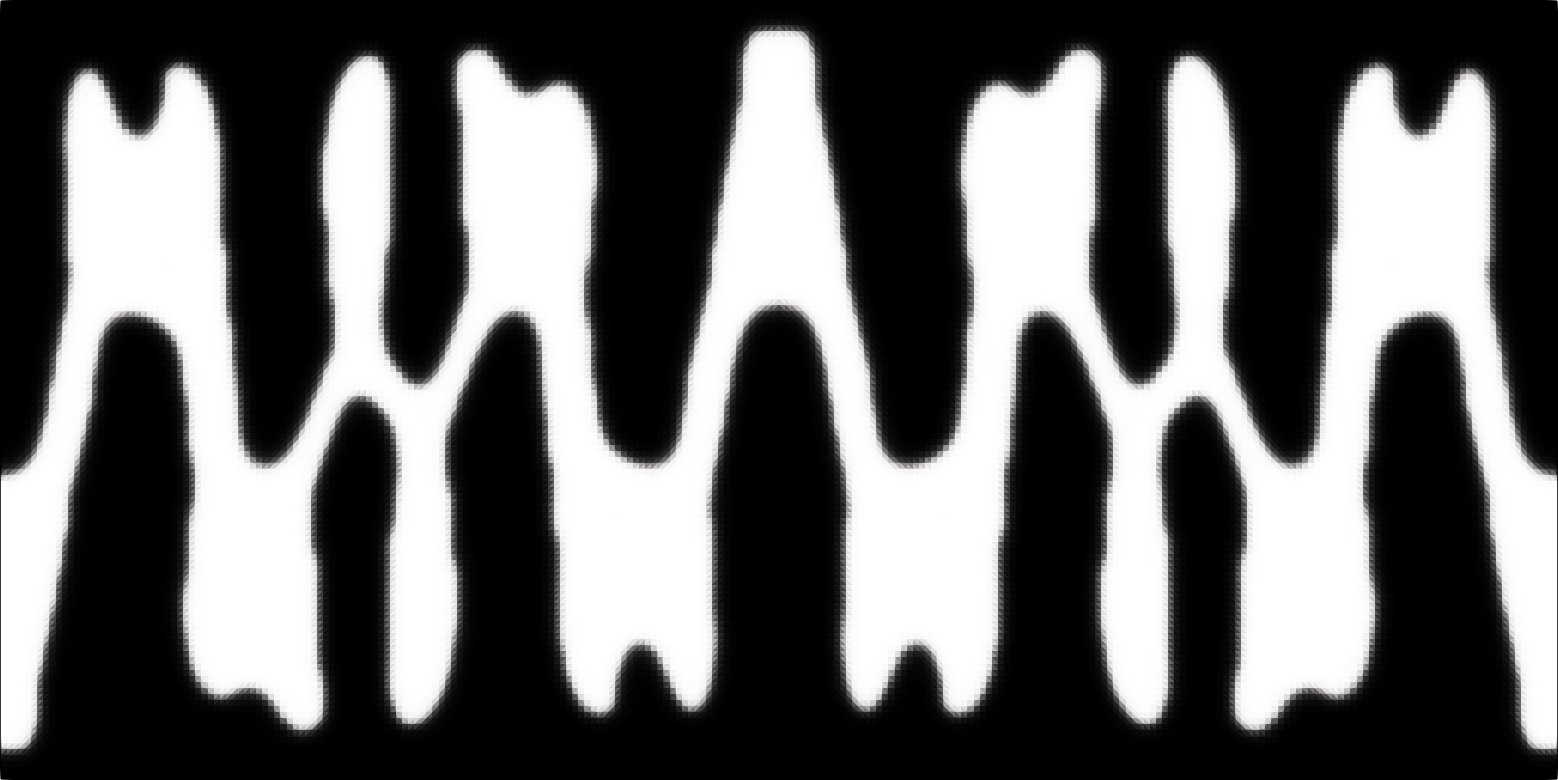}
    }{
        \includegraphics[width=\linewidth]{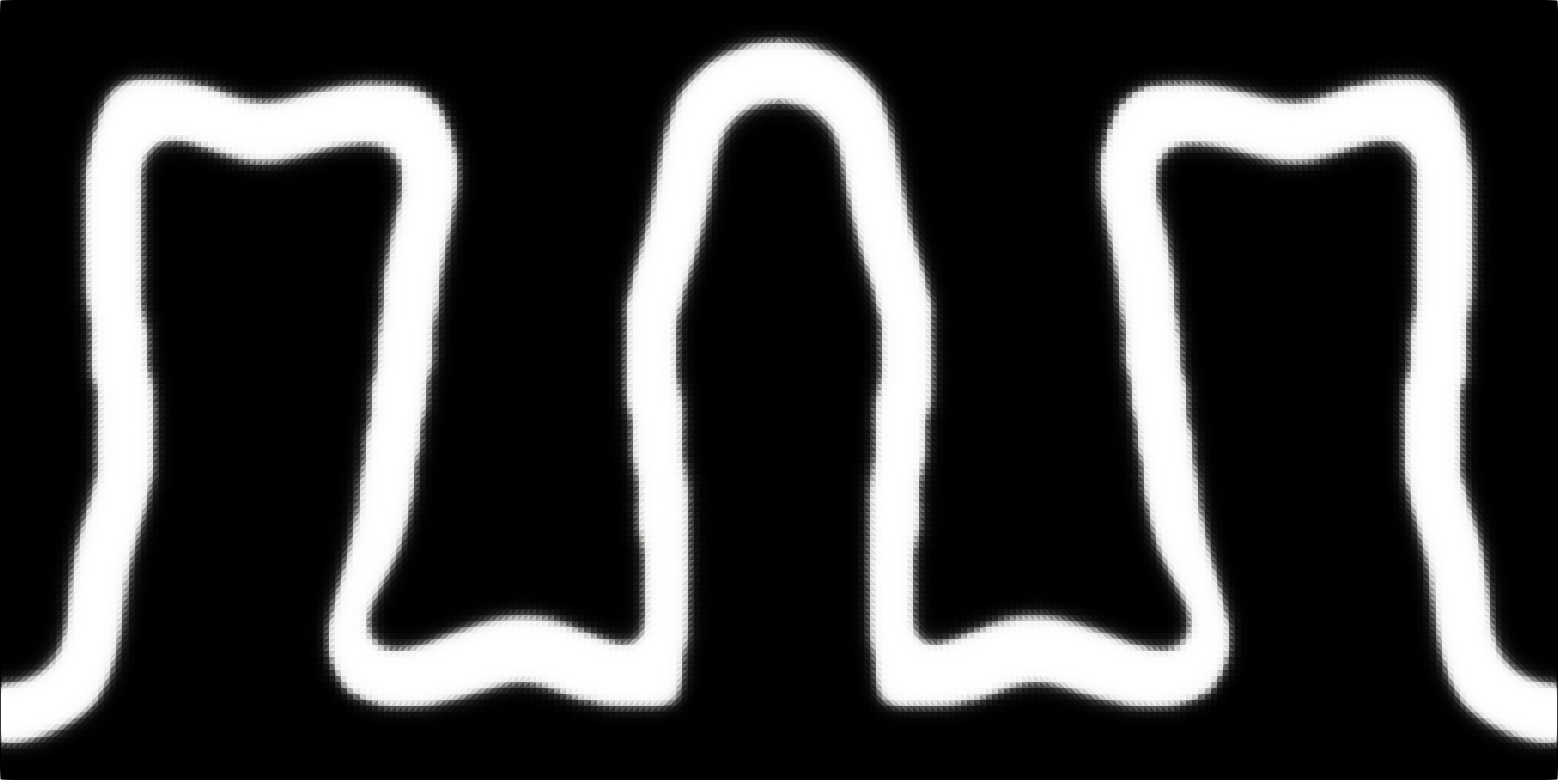}
    }{
        \includegraphics[width=\linewidth]{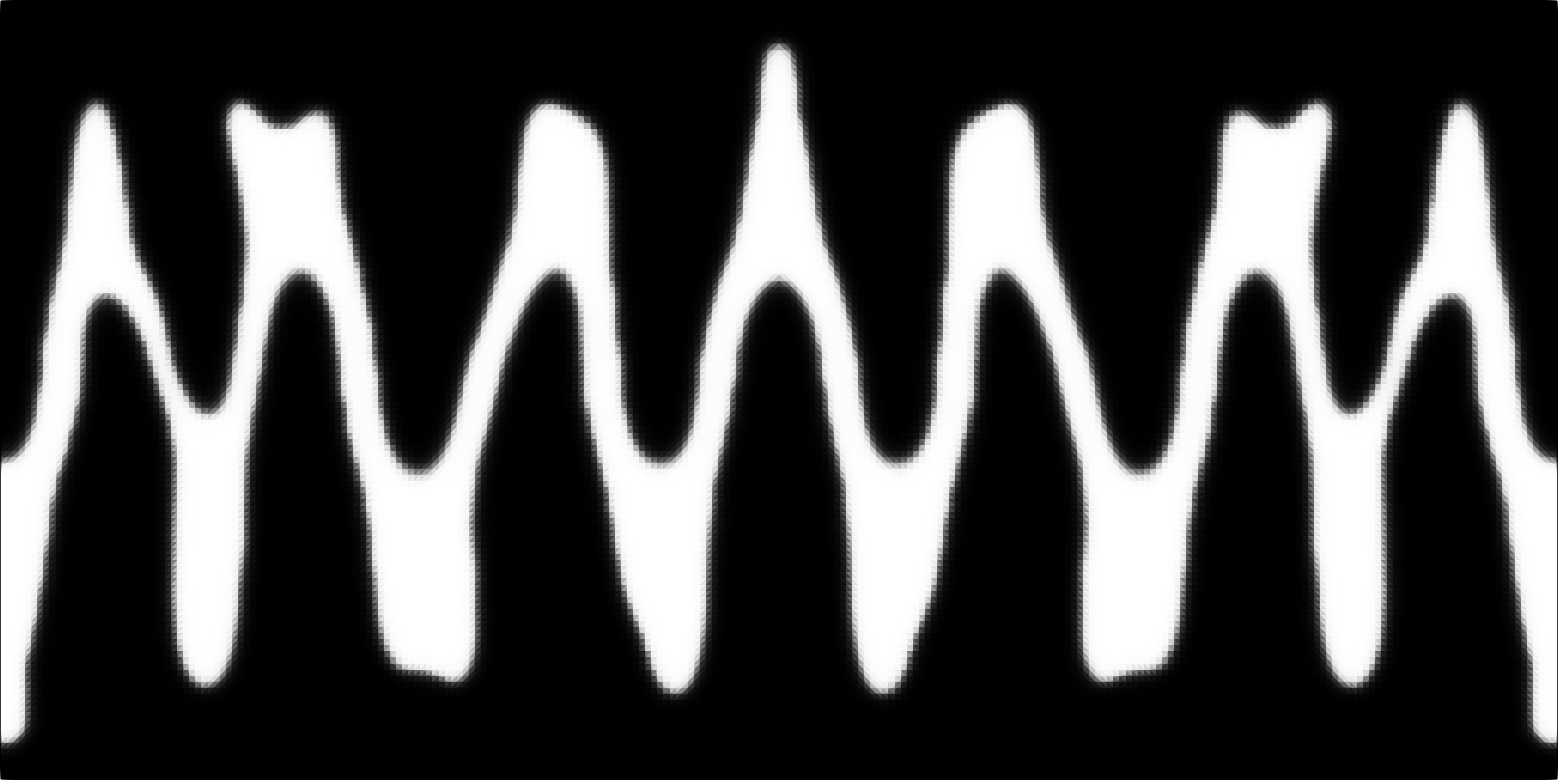}
    }{
        \includegraphics[width=\linewidth]{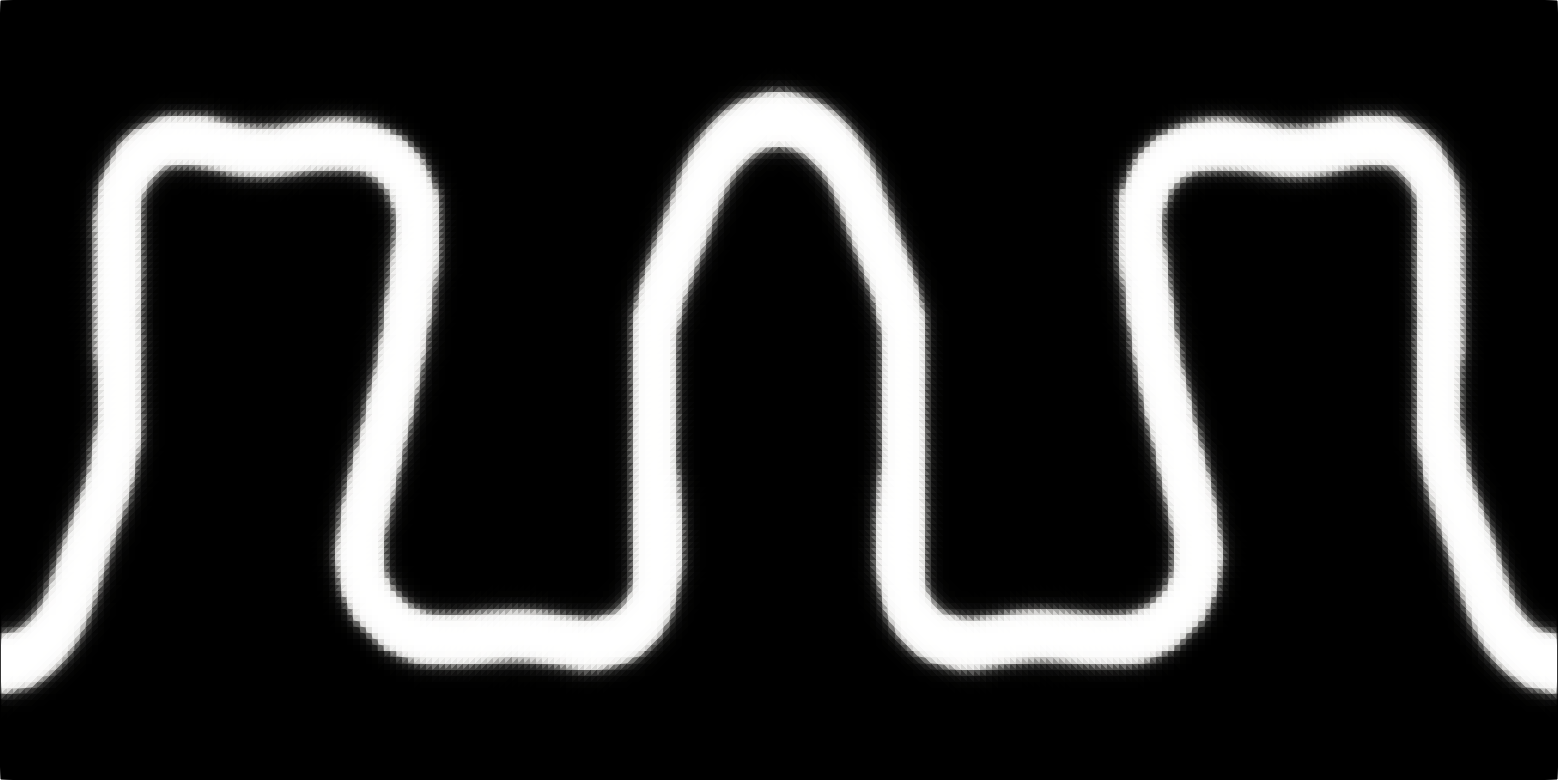}
    }{
        \includegraphics[width=\linewidth]{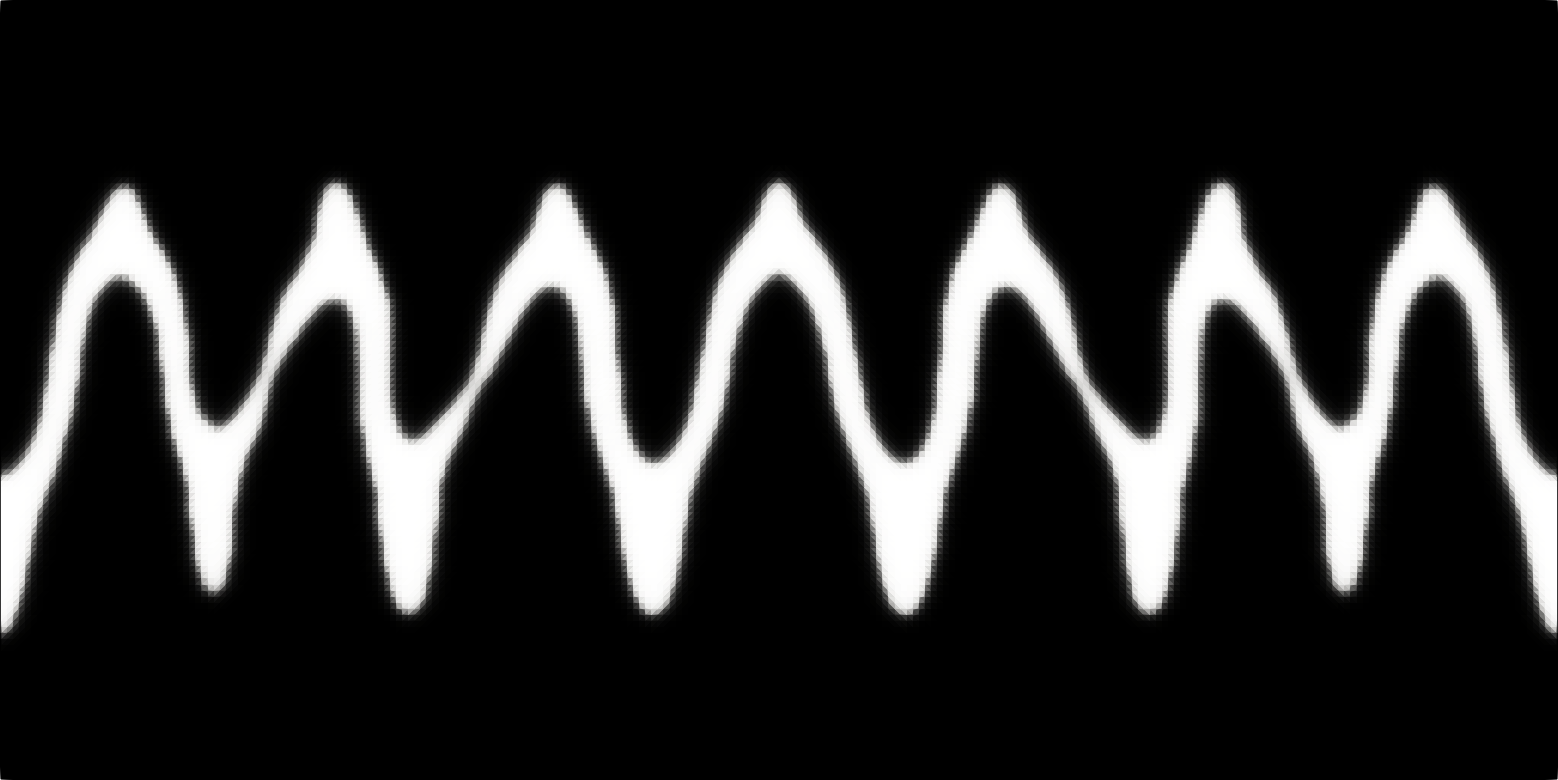}
    }{.0910\,(179\%), .0189\,(40\%)}{.0731\,(136\%), .0337\,(66\%)}{.1040\,(163\%), .0153\,(39\%)}{.0775\,(118\%), .0283\,(68\%)}{.1092\,(135\%), .0145\,(47\%)}{.0846\,(106\%), .0250\,(78\%)} %
    {\phantomsubcaption \label{tab:pe_oribrug2a}}
    {\phantomsubcaption \label{tab:pe_oribrug2b}}
    {\phantomsubcaption \label{tab:pe_oribrug2c}}
    {\phantomsubcaption \label{tab:pe_oribrug2d}}
    {\phantomsubcaption \label{tab:pe_oribrug2e}}
    {\phantomsubcaption \label{tab:pe_oribrug2f}}
    \caption{Optimized porous electrode designs modeled with the original Bruggeman correlation, $\delta=2$, and varying $\lambda$ and $\gamma$. Black is $\bar{\rho} = 1$; white is $\bar{\rho} =0$.}
    \label{tab:pe_oribrug2}
\end{figure*}

Shown in Fig.~\ref{tab:pe_oribrug1} is the first series of designs generated using the original Bruggeman correlation, a fixed $\lambda=0.01$, and varying $\delta$ and $\gamma$.
All optimized designs possess an interdigitated structure, which was found by the optimization algorithm thanks to the continuation schemes described in Sect.~\ref{sec:optproblem}. 
The interdigitation allows reactions to happen closer to the current collectors, boosting current density.
The structure also reduces ohmic losses by promoting more efficient ionic transport.
It is important to note that, if we simply force electrode separation by increasing the weight $w_{SC}$ on the short circuit intensity penalty in \eqref{eq:min_problem}, we obtain near monolithic designs with a large distance between electrodes.

In kinetic resistance limited cases ($\delta = 0.5$), there are no notable differences between the designs obtained with different $\gamma$ values (see left column in Fig.~\ref{tab:pe_oribrug1}.).
In other words, the bulk geometry is not sensitive to the partitioning between capacitive and redox reaction due to reactions being more uniform.
Indeed, the designs are driven by the heterogeneous ohmic losses $E_{\mrm{ohm}}$.
In these cases, the interdigitated optimized designs achieve a maximum energy storage increase of 10\%, and a maximum ohmic loss decrease of 60\% over the monolithic design.

For the ohmic resistance limited cases ($\delta= 5$), we notice obvious variations among the designs (see the right column in Fig.~\ref{tab:pe_oribrug1}).
The periodicity of the designs remains the same for varying $\gamma$ values.
Since the current model does not incorporate surface concentration that limits the maximum ion adsorption, the overall concentration decreases significantly for capacitor electrodes ($\gamma = 0$).
Therefore, for capacitive-dominated reactions (small $\gamma$), more electrolyte is needed to support the reaction and prevent reactant starvation and dramatic energy dissipation.
This explains the wider white electrolyte region between the electrodes in the $\gamma = 0$ design.
As the charging becomes more redox dominated ($\gamma \rightarrow 1$), the designs' interdigitation depth increases, whereas the electrolyte width decreases.
Such an effect is only obvious in these ohmic resistance limited cases.
A smaller separation distance results in a faster ion exchange between the electrodes and hence sustains high reaction rates.
Additionally, more electrode material is placed close to the current collectors to support the reaction on the electrode tips.
For these $\delta = 5$ designs, we observe approximately 100\% increase in energy storage and 40\% reduction in ohmic loss over the monolithic design.

The first benefit of interdigitation is a more uniform concentration distribution.
We quantify the concentration variation by
\begin{equation}
    \hat{c}_{\mrm{var}} = |\hat{\Omega}|^{-\frac{1}{2}}\|\hat{c} - \bar{c}\|_{L^2(\hat{\Omega})},
\end{equation}
where $\bar{c} = 1$ is the average dimensionless concentration in the system.
As illustrated in Fig.~\ref{fig:oribrug_analysis_comp}, using the $\delta = 5$ and $\gamma = 1$ case as an example, the interdigitated design decreases the variation in concentration by more than 50\% over the monolithic design.
A more uniform concentration distribution ensures better ion adsorption and redox reaction, since small concentrations hinder the adsorption/reaction rate, while large localized concentrations indicate poor ion utilization.

In the case of redox-active materials ($\gamma=1$), the interdigitation also increases the interface area between the electrode and the pure electrolyte regions, facilitating ion transport between them.
For an applied voltage boundary condition, interdigitation reduces the cell resistance and hence increases the current output.
Summarizing, by facilitating ion transport in the electrolyte, the interdigitated design reduces the energy dissipation and improves cell efficiency.

For small $\lambda$, energy dissipation mainly results from ion transport and is proportional to the magnitude of the ionic-potential gradient.
We compute the average gradient magnitude of ionic potential by
\begin{equation}
    |\hat{\nabla} \hat{\Phi}_2|_{\mrm{avg}} = |\hat{\Omega}|^{-1} \|\hat{\nabla}\hat{\Phi}_2\|_{L^2(\hat{\Omega})}.
\end{equation}
Fig.~\ref{fig:oribrug_analysis_comp} shows a more uniform ionic potential distribution in the optimized design compared to that in the monolithic design, and a $40\%$ decrease in the average gradient.

Next, we study the effect of varying $\lambda$ and $\gamma$ for a fixed $\delta=2$.
A small $\lambda$ indicates larger relative electronic conductivity, which benefits designs with more intensive interdigitation. 
A small $\lambda$ is characteristic of Graphite-based anodes with large electronic conductivity, on the order of \qty{1000}{mS\per cm}, and of porous cathodes that incorporate carbon additives to enhance the average electric conductivity.
On the other hand, the average room-temperature conductivity of liquid electrolyte materials is on the order of \qty{10}{mS\per cm}.
Overall, it is reasonable to assume that the electronic conductivity is at least one or two orders of magnitude larger than the ionic conductivity.
This corresponds to $\lambda$ values ranging from 0.01 to 0.1.

As shown in Fig.~\ref{tab:pe_oribrug2}, we observe long and bulky columns in the $\lambda = 0.01$ designs.
These designs achieve substantial current densities over the electrode tip due to the larger over-potential.
The ohmic loss associated with transporting the electrons to and from the current collector is negligible, and thus results in more efficient designs.

Comparing the $\lambda = 0.1$ cases (right column in Fig.~\ref{tab:pe_oribrug2}) with the $\lambda = 0.01$ cases (left column in Fig.~\ref{tab:pe_oribrug2}), we notice that an increase in $\lambda$ favors designs with more complexity.
This can be attributed to the fact that as $\lambda$ approaches $0.5$, electron and ion transport become equally limiting.
Consequently, smaller electrode columns are needed to shorten the electron transport distance. 
This explain why the $\lambda=0.1$ designs have more, thinner columns.
And unlike the sharp teeth-like structures observed in \cite{roy2022tope}, the features here have round corners since smaller curvature enhances ion diffusion, and prevents electric field congestion.
Deep interdigitation provides smaller benefits as compared to the previous Fig.~\ref{tab:pe_oribrug1} and Fig.~\ref{tab:pe_oribrug2} (left column) cases.
However, maximizing the electrode/pure electrolyte interfacial area is still desirable as it introduces additional pathways for ionic current to enter and leave the pure electrolyte region.
Therefore, the complex bulk geometry enhances the ionic transport within the electrode region, and thus reduces local ion depletion.
\rone{A more comprehensive study on the impact of column sizes on device performance is presented in \cite{lin2024shape}.}

The optimization mainly tries to reduce energy dissipated due to ohmic resistance in these $\delta = 2$ designs.
Therefore, the $\lambda = 0.1$ designs with increased electronic resistance generally store less energy.
Although all optimized designs outperform the monolithic design, the designs generated with $\lambda = 0.01$ provide an additional $40\%$ increase in energy storage and $30\%$ reduction in ohmic loss, as compared to the corresponding $\lambda = 0.1$ designs.

\subsubsection{Modified Bruggeman correlation}
We now compare the results generated using the modified Bruggeman correlation with $f_m = 0.02$, which is equivalent to a $98\%$ ionic diffusion reduction inside the porous electrode.
As a result, the penetration depth of the reaction front in the monolithic design is greatly diminished.
We first regenerated the Fig.~\ref{tab:pe_oribrug1} designs with the modified Bruggeman correlation cf. Fig.~\ref{tab:pe_modbrug1}.
To compensate for extremely slow ionic diffusion inside the electrode phase, all designs form spikey teeth-like structures to increase the electrode/pure electrolyte interfacial area. 

Similar to the observations made from the original Bruggeman cases, among the $\delta = 0.5$ designs, changing $\gamma$ does not noticeably change the design because it does not improve the uniformity of current distribution.
For the $\delta = 5$ designs, capacitor electrodes ($\gamma = 0$) are thinner with shorter spikes as compared to redox electrodes ($\gamma = 1$) due to the depletion of ions in the electrolyte that causes substantial increases in ohmic loss.

By increasing $\delta$ (see right column of Fig.~\ref{tab:pe_modbrug1}), we observe a marked increase in the number of spikes, and the teeth features protrude deeper into the electrode columns.
The limiting kinetics of localized reactions is compensated by the increased electrode/pure electrolyte interfacial area, allowing for better ion transport.

Conversely, in monolithic designs, electrode utilization is poor.
Charging occurs only in the regions immediately adjacent to the electrode/pure electrolyte interface, leading to minimal energy storage, as further discussed in Sect.~\ref{sec:performance}.
Therefore, we observe exceptional improvements on energy storage across all cases, ranging from 350\% to 750\%.
And despite significantly larger charging currents, the ohmic loss is only increased by a maximum of 200\%.
Considering that the energy input is much larger due to a lower overall cell resistance, the energy efficiency for the optimized designs improves significantly.

We now regenerate the Fig.~\ref{tab:pe_oribrug2} designs with the modified Bruggeman correlation cf. Fig.~\ref{tab:pe_modbrug2}.
Similar to the observations made in the corresponding original Bruggeman designs (Fig.~\ref{tab:pe_oribrug2}), going from $\lambda = 0.01$ to $\lambda = 0.1$, the design complexity increases.
Interestingly, the frequency of the microscale teeth remains roughly the same between the $\lambda = 0.01$ and $\lambda = 0.1$ designs with the same $\gamma$.
The teeth in the $\lambda = 0.01$ designs are enlarged proportionally with the macroscopic structures.
We observe $500\%$ to $750\%$ improvement in energy storage with only $\sim 50\%$ increase in ohmic loss.

\subsubsection{Summary}
To summarize the effect of different dimensionless groups on the performance and optimized design:
\begin{itemize}
    {\small \item The energy gain is more significant in systems limited by ohmic resistance (mainly mass transport induced resistance), corresponding to large $\delta$ values, with $\delta$ being the inverse Wagner number.}
    {\small \item Macroscale features are most effective in enhancing overall ionic transport for porous electrodes that have high electronic conductivity corresponding to small $\lambda$ values.}
    {\small \item In extreme tortuosity systems characterized by the modified Bruggeman case, features with spikey teeth are advantageous because they increase the electrode/pure electrolyte interfacial area.}
    {\small \item Interdigitated electrodes reduce the transport distance and increase active-material utilization.}
\end{itemize}
\rone{Additionally, we observe similarities between the cathode and the anode design. This is due to the simplifying usage of the same material parameters for both electrodes in this theoretical study. In realistic scenarios, the anode and the cathode are often constructed with different materials, which will likely result in asymmetric designs.}

\subsection{Performance improvements}
\label{sec:performance}
\begin{figure*}[!htbp]
    \centering
    \begin{multicols}{2}
        \centering
        \begin{subfigure}[b]{\linewidth}
            \centering
            Monolithic electrodes \\
            \includegraphics[width=\linewidth]{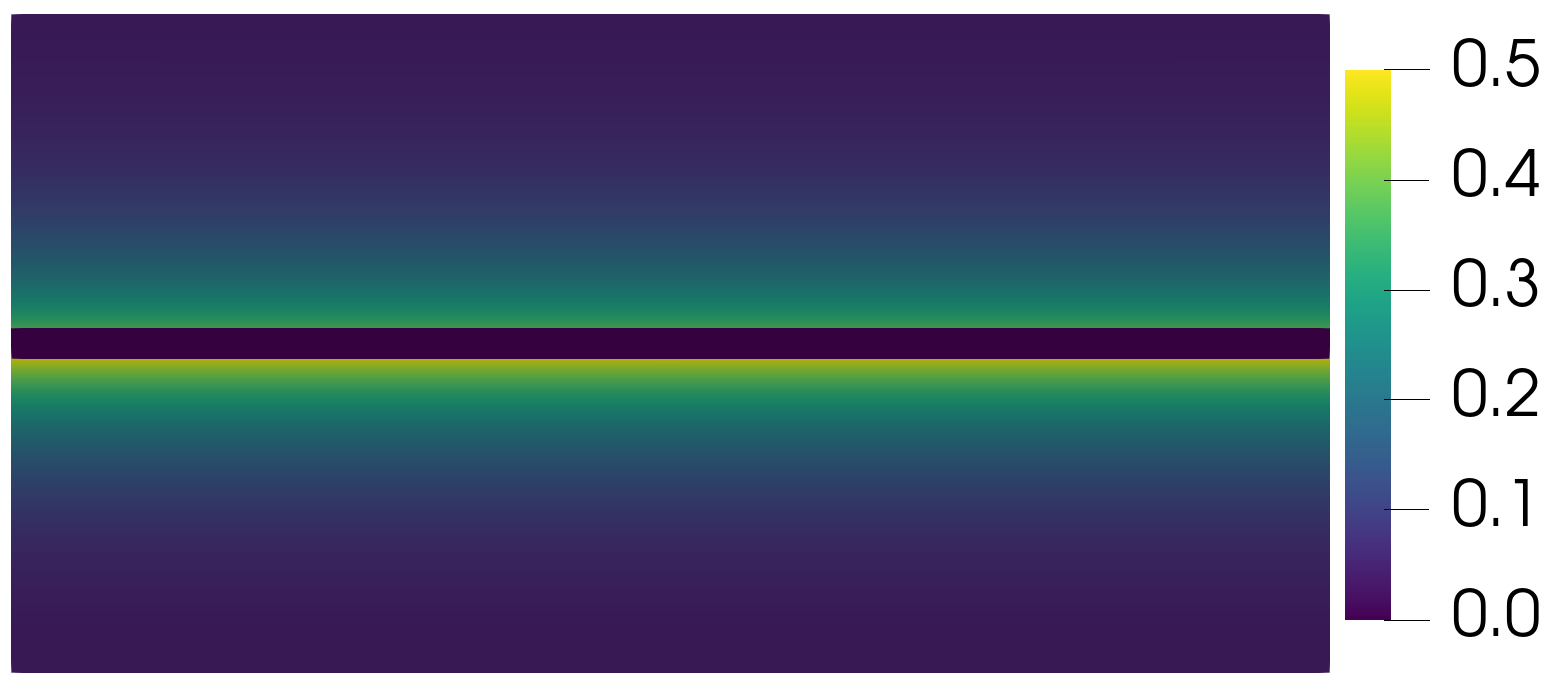}
            \caption{$E_{\mrm{kin}}=1.1712\times 10^{-1},\ E_{\mrm{var}}=0.8077$}
            \label{fig:oribrug_mono_engy}
        \end{subfigure}
        \begin{subfigure}[b]{\linewidth}
            \includegraphics[width=\linewidth]{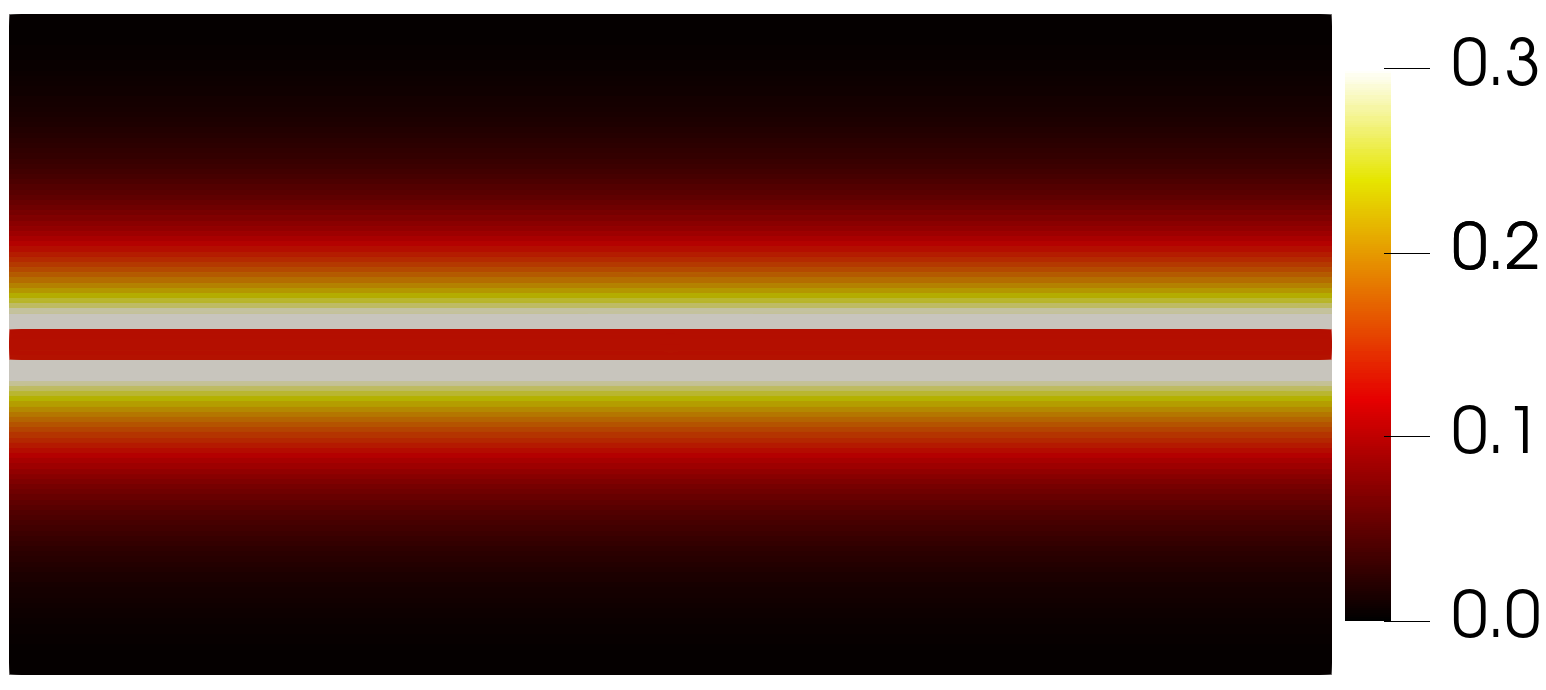}
            \caption{$E_{\mrm{ohm}}=8.4885\times 10^{-2},\ R_{\mrm{avg}}=161.29$}
            \label{fig:oribrug_mono_loss}
        \end{subfigure}
        \begin{subfigure}[b]{\linewidth}
            \centering
            Optimized electrodes \\
            \includegraphics[width=\linewidth]{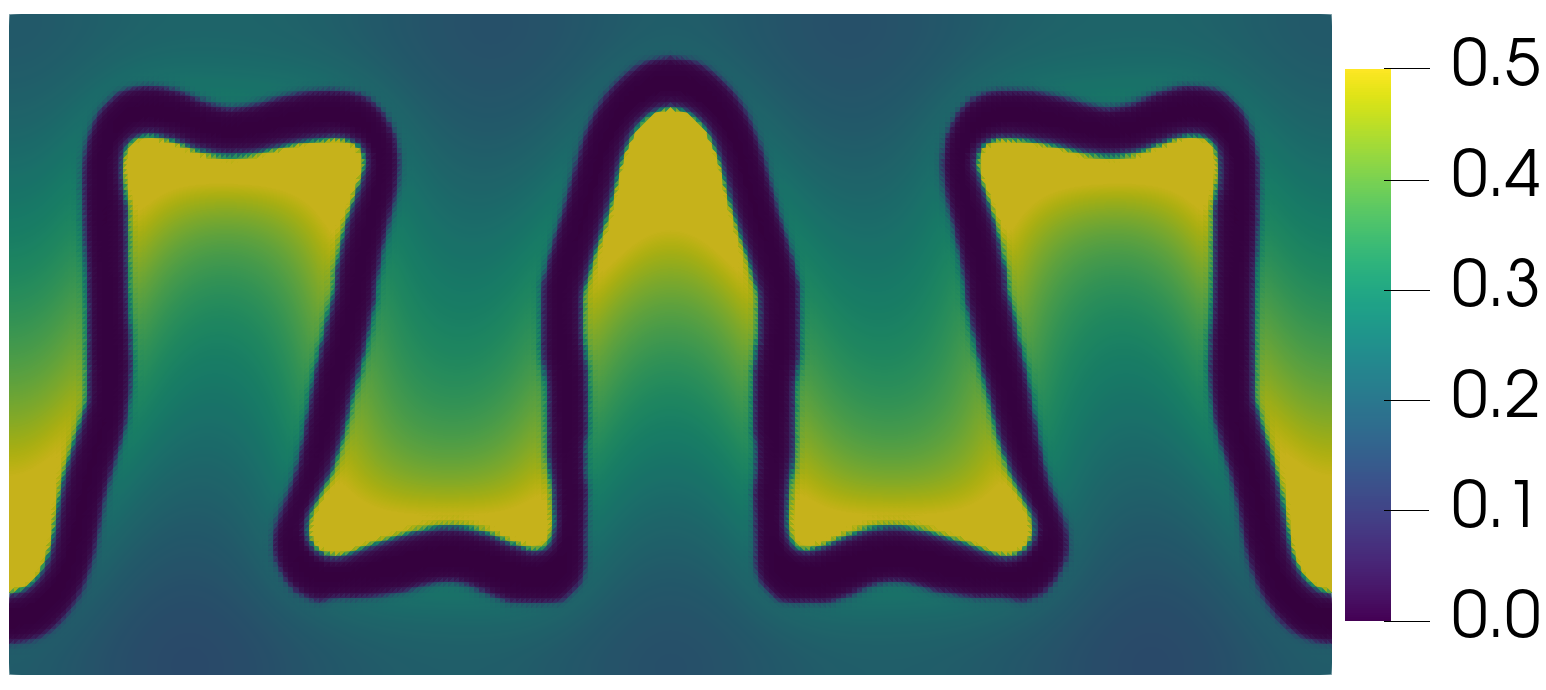}
            \caption{$E_{\mrm{kin}}=2.2538\times 10^{-1},\ E_{\mrm{var}}=0.3623$}
            \label{fig:oribrug_opt_engy}
        \end{subfigure}
        \begin{subfigure}[b]{\linewidth}
            \includegraphics[width=\linewidth]{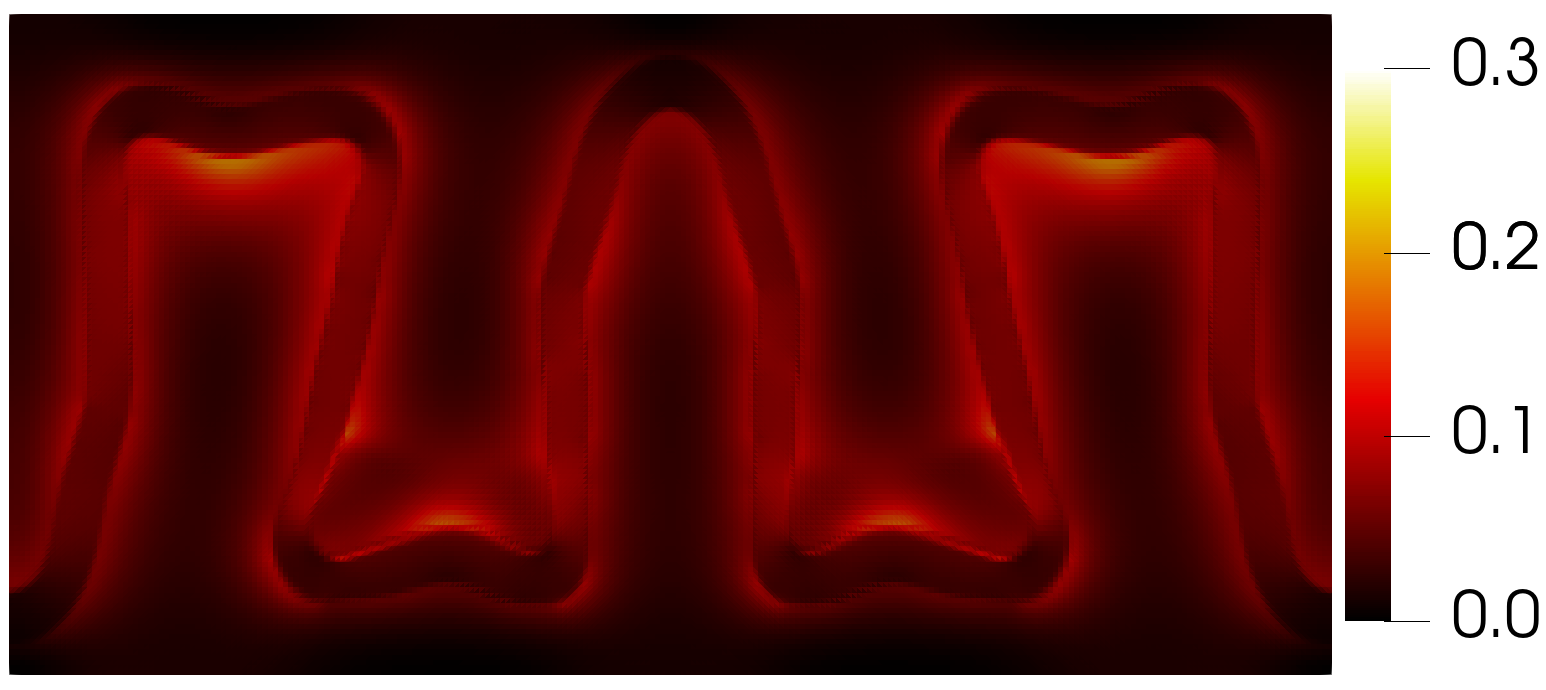}
            \caption{$E_{\mrm{ohm}}=4.9681\times 10^{-2},\ R_{\mrm{avg}}=121.98$}
            \label{fig:oribrug_opt_loss}
        \end{subfigure}
    \end{multicols}
    \setlength{\abovecaptionskip}{-10pt}
    \caption{Energy storage $e_{\mrm{kin}}(\hat{x})$ (top) and ohmic loss $e_{\mrm{ohm}}(\hat{x})$ (bottom) fields of monolithic (left) and optimized (right) electrodes with $\delta=5$, $\gamma=1$, $\lambda=0.01$, and the original Bruggeman correlation.}
    \label{fig:oribrug_compare}
\end{figure*}
\begin{figure*}[!htbp]
    \centering
    \begin{multicols}{2}
        \centering
        \begin{subfigure}[b]{\linewidth}
            \centering
            Monolithic electrodes \\
            \includegraphics[width=\linewidth]{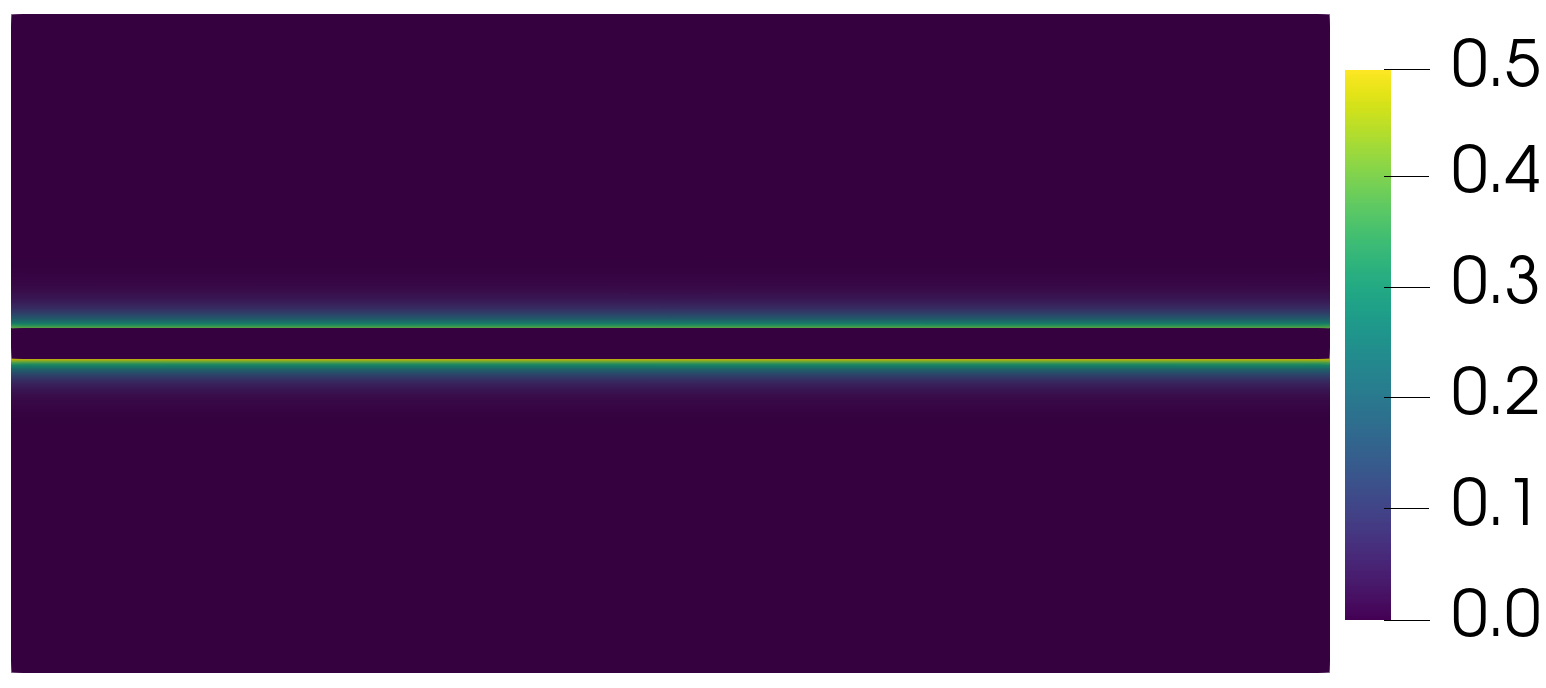}
            \caption{$E_{\mrm{kin}}=1.5622\times 10^{-2},\ E_{\mrm{var}}=3.6411$}
            \label{fig:modbrug_mono_engy}
        \end{subfigure}
        \begin{subfigure}[b]{\linewidth}
            \includegraphics[width=\linewidth]{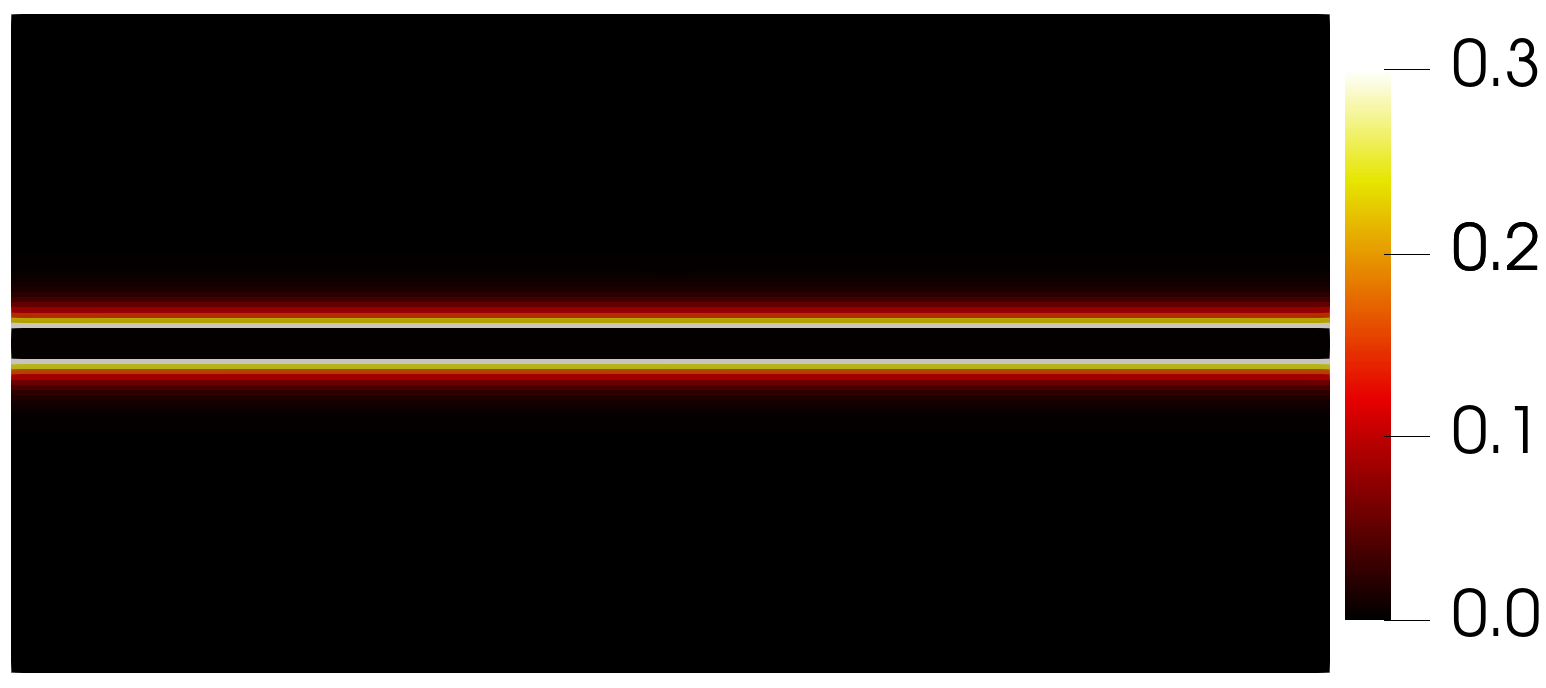}
            \caption{$E_{\mrm{ohm}}=1.5616\times 10^{-2},\ R_{\mrm{avg}}=920.62$}
            \label{fig:modbrug_mono_loss}
        \end{subfigure}
        \begin{subfigure}[b]{\linewidth}
            \centering
            Optimized electrodes \\
            \includegraphics[width=\linewidth]{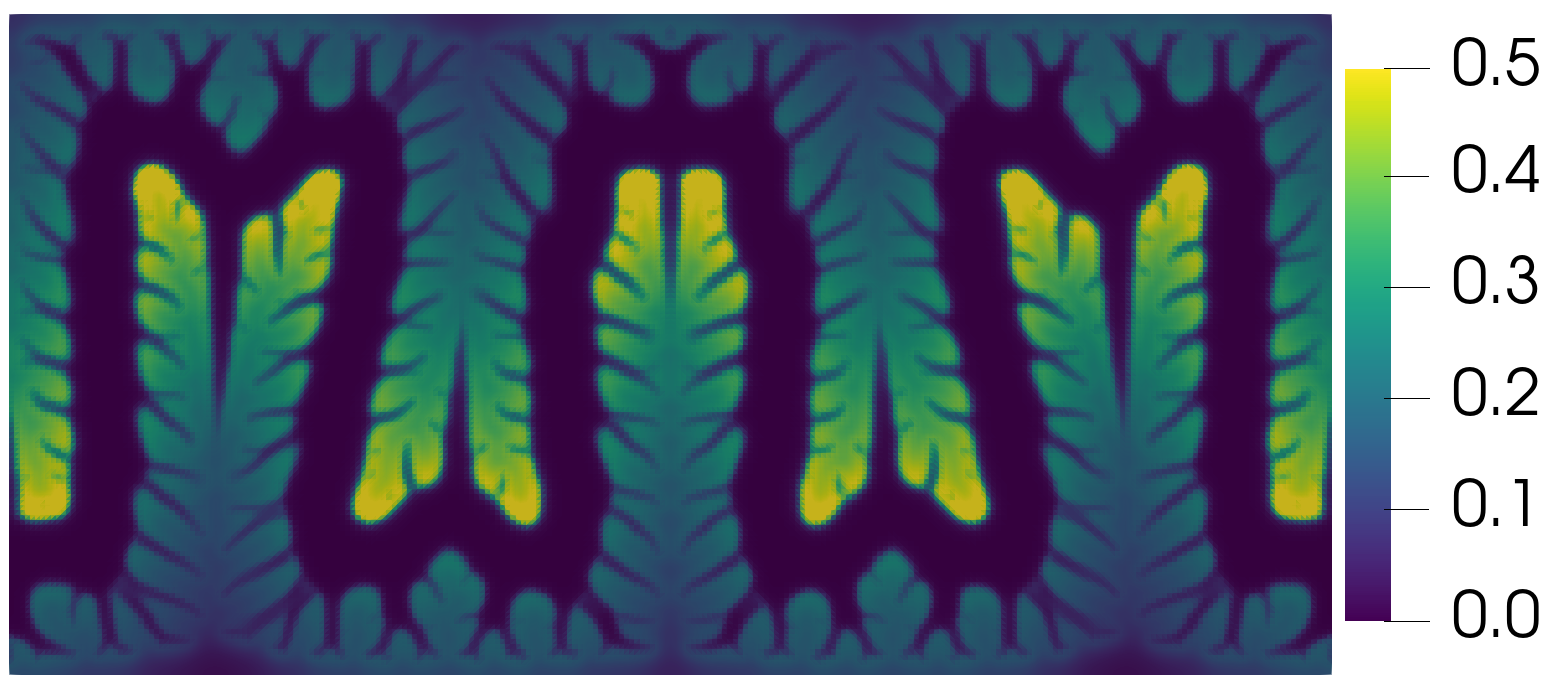}
            \caption{$E_{\mrm{kin}}=1.3296\times 10^{-1},\ E_{\mrm{var}}=0.3164$}
            \label{fig:modbrug_opt_engy}
        \end{subfigure}
        \begin{subfigure}[b]{\linewidth}
            \includegraphics[width=\linewidth]{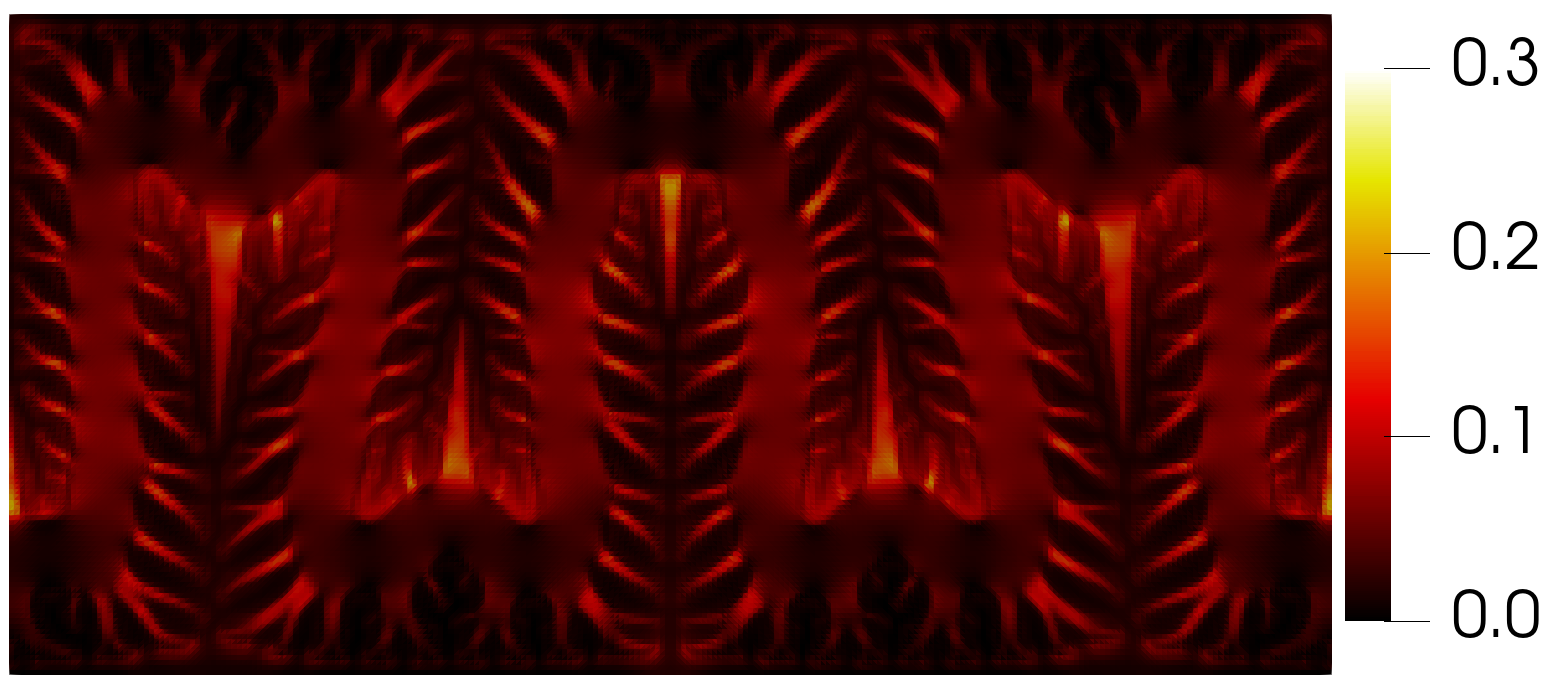}
            \caption{$E_{\mrm{ohm}}=4.1372\times 10^{-2},\ R_{\mrm{avg}}=192.77$}
            \label{fig:modbrug_opt_loss}
        \end{subfigure}
    \end{multicols}
    \setlength{\abovecaptionskip}{-10pt}
    \caption{Energy storage $e_{\mrm{kin}}(\hat{x})$ (top) and ohmic loss $e_{\mrm{ohm}}(\hat{x})$ (bottom) between monolithic (left) and optimized (right) electrodes with $\delta=5$, $\gamma=1$, $\lambda=0.01$, and the modified Bruggeman correlation.}
    \label{fig:modbrug_compare}
\end{figure*}

In this section, we compare the energy storage and ohmic loss between the optimized and monolithic designs to illustrate the improved material utilization and ionic transport of the optimized cells.
We first define the energy storage and ohmic loss fields as
\begin{subequations}
  \label{eq:energy_fields}
  \begin{align}
    e_{\mrm{kin}}(\hat{x}) =& \delta_r \int_0^{\hat{T}_{\mrm{f}}}\hat{a}\hat{i}_n(\hat{\Phi}_1-\hat{\Phi}_2) \diff\hat{t}\nonumber \\
    + &\delta_c \int_0^{\hat{T}_{\mrm{f}}}\hat{a}\hat{i}_c(\hat{\Phi}_1 - \hat{\Phi}_2) \diff\hat{t},\label{eq:storage_distribution}\\
    e_{\mrm{ohm}}(\hat{x})=&\frac{1}{\lambda} \int_0^{\hat{T}_{\mrm{f}}} \hat\sigma  \hat{\nabla}\hat{\Phi}_1 \cdot \hat{\nabla}\hat{\Phi}_1 \diff\hat{t}\nonumber \\
    + \frac{1}{1-\lambda} &\int_0^{\hat{T}_{\mrm{f}}}\hat D \hat{c} \hat{\nabla} \hat{\Phi}_2 \cdot \hat{\nabla} \hat{\Phi}_2 \diff\hat{t}\nonumber \\
    + \frac{1}{1-\lambda}&\left(\frac{t_{+}}{z_+} + \frac{t_{-}}{z_-}\right)\int_0^{\hat{T}_{\mrm{f}}} \hat D \hat{\nabla}\hat{c} \cdot \hat{\nabla} \hat{\Phi}_2 \diff\hat{t}\label{eq:ohmic_distribution}.
    \end{align}
\end{subequations}
Then the material utilization is quantified by the normalized variation in the energy storage field
\begin{equation}
    E_{\mrm{var}}=\left\|\bar\rho\left(\frac{e_{\mrm{kin}}(\hat{x})}{\bar{E}} - 1\right)\right\|_{L^2(\hat{\Omega})},
\end{equation}
where
\begin{equation}
    \bar{E} = \left(\int_{\hat{\Omega}} \bar\rho \diff \hat{x}\right)^{-1} E_{\mrm{kin}}
\end{equation}
is the average energy density.
The effectiveness of ionic transport is quantified by the average cell resistance
\begin{equation}
    R_{\mrm{avg}}=\hat{T}_{\mrm{f}}^{-1} \int_0^{\hat{T}_{\mrm{f}}} \frac{\xi \hat{t}}{\int_{\hat{\Gamma}_c}\hat{\sigma} \hat{\nabla} \hat{\Phi}_1\cdot \mb{n} \diff \hat{x}} \diff \hat{t}.
\end{equation}
With a fixed boundary potential across all designs, a smaller cell resistance results in an increased charging current and thus more energy storage.

The energy storage $e_{\mrm{kin}}(\hat{x})$ and ohmic loss $e_{\mrm{ohm}}(\hat{x})$ fields for the $\delta=5$, $\gamma = 1$, and $\lambda = 0.01$ designs with the original and modified Bruggeman correlations are illustrated in Figs.~\ref{fig:oribrug_compare} and \ref{fig:modbrug_compare}.
First, looking at the original Bruggeman cases in Fig.~\ref{fig:oribrug_compare}, notably better electrode utilization is seen in the optimized vs. monolithic design.
Regions near the current collector in the monolithic electrodes have little energy storage, as opposed to the optimized electrodes.
The optimized designs also achieve a larger energy density at the electrode tips due to their increased over-potential, which improves their material utilization by approximately 50\%.
The monolithic design also exhibits significant ohmic loss, in the electrode/pure electrolyte interfacial region, since the ionic currents must pass the region to cross the gap.
The ohmic loss density in the optimized electrodes is similar, however, its net effect is less since the interdigitated structure provides more than twice as much electrode/pure electrolyte interfacial area than the monolithic design.
The increased area provides additional pathways for the currents to cross the pure electrolyte gap.
Consequently, the ionic current is more uniformly distributed, resulting in less ohmic loss.
This is also reflected in the 25\% decrease in average cell resistance.

Analogous $e_{\mrm{kin}}(\hat{x})$ and $e_{\mrm{ohm}}(\hat{x})$ plots for designs obtained with modified Bruggeman correlation are provided in Fig.~\ref{fig:modbrug_compare}.
We again observe poor material utilization by the monolithic electrodes wherein most regions are considered ``dead zones" with negligible energy storage.
On the other hand, the spikey teethed optimized structures promote ion exchange, thereby the energy density is more uniformly distributed throughout the electrodes.
Therefore, the overall storage is enhanced by an order of magnitude and the material utilization is enhanced by $\sim 90\%$.
The ohmic loss distribution is similar to the original Bruggeman case.
Although the maximum loss per unit area is significantly less for the optimized design, the total ohmic loss increases by  $\sim 150\%$ due to the substantial increase in energy input.
Nevertheless, we observe an $80\%$ reduction in average cell resistance.
Also, the energy efficiency $1-E_{\mrm{ohm}}/E_{\mrm{in}}$ reaches 76\%, a 26\% improvement over the monolithic design.

\subsection{3D example}
\begin{figure}
    \centering
    \includegraphics[width=\linewidth]{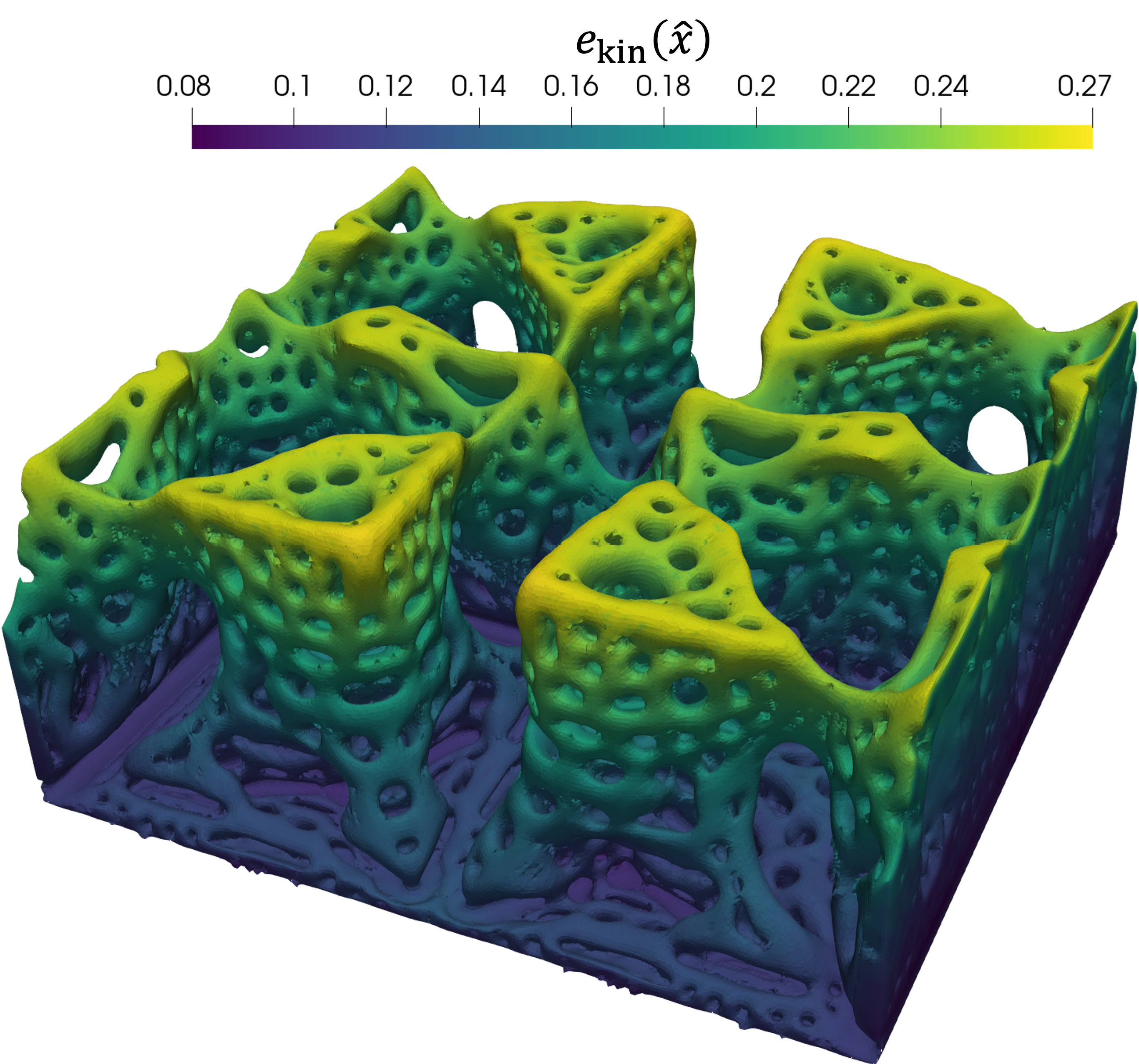}
    \caption{Optimized anode and its energy storage field.}
    \label{fig:3D_Estore}
\end{figure}

\begin{figure}
    \centering
    \begin{subfigure}{\linewidth}
        \centering
        \includegraphics[width=\linewidth]{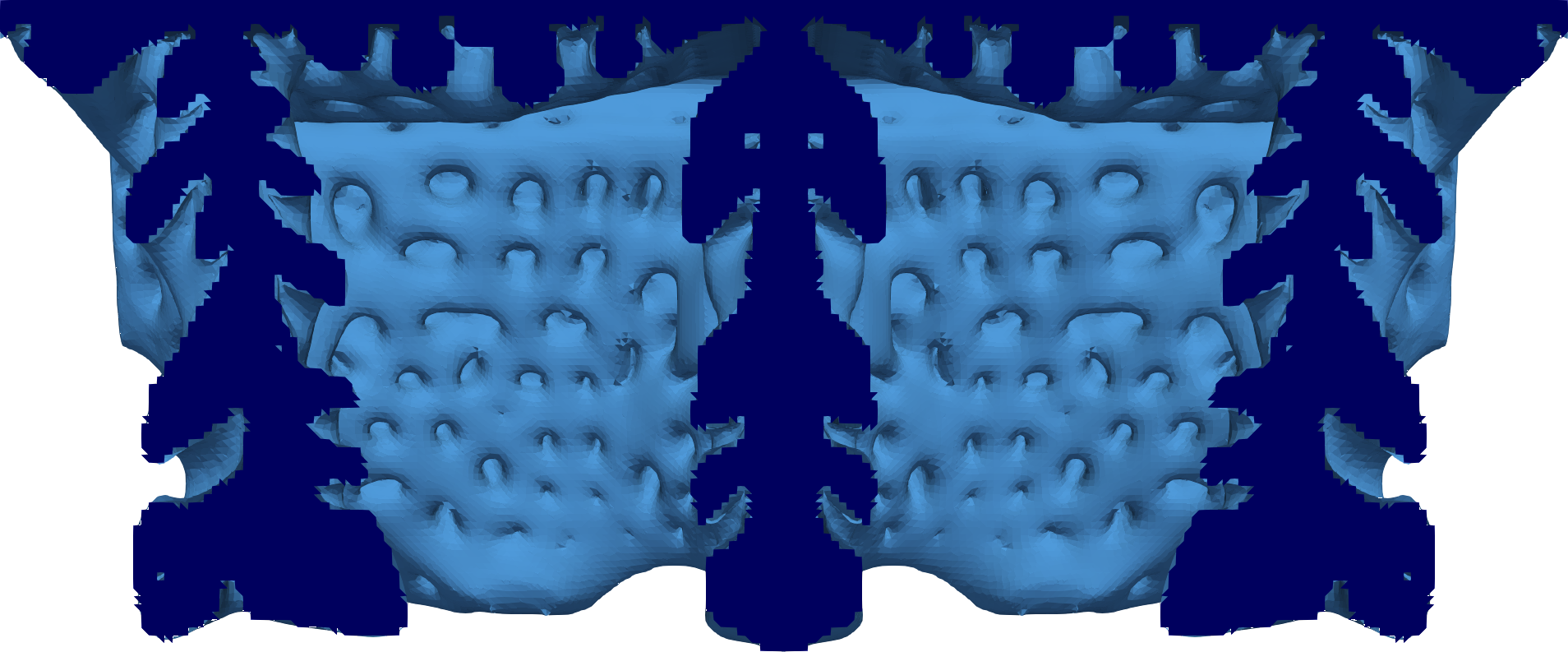}
    \end{subfigure}
    
    \vspace{8pt}
    
    \begin{subfigure}{\linewidth}
        \centering
        \includegraphics[width=\linewidth]{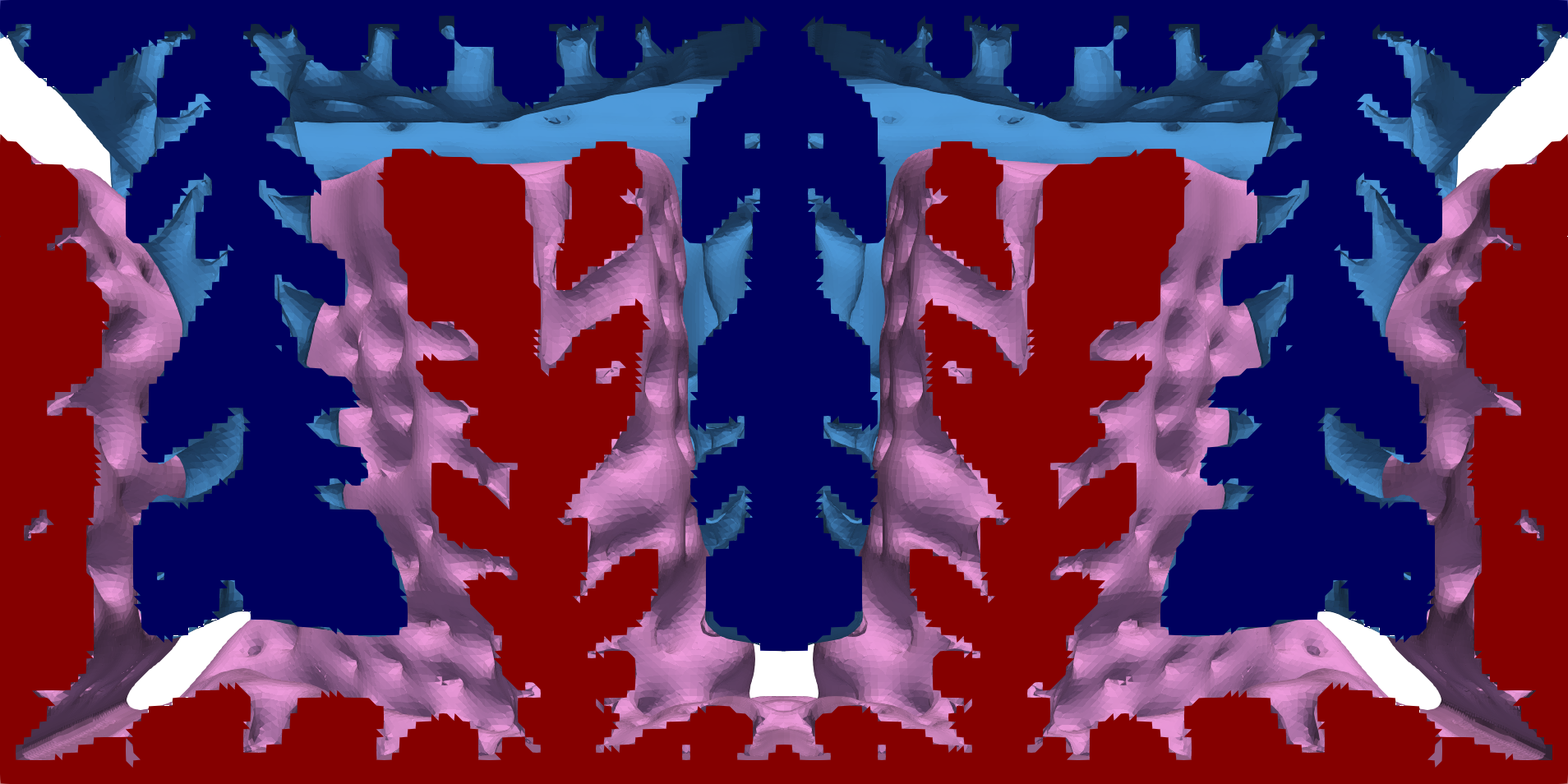}
    \end{subfigure}

    \vspace{8pt}
    
    \begin{subfigure}{\linewidth}
        \centering
        \includegraphics[width=\linewidth]{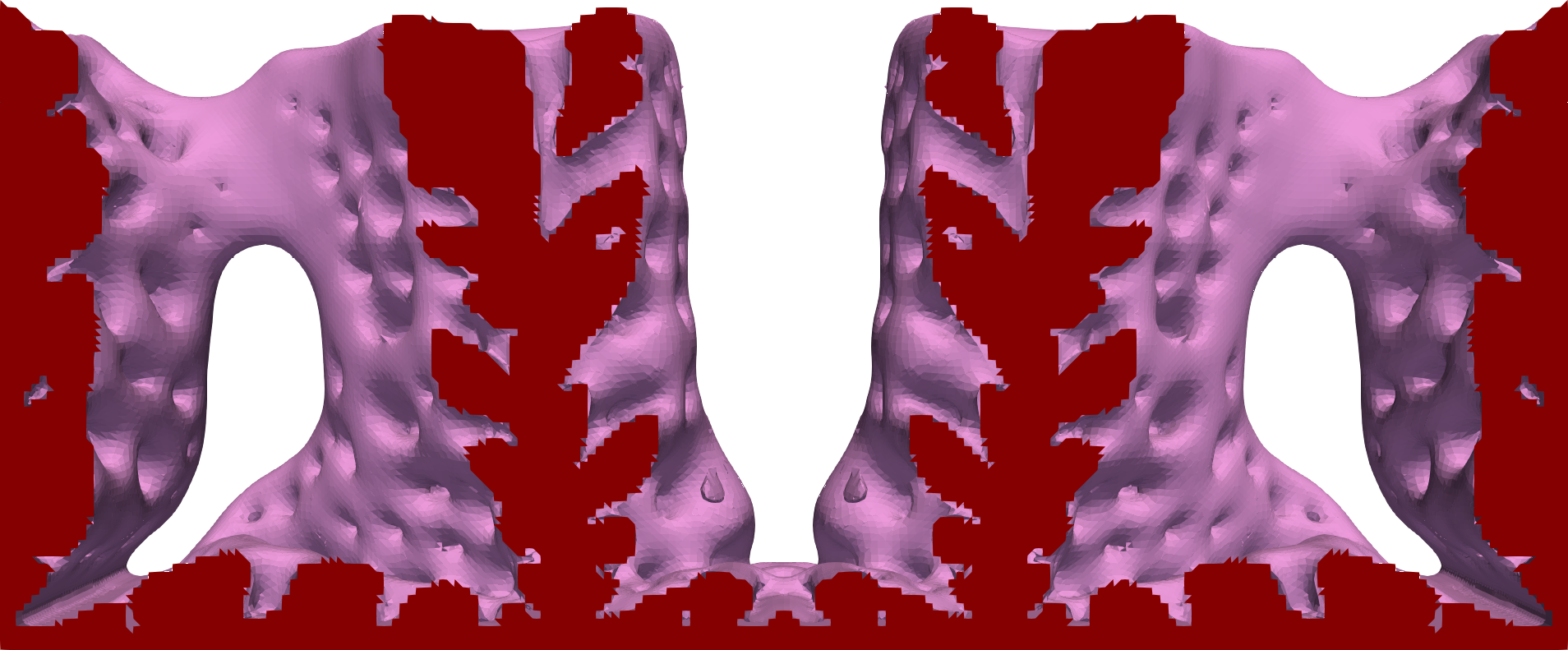}
    \end{subfigure}
    \caption{Cross section of the 3D optimized designs: cathode (top), anode (bottom), and combined cathode/anode (middle).}
    \label{fig:cross-sections}
\end{figure}

\begin{figure}
    \centering
    \begin{subfigure}{0.49\linewidth}
        \includegraphics[width=\linewidth]{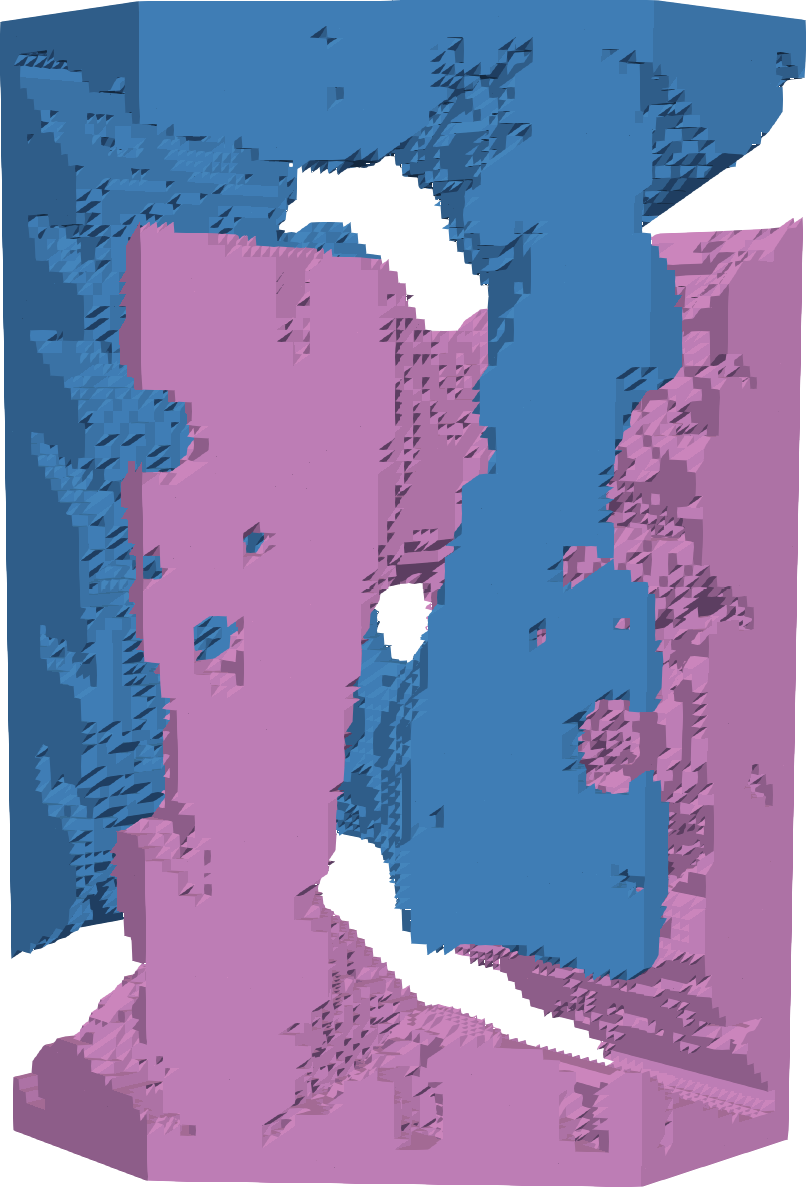}
    \end{subfigure}
    \hfill
    \begin{subfigure}{0.49\linewidth}
        \includegraphics[width=\linewidth]{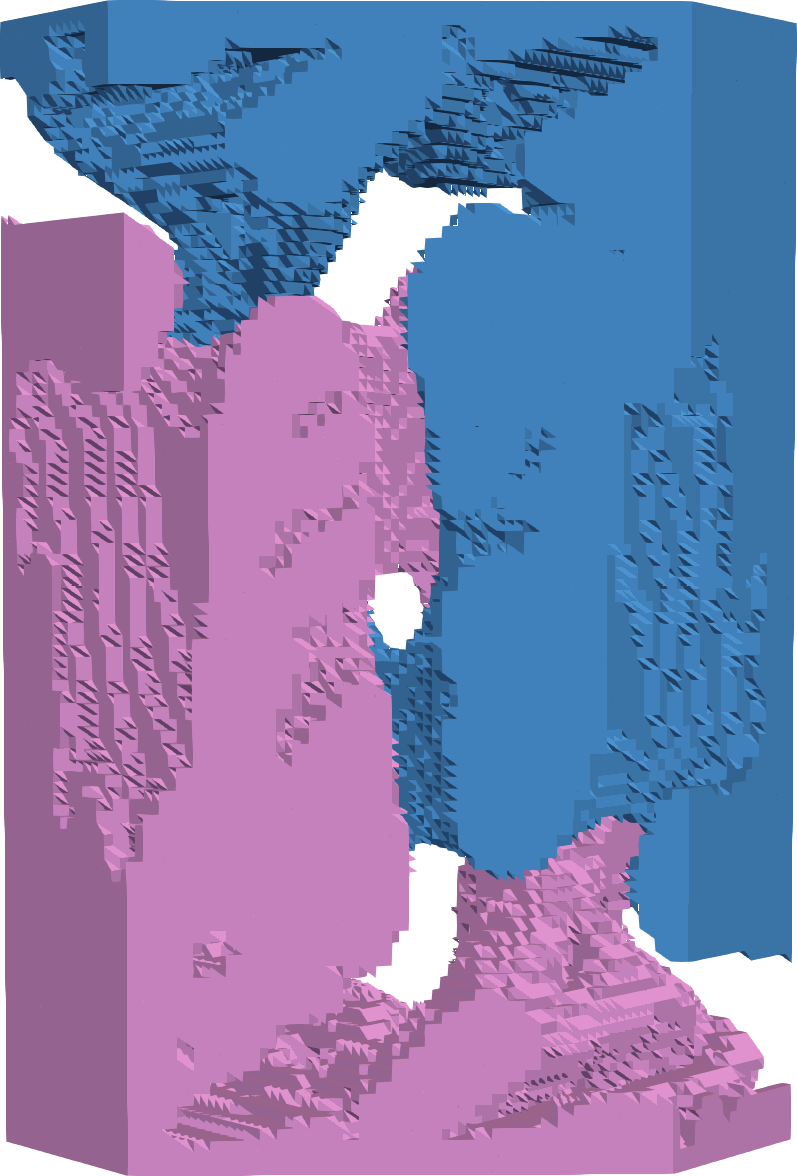}
    \end{subfigure}  
        \caption{Zoomed in view of the interlocked structure formed by the electrode arches.}
    \label{fig:interlock}
\end{figure}

\begin{figure}
    \centering
    \includegraphics[width=\linewidth]{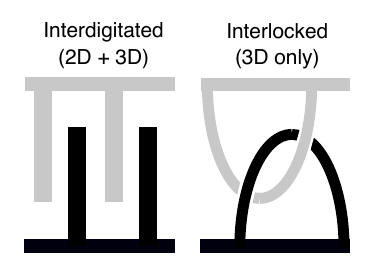}
    \caption{Schematic of electrode structures. Both 2D and 3D designs form interdigitated columns. In addition, 3D designs also form interlocked structures.}
    \label{fig:structure_schematic}
\end{figure}

\begin{figure}
    \centering
    \begin{subfigure}{\linewidth}
        \centering
        \includegraphics[width=\linewidth]{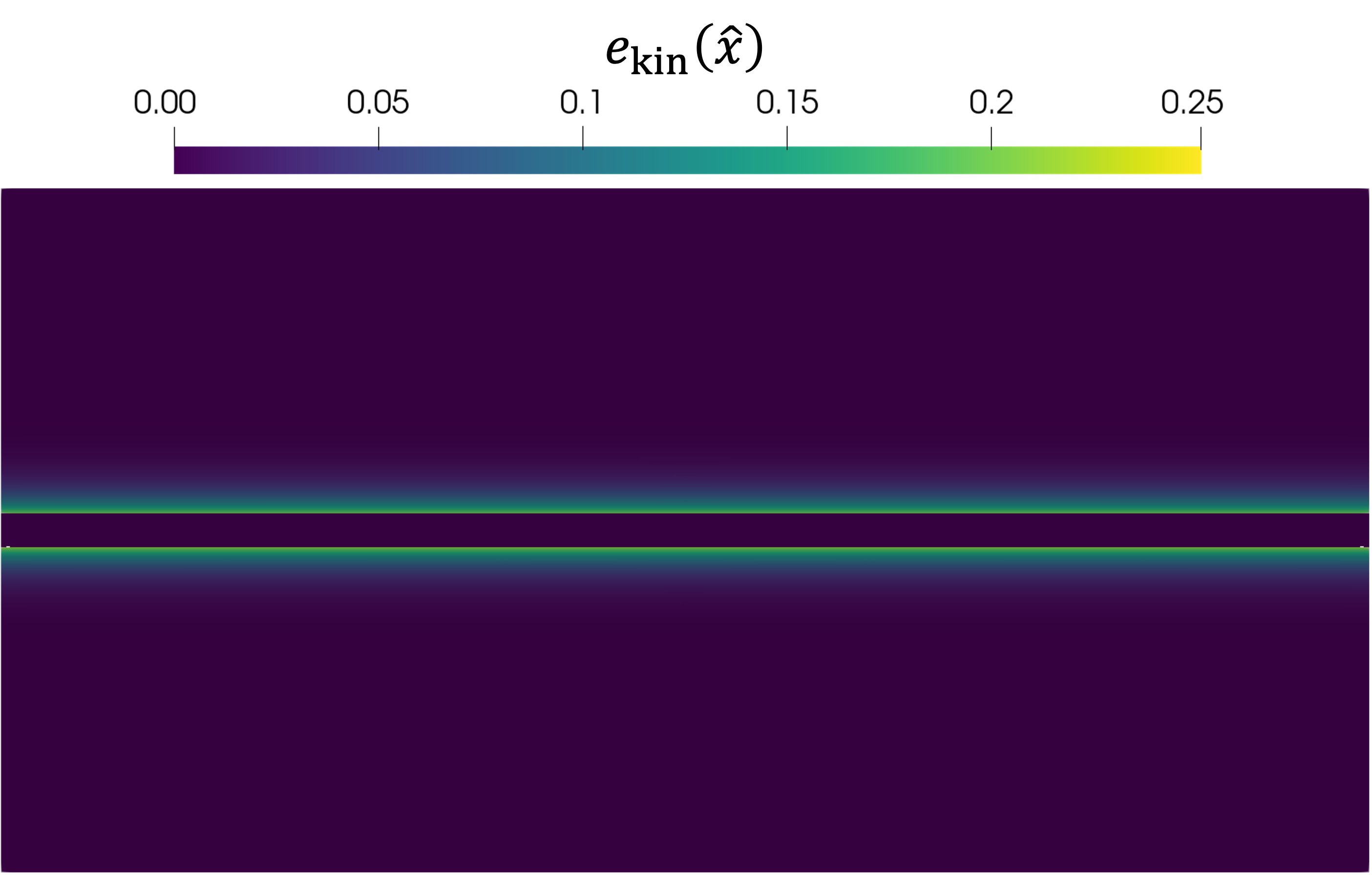}
        \caption{Monolithic design}
    \end{subfigure}
    \begin{subfigure}{\linewidth}
        \centering
        \includegraphics[width=\linewidth]{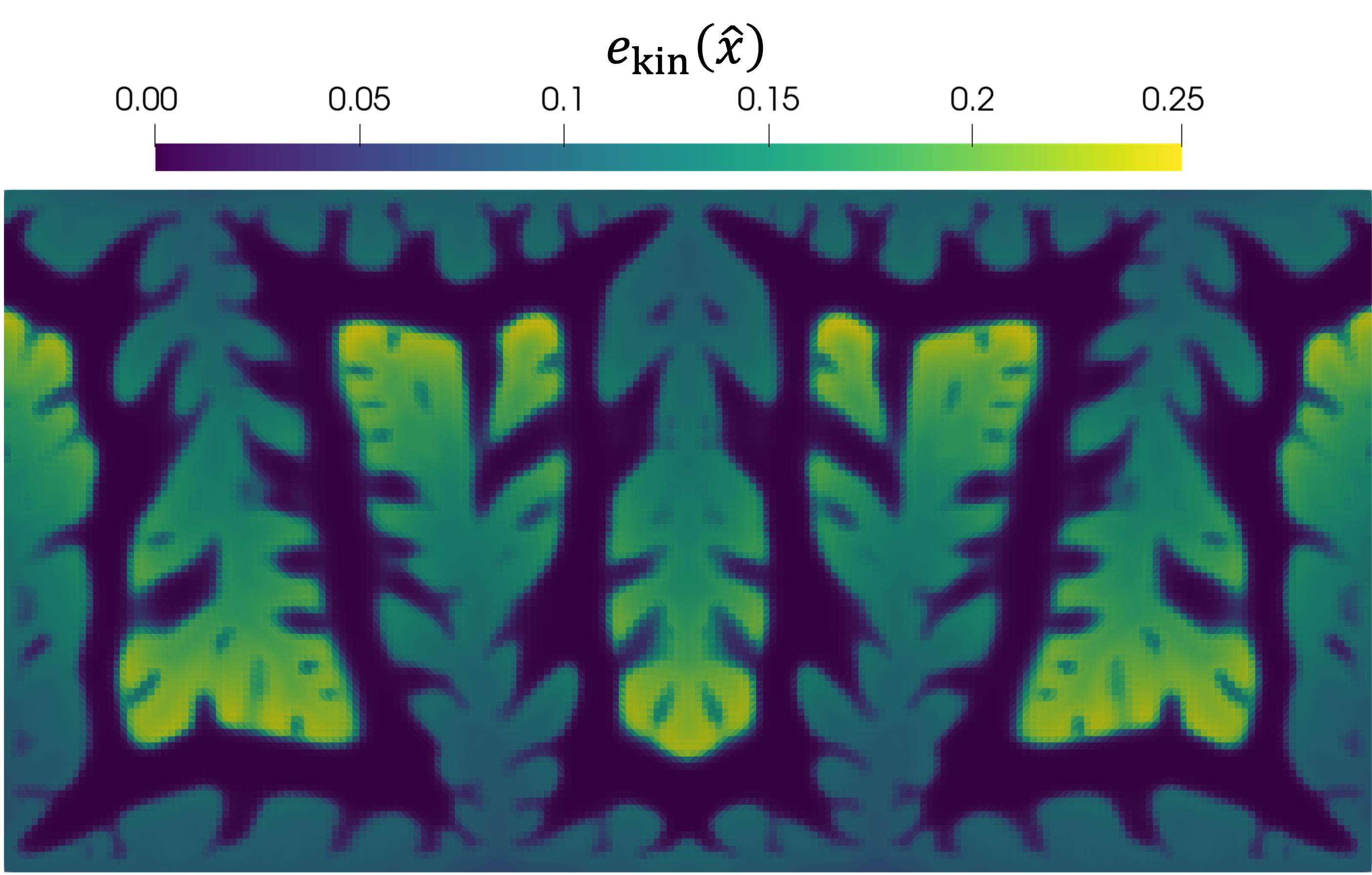}
        \caption{Optimized design}
    \end{subfigure}  
        \caption{Cross sections of the energy storage distribution in the 3D monolithic (a) and optimized (b) designs.}
    \label{fig:E_store_cross}
\end{figure}

In our final example, we optimize a 3D design with $\delta = 2$, $\gamma = 0.5$, $\lambda = 0.01$, and modified Bruggeman correlation.
The two optimized electrodes are similar and so we only illustrate the anode and its energy storage distribution, cf. Fig.~\ref{fig:3D_Estore}.
Similar protruding features of the cathode are interdigitated with these of the anode.
Throughout the design of both electrodes, we observe small penetrating holes, which increase interfacial area similar to the spikey teeth from the 2D designs.
From the cross sections of both electrodes plotted in Fig.~\ref{fig:cross-sections}, we observe the interdigitation similar to the 2D designs.
In addition, connections between the main pillar and the side are observed in certain locations, resulting in an arch-like structure, as observed in the cross section of the anode.
The region spanned by below the arch is filled with an arch from the other electrode, creating an interlocked structure as illustrated by the zoomed in view of Fig.~\ref{fig:interlock}.
A schematic summarizing the different design features is presented in Fig.~\ref{fig:structure_schematic}

To demonstrate superior material utilization of the optimized design, the energy storage distribution over a cross section is presented in Fig.~\ref{fig:E_store_cross}.
Similar to the 2D observations, only the regions neighboring the electrode/pure electrolyte interface have noticeable energy storage in the monolithic design.
Better material utilization is observed in the optimized design, wherein the stored energy is more uniformly distributed.
The total energy storage and ohmic loss is $814\%$ and $145\%$ greater than the monolith, which is slightly better than the corresponding 2D case ($798\%$ and $159\%$), possibly due to three-dimensional effects such as interlocking.

\rone{Obviously, the performance difference between the 3D and extruded 2D designs with the same physical inputs varies on a case-by-case basis. The 2D study investigates the effects of dimensionless parameters on the optimized design while the 3D study  demonstrates the scalability of our computational implementation. A more comprehensive 3D study is necessary to determine how the parameter choices affect the optimized 3D designs and the benefits of 3D versus extruded 2D designs.}

\section{Conclusion}
\label{sec:conclusion}
In this work, we present a density-based topology optimization strategy for the design of porous electrodes in electrochemical energy storage devices with Faradaic reactions and capacitive storage.
A full-cell model is utilized to simultaneously optimize the cathode and anode.

We present various 2D optimized designs to observe the changes in their features as we vary the value of dimensionless groups.
Interdigitation is observed in all optimized designs, which unlocks additional energy storage by allowing anode reactions to occur closer to the cathode current collector and vice versa and by enhancing ionic transport between the electrodes.
In general, larger energy storage improvement is acquired through electrode structure for $\delta = 5$ designs where reactions are localized near the electrode/pure electrolyte interface.
Among these cases, the designs also vary noticeably when changing between Faradaic ($\gamma = 1$) and capacitive storage-mechanisms ($\gamma = 0$).
$\lambda$, i.e., the ratio between ionic and total conductivity, controls the interdigitation depth.
We observe more shorter protrusions in designs with lower electronic conduction, i.e., as $\lambda \rightarrow 0.1$.
Designs generated using the standard Bruggeman correlation for effective transport possess smooth surfaces. 
The designs generated with modified Bruggeman correlation, where effective ionic transport is much slower, possess spiky teeth cutting into the main pillars to further boost the ion exchange between the electrode and the pure electrolyte.

The key performance metrics of the optimized designs are compared against those from monolithic electrodes.
We observe increased energy storage for all cases with the best improvement exceeding 700\% from the modified Bruggeman correlation cases.
Comparisons of energy storage and ohmic loss distributions suggest superior material utilization in the optimized designs.
In addition, the ohmic loss is more evenly distributed and thus does not grow significantly, even when the total charging current increases by an order of magnitude due to the decreased overall cell resistance from the optimized designs.
Finally, a 3D design is generated to further demonstrate the utility of the proposed optimization strategy.
More complex features, such as channels, holes, and interlocked arches between the electrodes are observed in this design.

\rtwo{Fabricating the intricate designs proposed here is a challenge, but not an insurmountable one.
Projection Micro Stereolithography (P\textmu SL) has been used to produce topology optimized electrodes \citep{batista2023design}.
Electrodes with planar interdigitated concentric and periodic features \citep{long20203d}, have successfully been fabricated by, direct-ink write printing \citep{sun2013print}, laser ablation \citep{proll2014laser}, and sequential deposition \citep{talin2016fabrication}.
Where necessary, manufacturing constraints can be incorporated into the topology optimization framework.
For example, in this work, the minimum feature size is controlled via a filtering technique Eqn.~\eqref{eq:filter}.
Manufacturing constraints can also be incorporated by modeling the fabrication process \citep{langelaar2017am}, e.g. to prevent unsupported overhangs \citep{gaynor2016support}.
}

As mentioned above, our current work does not track surface or solid concentrations.
To address this, our future work may incorporate diffusion in the solid particles as in the Doyle-Fuller-Newman model \citep{doyle1993galvano}.
Typically used for a 1D geometry, this model includes diffusion inside spherical particles, making it a pseudo-2D model.
For 3D geometries, this becomes a pseudo-4D model, further exacerbating the computational effort.

\section*{Funding information}
This work was performed under the auspices of the U.S. Department of Energy by Lawrence Livermore National Laboratory under Contract DE-AC52-07NA27344 and was supported by the LLNL-LDRD program under project numbers 20-ERD-019 and 23-SI-002. LLNL Release Number LLNL-JRNL-862205.

\section*{Declarations}
\subsection*{Conflict of interest}
On behalf of all authors, the corresponding author states that there is no conflict of interest.
\subsection*{Replication of results}
Sufficient information is presented within the manuscript for the readers to replicate the results.
Computer codes are subjected to internal control at the moment but will be open-source in the near future.

\appendix
\section{Energy balance derivation}
\label{sec:energy_balance}
We derive the energy balance for the dimensionless system with energy input, energy storage, and ohmic loss.
We first scale the electric and ionic potential equations \eqref{eq:electric_dimless} and \eqref{eq:ionic_dimless} such that the source terms are the same (modulo their signs):
\begin{equation}
  \label{eq:potential_scale}
  \begin{aligned}
  -\frac{1}{\lambda} \hat{\nabla}\cdot \big(\hat{\sigma}\hat{\nabla}\hat{\Phi}_1\big) =& -\delta_r \hat{a}\hat{i}_n - \delta_c \hat{a} \hat{i}_c, \\
  -\frac{1}{1-\lambda}\hat{\nabla}\cdot (\hat{D} \hat{c}\hat{\nabla} \hat{\Phi}_2) - \frac{1}{1-\lambda}&\left(\frac{t_{0,+}}{z_+} + \frac{t_{0,-}}{z_-}\right) \\
  \hat{\nabla}\cdot(\hat{D} \hat{\nabla}\hat{c}) =& \delta_r \hat{a}\hat{i}_n +\delta_c \hat{a} \hat{i}_c.
  \end{aligned}
\end{equation}
Since $\hat{\Phi}_1=0$ on $\hat{\Gamma}_a$ and $\nabla \hat{\Phi}_1 \cdot \mb{n}= 0$  on $\partial\hat{\Omega} \setminus (\hat{\Gamma}_a \bigcup \hat{\Gamma}_c)$ the dimensionless energy input can be expressed as
\begin{equation}
  \label{eq:energy_in}
  \begin{aligned}
  E_{\mrm{in}} &= \frac{1}{\lambda} \int_0^{\hat{T}_{\mrm{f}}}\int_{\hat{\Gamma}_c} \hat{\sigma} \hat{\nabla}\hat{\Phi}_1 \cdot \mb{n}\hat{\Phi}_1 \diff\hat{s} \diff\hat{t} \\
  &=\frac{1}{\lambda} \int_0^{\hat{T}_{\mrm{f}}}\int_{\partial\hat{\Omega}} \hat{\sigma} \hat{\nabla}\hat{\Phi}_1 \cdot \mb{n}\hat{\Phi}_1 \diff\hat{s} \diff\hat{t},
  \end{aligned}
\end{equation}
which extends integral from the cathode current collector $\hat{\Gamma}_c$ to the entire boundary $\partial\hat{\Omega}$.
With $\hat{\nabla} \hat{\Phi}_2 \cdot \mb{n} =0$ and $\hat{\nabla}\hat{c}\cdot \mb{n} =0$ on $\partial\hat{\Omega}$, the dimensionless energy input can be further extended as
\begin{equation}
  \label{eq:energy_in_full}
  \begin{aligned}
  &E_{\mrm{in}} = \frac{1}{\lambda} \int_0^{\hat{T}_{\mrm{f}}}\int_{\partial\hat{\Omega}} \hat{\sigma} \hat{\nabla}\hat{\Phi}_1 \cdot \mb{n}\hat{\Phi}_1 \diff\hat{s} \diff\hat{t} \\
  &+ \frac{1}{1-\lambda}\int_0^{\hat{T}_{\mrm{f}}}\int_{\partial\hat{\Omega}}\hat{D} \hat{c} \hat{\nabla} \hat{\Phi}_2\cdot \mb{n}\hat{\Phi}_2 \diff\hat{s} \diff\hat{t} \\
  &+ \frac{1}{1-\lambda}\left(\frac{t_{0,+}}{z_+} + \frac{t_{0,-}}{z_-}\right)\int_0^{\hat{T}_{\mrm{f}}}\int_{\partial\hat{\Omega}}\hat{D} \hat{\nabla}\hat{c} \cdot \mb{n}\hat{\Phi}_2 \diff\hat{s} \diff\hat{t},
  \end{aligned}
\end{equation}
Applying the divergence theorem we obtain
\begin{equation}
  \label{eq:div_balance}
  \begin{aligned}
  &E_{\mrm{in}} = \frac{1}{\lambda} \int_0^{\hat{T}_{\mrm{f}}}\int_{\hat{\Omega}} \hat{\nabla}\cdot\left(\hat{\sigma} \hat{\nabla}\hat{\Phi}_1 \hat{\Phi}_1\right) \diff\hat{x} \diff\hat{t} \\
  &+ \frac{1}{1-\lambda} \int_0^{\hat{T}_{\mrm{f}}}\int_{\hat{\Omega}} \hat{\nabla}\cdot \left(\hat{D} \hat{c} \hat{\nabla} \hat{\Phi}_2 \hat{\Phi}_2\right)\diff\hat{x} \diff\hat{t} \\
  &+ \frac{1}{1-\lambda}\left(\frac{t_{0,+}}{z_+} + \frac{t_{0,-}}{z_-}\right) \int_0^{\hat{T}_{\mrm{f}}}\int_{\hat{\Omega}} \hat{\nabla}\cdot \left(\hat{D} \hat{\nabla}\hat{c}\hat{\Phi}_2\right)\diff\hat{x} \diff\hat{t}.
  \end{aligned}
\end{equation}
An application of the product rule subsequently gives
\begin{equation}
\label{eq:product_balance}
    \begin{aligned}
    &E_{\mrm{in}} =\frac{1}{\lambda} \int_0^{\hat{T}_{\mrm{f}}}\int_{\hat{\Omega}} \hat{\sigma} \hat{\nabla}\hat{\Phi}_1 \cdot \hat{\nabla}\hat{\Phi}_1 \diff\hat{x} \diff\hat{t} \\
    &+ \frac{1}{\lambda} \int_0^{\hat{T}_{\mrm{f}}}\int_{\hat{\Omega}} \hat{\nabla}\cdot(\hat{\sigma} \hat{\nabla}\hat{\Phi}_1) \hat{\Phi}_1 \diff\hat{x} \diff\hat{t} \\
    &+ \frac{1}{1-\lambda} \int_0^{\hat{T}_{\mrm{f}}}\int_{\hat{\Omega}}\hat{D} \hat{c} \hat{\nabla} \hat{\Phi}_2 \cdot \hat{\nabla} \hat{\Phi}_2 \diff\hat{x} \diff\hat{t} \\
    &+ \frac{1}{1-\lambda} \int_0^{\hat{T}_{\mrm{f}}}\int_{\hat{\Omega}} \hat{\nabla}\cdot (\hat{D} \hat{c} \hat{\nabla} \hat{\Phi}_2) \hat{\Phi}_2 \diff\hat{x} \diff\hat{t} \\
    &+ \frac{1}{1-\lambda}\left(\frac{t_{0,+}}{z_+} + \frac{t_{0,-}}{z_-}\right)\int_0^{\hat{T}_{\mrm{f}}}\int_{\hat{\Omega}} \hat{D} \hat{\nabla}\hat{c} \cdot \hat{\nabla} \hat{\Phi}_2 \diff\hat{x} \diff\hat{t}\\
    &+\frac{1}{1-\lambda}\left(\frac{t_{0,+}}{z_+} + \frac{t_{0,-}}{z_-}\right)\int_0^{\hat{T}_{\mrm{f}}}\int_{\hat{\Omega}} \hat{\nabla}\cdot(\hat{D} \hat{\nabla}\hat{c})\hat{\Phi}_2 \diff\hat{x} \diff\hat{t}.
    \end{aligned}
\end{equation}
Combining \eqref{eq:potential_scale} and \eqref{eq:product_balance} yields
\begin{equation}
  \label{eq:energy_balance}
  \begin{aligned}
  &E_{\mrm{in}} =\frac{1}{\lambda} \int_0^{\hat{T}_{\mrm{f}}}\int_{\hat{\Omega}} \hat\sigma  \hat{\nabla}\hat{\Phi}_1 \cdot \hat{\nabla}\hat{\Phi}_1 \diff\hat{x} \diff\hat{t} \\
  &+ \frac{1}{1-\lambda} \int_0^{\hat{T}_{\mrm{f}}}\int_{\hat{\Omega}}\hat D \hat{c} \hat{\nabla} \hat{\Phi}_2 \cdot \hat{\nabla} \hat{\Phi}_2 \diff\hat{x} \diff\hat{t} \\
  &+ \frac{1}{1-\lambda}\left(\frac{t_{+}}{z_+} + \frac{t_{-}}{z_-}\right)\int_0^{\hat{T}_{\mrm{f}}}\int_{\hat{\Omega}} \hat D \hat{\nabla}\hat{c} \cdot \hat{\nabla} \hat{\Phi}_2 \diff\hat{x} \diff\hat{t} \\
  &+\delta_r \int_0^{\hat{T}_{\mrm{f}}}\int_{\hat{\Omega}} \hat{a}\hat{i}_n(\hat{\Phi}_1-\hat{\Phi}_2)\diff\hat{x} \diff\hat{t} \\
  &+\delta_c \int_0^{\hat{T}_{\mrm{f}}}\int_{\hat{\Omega}} \hat{a}\hat{i}_c(\hat{\Phi}_1-\hat{\Phi}_2)\diff\hat{x} \diff\hat{t},
  \end{aligned}
\end{equation}
which is equivalent to $E_{\mrm{in}}=E_{\mrm{kin}}+E_{\mrm{ohm}}$.

\section{Mixed discretizations}
\label{sec:mixed}
A mixed formulation is utilized to solve the filter problem \eqref{eq:filter} to obtain piecewise constant $\tilde\rho$ over triangular and tetrahedral meshes.
In this way, we are sure that $\tilde \rho (\mb x)\in [0,1]$.
Indeed, the mixed method enforces flux continuity across the element interfaces, which prevents oscillations in the filtered field as the control field goes to $\{0,1\}$ \citep{Salazar_de_Troya_2020}.
As opposed to the two-point flux approximation scheme from \cite{roy2022tope}, no orthogonality condition is required on the mesh regularity.

With $r^2 > 0$, we define the auxiliary flux term as
\begin{equation}
    r^{-2}\boldsymbol{u}_f = -\nabla \tilde\rho.
\end{equation}
Then the filter problem \eqref{eq:filter} is rewritten as
\begin{equation}
    \begin{aligned}
        r^{-2}\boldsymbol{u}_f + \nabla \tilde\rho &= 0 \quad &&\text{in }\Omega, \\
        -\nabla\cdot \boldsymbol{u}_f - \tilde\rho &= -\rho \quad &&\text{in }\Omega, \\
        \boldsymbol{u}_f \cdot \mb{n} &= 0 \quad && \text{on }\partial\Omega.
    \end{aligned}
\end{equation}
The function spaces for the mixed weak variational form are
\begin{equation}
    \boldsymbol{V} = \left\{\boldsymbol v \in H(\mrm{div},\Omega)\; \middle| \; \boldsymbol v \cdot \mb n = 0 \text{ on } \partial \Omega\right\},\;\; W = L^2(\Omega),
\end{equation}
where the Neumann flux boundary condition are imposed strongly.
We seek to find $\left(\boldsymbol{u}_f, \tilde\rho \right)  \in \boldsymbol{V}\times W$ such that
\begin{equation}
    \begin{aligned}
        (r^{-2}\boldsymbol{u}_f, \boldsymbol{v})_{\Omega} &- (\tilde\rho, \nabla\cdot \boldsymbol{v})_{\Omega} = 0,\\
        -(\nabla\cdot \boldsymbol{u}_f, w)_{\Omega} &- (\tilde\rho, w)_{\Omega} = -(\rho, w)_{\Omega}, 
    \end{aligned}
    \label{eq:mixed_variational_filter}
\end{equation}
$\forall \left(\boldsymbol{v}, w\right) \in \boldsymbol{V}\times W$.

A similar mixed formulation is used for solving the boundary propagation PDE \eqref{eq:bd_prop}.
From \eqref{eq:bd_prop}, since $\bar\rho$ is strictly positive, we define the auxiliary flux term as
\begin{equation}
    \bar\rho^{-1} \boldsymbol{u}_b = -\nabla \beta.
\end{equation}
Then the continuous boundary propagation problem can be reformulated as
\begin{equation}
    \begin{aligned}
        \bar\rho^{-1} \boldsymbol{u}_b + \nabla \beta &= 0 \quad &&\text{in }\Omega, \\
        -\nabla\cdot \boldsymbol{u}_b - (1-\bar\rho)\beta &= 0 \quad &&\text{in }\Omega, \\
        \beta &= 1 \quad &&\text{on }\Gamma_\mathrm{a}, \\
        \beta &= -1 \quad &&\text{on }\Gamma_\mathrm{c}, \\
        \boldsymbol{u}_b \cdot \mb{n} &= 0 \quad &&\text{on }\partial\Omega/\Gamma_\mathrm{c}\cup\Gamma_\mathrm{a}.
    \end{aligned}
\end{equation}
Let $\Gamma_D = \Gamma_a \cup \Gamma_c$.
We introduce the function space
\begin{equation}
    \boldsymbol{V}_D = \left\{\boldsymbol v \in H(\mrm{div},\Omega) \middle| \boldsymbol v \cdot \mb n = 0 \text{ on } \Gamma_D \right\}.
\end{equation}
We seek to find $\left(\boldsymbol{u}_b, \beta\right) \in \boldsymbol{V}_D\times W$ such that
\begin{equation}
    \begin{aligned}
        (\bar\rho^{-1} \boldsymbol{u}_b, \boldsymbol{v})_{\Omega} - (\beta,\nabla\cdot\boldsymbol{v})_{\Omega} &=  \\
        -\langle 1,\boldsymbol{v}\cdot\mb{n}\rangle_{\Gamma_\mathrm{a}} -& \langle -1,\boldsymbol{v}\cdot\mb{n}\rangle_{\Gamma_\mathrm{c}},\\
        -(\nabla\cdot \boldsymbol{u}_b, w)_{\Omega} - ((1-\bar\rho)\beta, w)_{\Omega} &= 0, 
    \end{aligned}
    \label{eq:mixed_variational_prop}
\end{equation}
$\forall \left(\boldsymbol{v}, w \right) \in \boldsymbol{V}_D \times W$.
Both problems use Raviart-Thomas basis functions with the lowest order to obtain the discretized solution, which results in piecewise constant $\tilde\rho$ and $\beta$ and piecewise linear $\boldsymbol{u}_f$ and $\boldsymbol{u}_b$ with continuous normal flux across the element interfaces.

For the three-dimensional case, we use an iterative method to solve the linear systems resulting from the weak forms described by \eqref{eq:mixed_variational_filter} and \eqref{eq:mixed_variational_prop}.
These systems have symmetric block matrices of the form
\begin{equation}\label{eq:saddle}
    \begin{bmatrix}
        A & B\\
        B^\top & D\\
    \end{bmatrix}
\end{equation}
where $A$ represents a weighted mass matrix for the flux, $D$ is a weighted mass matrix for $\tilde\rho$ or $\beta$, and $B$ is the coupling term.
The approach introduced in \cite{wathen1994precondition} is used to solve the block system efficiently.
The block matrix \eqref{eq:saddle} admits a factorization of the type
\begin{equation}
    \begin{bmatrix}
        A & B\\
        B^\top & D\\
    \end{bmatrix}
    =
    \begin{bmatrix}
        I & 0 \\
        B^\top A^{-1} & I 
    \end{bmatrix}
    \begin{bmatrix}
        A & 0 \\
        0 & S 
    \end{bmatrix}
    \begin{bmatrix}
        I & A^{-1} B \\
        0 & I
    \end{bmatrix},
\end{equation}
where $S = D - C A^{-1} B$ is called the \emph{Schur complement}.
The inverse of the block matrix is
\begin{multline}
    \begin{bmatrix}
        A & B\\
        B^\top & D\\
    \end{bmatrix}^{-1}
    \\=
    \begin{bmatrix}
        I & -A^{-1} B \\
        0 & I
    \end{bmatrix}
    \begin{bmatrix}
        A^{-1} & 0 \\
        0 & S^{-1} 
    \end{bmatrix}
    \begin{bmatrix}
        I & 0 \\
        -B^\top A^{-1} & I 
    \end{bmatrix},
\end{multline}
which requires inverting the blocks $A$ and $S$.
Our preconditioner approximates the blocks of this factorization.
We approximate the Schur complements using the diagonal of $A$, i.e., we use the sparse $\tilde S = D - C \mathrm{diag}(A)^{-1} B$.
The inverse of $\tilde S$ is approximated by Algebraic Multigrid \citep{ruge1987algebraic}, and the inverse of $A$ is approximated by a block incomplete LU factorization with 0 fill (ILU0) \citep{meijerink1977iterative}.
The \texttt{PETSc} solver options used for the three-dimensional simulations are given in Appendix \ref{sec:petsc}.

\section{PETSc solver options}
\label{sec:petsc}
Here we provide he \texttt{PETSc} solver options used for the three-dimensional case.
The options for solving the governing physical system are:
\begin{lstlisting}
    "snes_rtol": 0,
    "snes_atol": 1e-4,
    "mat_type": "aij",
    "ksp_type": "gmres",
    "ksp_rtol": 1e-4,
    "ksp_gmres_restart": 30,
    "pc_type": "fieldsplit",
    "pc_fieldsplit_type": "multiplicative",
    "fieldsplit_pc_type": "hypre",
    "fieldsplit_pc_hypre_boomeramg": {
        "strong_threshold": 0.7,
        "coarsen_type": "HMIS",
        "agg_nl": 3,
        "interp_type": "ext+i",
        "agg_num_paths": 5,
        },
\end{lstlisting}
The options for the mixed formulation used for the filtering and boundary propagation problems are:
\begin{lstlisting}
    "mat_type": "nest",
    "ksp_type": "gmres",
    "ksp_rtol": 1e-4,
    "pc_type": "fieldsplit",
    "pc_fieldsplit_type": "schur",
    "pc_fieldsplit_schur_fact_type": "full",
    "fieldsplit_0_ksp_type": "preonly",
    "fieldsplit_0_pc_type": "bjacobi",
    "fieldsplit_0_sub_pc_type": "ilu",
    "fieldsplit_1_ksp_type": "preonly",
    "pc_fieldsplit_schur_precondition": "selfp",
    "fieldsplit_1_pc_type": "hypre",
    "fieldsplit_1_pc_hypre_boomeramg": {
        "strong_threshold": 0.7,
        "coarsen_type": "HMIS",
        "agg_nl": 3,
        "interp_type": "ext+i",
        "agg_num_paths": 5,
        }.
\end{lstlisting}

\bibliography{ref.bib}

\end{document}